\documentclass[12pt,titlepage]{book}

\usepackage{a4wide,epsfig,pstricks,amssymb}

\usepackage{times}

\usepackage[english,polish]{babel}

\newcommand{\prtext}[1]{\mbox{\rm #1}}
\newcommand{\mirror}{\psline[linewidth=2pt](-.7,0)(.7,0)}
\newcommand{\beamsplitter}{\psline[linewidth=2pt,linecolor=gray]%
(-.7,0)(.7,0)}

\begin{document}

\selectlanguage{english}

\renewcommand{\figurename}{\small Figure}

\addtolength{\baselineskip}{0.1\baselineskip}

\begin{titlepage}
\begin{center}

\vspace*{1.5cm}

\epsfig{file=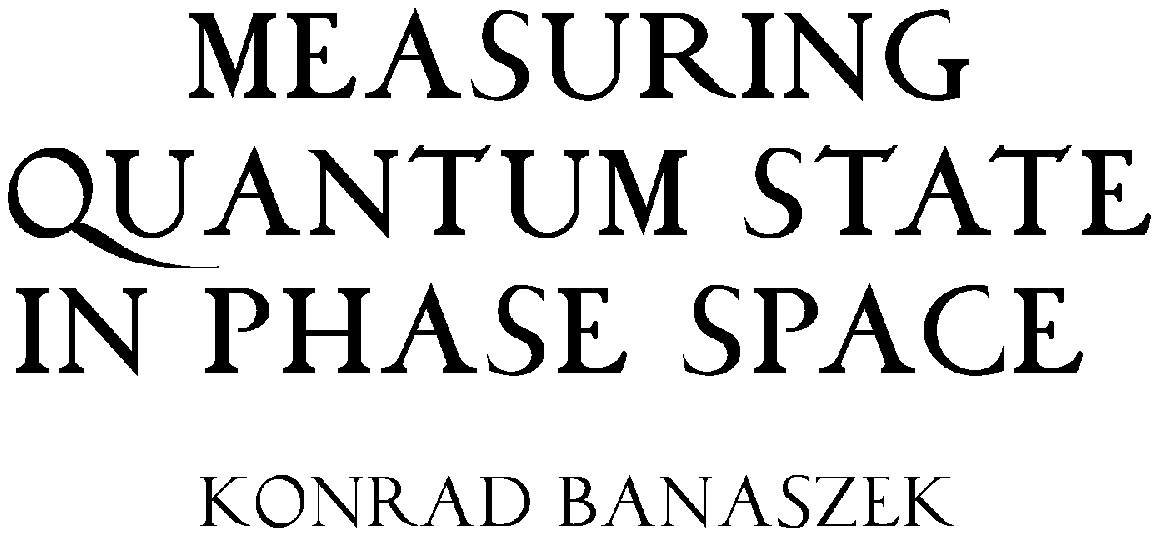}

\end{center}

\vspace*{20mm}

\noindent\hspace{8cm}\begin{minipage}{8cm}
Rozprawa doktorska przygotowana\\
w Instytucie Fizyki Teoretycznej\\
Uniwersytetu Warszawskiego\\
pod kierunkiem Prof.\ Krzysztofa\\
W\'{o}dkiewicza
\end{minipage}

\vspace*{\fill}

\begin{center}
\epsfig{file=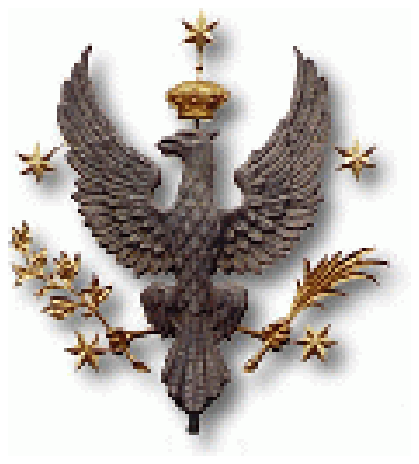}

\vspace{5pt}

{\large Wydzia{\l} Fizyki\\
\large Uniwersytet Warszawski\\[2pt]
\large Warszawa 1999}
\end{center}

\end{titlepage}

\thispagestyle{empty}\vspace*{\fill}

\thispagestyle{empty}\vspace*{\fill}

\setcounter{page}{0}

\cleardoublepage

\markboth{MEASURING QUANTUM STATE IN PHASE SPACE}{CONTENTS}

\tableofcontents

\chapter{Introduction}

\markright{CHAPTER \thechapter . \uppercase{Introduction}}

The development of quantum mechanics was connected with one of
the greatest conceptual leaps in theoretical physics. In order
to describe properly microscopic phenomena, it was necessary to
abandon the classical notion of a physical property, and to
resort to a completely new formalism representing microobjects.
This formalism lies at the heart of quantum mechanics.
It makes a clear distinction between the state of a
physical system, and the performed observation. The state is
characterized by a wave function, whose time evolution is
governed by an appropriate equation of motion (e.g.\ the
Schr\"{o}dinger equation for a nonrelativistic particle). If we
want to relate the state to  quantities observed in an
experiment, we need to use the second element of the quantum
mechanical formalism, i.e.\ the representation of the measuring
apparatus in terms of operators acting on the wave function.
According to the Born interpretation,
quantum mechanics provides probabilistic predictions concerning
the behaviour of a quantum system \cite{BornZfP26}.
Quantum description of an
experiment specifies only the chance that the measurement will
yield a given outcome. In order to verify such predictions, we
need to repeat the measurement many times on identically prepared
systems, and then to compare the histogram of experimental
outcomes with the probability distribution calculated from the
theory. The result of a single measurement cannot be described
in a deterministic way. This randomness seems to be a very
fundamental feature of the microworld. So far, all attempts to
introduce deterministic description of microscopic phenomena
have failed, and experiments have ruled out whole classes of
theories alternative to quantum mechanics.

Pictorially speaking, the click on a measuring apparatus is only
a faint shadow of the quantum state, which stays hidden behind
the scene, though being the main actor.
However, one might ask if it is possible to reveal experimentally
the complete information on the state of a quantum system.
This cannot be the case if we are given only a single copy
of the system. Any measurement consists in a certain kind of
interaction between the system and the detecting apparatus. This
interaction maps some properties of the measured system onto the
state of the apparatus, and makes them accessible to our
cognition. After the measurement, the state of the system is
perturbed by the interaction with the detector, and it is no
longer described by the original wave function. For example,
observation of the position of a particle by scattering photons
inevitably modifies its momentum. The deleterious character of
quantum measurement was realized very early in the development
of quantum mechanics, and it is closely related to
the Heisenberg uncertainty principle \cite{Heisenberg}.
Thus, our system can
be detected only once, and we cannot gain more information by
repeating the measurement. One might try to circumvent this
difficulty by designing a
single measurement that would yield complete information on the
quantum state. Such a measurement would map each possible state
of the system onto a different, fully distinguishable state of
the apparatus. This means that all the final states of the
apparatus would have to be mutually orthogonal, even these
corresponding to nonorthogonal initial states of the measured
system. Of course, this would violate unitarity of quantum
evolution, and such a measurement does not exist. One could also
consider cloning of quantum states, i.e.\ an operation that would
generate two or more identical copies from a given system. Then,
one could increase the amount of information by performing
measurements on the reproduced copies. Such a strategy fails
because of the no-cloning theorem \cite{WootZureNAT82},
which is a simple consequence
of the quantum superposition principle.

Nevertheless, no principles of quantum mechanics prevent us from
characterizing the quantum state of an {\em ensemble} of
physical systems. By repeating the measurement many times on
individual copies, we may arrive at reliable information on the
properties of the ensemble. The aim of quantum state measurement
can be formulated in technological terms: suppose we have a
machine producing identically prepared copies of a quantum
system. When delivering this output for some application, we
should be able to provide its specification, which on the most
complete level means a full characterization of the quantum state.
Such a problem, apart from interesting fundamental aspects, is
currently of practical interest in many areas of science.  This
is due to extensive studies devoted presently to preparation,
manipulation, and control of quantum systems. The motivation of
this research is to overcome current technological limitations
by exploiting fully possibilities offered by quantum mechanics.
Let us mention just few examples. The yield of chemical
reactions can be increased by controlling the quantum state
of reactants \cite{ZareSCI98}. Application of so-called
squeezed states of light improves precision of interferometric
measurements, which can be used to enhance sensitivity of
gravitational wave detectors \cite{WallsMilburnGravitation}.
Coherent preparation and manipulation of entangled multiparticle
systems can be used to solve computational problems intractable
by classical computers \cite{DeutEkerPW98}.
An important matter in developing these
and other technologies is the possibility to gain extensive and
reliable information on the state of quantum systems. The
ultimate tool for this purpose is the measurement of the
complete quantum state.

When characterizing ensembles, we usually need to take into
account the possibility of statistical fluctuations, and to
describe their state using the more general concept of the
density matrix rather that a wave function. From the formal
point of view, the task of characterizing the quantum state can
be accomplished by a set of appropriately chosen measurements,
which would extract unambiguous information on all the elements
of the density matrix.  The crucial question, however, is: how
to do this in practice? The challenge of quantum state
measurement has several important aspects. The first one is
design and realisation of measurement schemes that yield
complete characterization of the quantum state. Further, there
is a nontrivial task of extracting precise information on the
quantum state from data collected in a realistic, imperfect
experimental setup. Finally, we have said above that expectation
values of a sufficiently large set of observables contain
complete information on the density matrix. The problem is that
true expectation values are obtained only in the limit of the
infinite number of measurements. In a real laboratory we always
deal with {\em finite} ensembles, and this is our source of
information on the quantum state, which we would like to use as
efficiently as possible. 

In quantum optics and related fields, several examples of simple quantum
systems have been thoroughly studied, including a single light mode,
a trapped ion, and a diatomic molecule. A lot of interest has been paid
to detection of subtle quantum statistical effects. It was therefore
natural that the domain of quantum state measurement has grown mainly on
the ground of advances in quantum optics. In 1993, the group of Michael
Raymer at the University of Oregon demonstrated complete experimental
characterization of the quantum state of a single light mode by means
of optical homodyne tomography \cite{SmitBeckPRL93}.
This seminal experiment was followed by extensive research in the
field of quantum state measurement. Over past several years, we have
witnessed a series of beautiful experiments with various quantum systems
\cite{FreyBardPW97}. The vibrational state of a diatomic molecule has been
reconstructed from measurements of the time-dependent fluorescence
spectrum \cite{DunnWalmPRL95}.  Optical homodyne tomography has
been applied to reconstruct a whole gallery of squeezed states of
light \cite{BreiSchiNAT97}. The motional state of a trapped ion has
been characterized using a very sophisticated technique based on the
monitoring of the fluorescence \cite{LeibMeekPRL96}. The tomographic
method has been used to measure the transverse motional state of an
atomic beam \cite{KurtPfauNAT97}. All these experiments were tightly
connected with the rapid theoretical development of the domain of quantum
state measurement. Numerous measurement schemes have been proposed and
analysed in detail. In particular, the role of statistical uncertainty
has been discussed, and various approaches to reconstructing quantum
state representations from experimental data have been described.

The subject of this thesis is the measurement of the quantum
state in the phase space. The concept of the phase space
provides a bridge between the quantum mechanical formalism and
classical physics. Predictions of quantum mechanics have
essentially statistical character. As a rule, one can predict
only probabilities of obtaining specific outcomes of
a measurement. Such a situation can be encountered in
classical mechanics as well. For example, if we deal with an
ensemble of classical systems, properties of a single copy
can be defined only in statistical terms. The state of the
ensemble is characterized by a phase space distribution, which
describes the probability of occupying a given volume element
by the system.  One may wonder whether this intuitive picture
of fluctuations has its counterpart in quantum physics. The
answer to this question is not straightforward. It is possible
to convert the quantum mechanical formalism into a form which
resembles a classical statistical theory.  The first phase
space representation of the quantum state was introduced in
1932 by Wigner \cite{WignPR32}.  However, such a phase space
representation is not unique: noncommutativity of quantum
observables leads to abundance of quantum analogs of the phase
space distribution, and none of them captures all the properties
of the classical object \cite{HillOConPRep84}. This is a
manifestation of the fact that quantum mechanics is essentially
different from a classical theory.  Nevertheless, quantum
phase space quasidistributions contain complete information on
the quantum state. The family of quasidistribution functions
provides a convenient framework for studying many quantum
optical problems. It is also a useful tool in visualising
quantum coherence and interference phenomena.

For a long time, quantum quasidistribution functions have been
considered mainly as a quite odd theoretical concept rather than
a quantity which can be measured in a feasible experimental
scheme. This perspective changed completely with the
demonstration of optical homodyne tomography, which brought
quasidistributions, in particular the Wigner function, to the
realm of a physical laboratory. Optical homodyne tomography is
based on the observation that marginal distributions of the
Wigner function of a light mode can be measured by means of
homodyne detection. The inverse problem, i.e.\ the retrieval of
the Wigner function from its projections, is similar to the
procedure used in medical imaging, where the spatial
distribution of the tissue is reconstructed from absorption
measured across the body. Practical implementation of the
reconstruction algorithm is a rather complex and delicate
matter: the back-projection transformation is singular, and its
application to experimental data has to be accompanied by a
special filtering procedure. Demonstration of optical homodyne
tomography was a successful combination of a precise
quantum optical measurement with sophisticated data processing.

In this work we develop a novel, entirely different approach
to measuring quasidistribution functions of light. We exploit
the fact that the value of a quasidistribution at a given point
of the phase space is itself a well defined quantum observable.
Motivated by this representation, we propose and demonstrate an
optical scheme for measuring {\em directly} quasidistribution
functions. This method, based on photon counting, avoids the
detour via complex numerical reconstruction algorithm. The
basic elements of our measurement scheme are very simple. The
light mode whose quantum state we want to measure is interfered
with an auxiliary coherent probe field, and a photon counting
detector is used to measure the photocount statistics of the
superposed fields. We show that a simple arithmetic operation
performed on the measured photocount statistics yields directly
the value of the quasidistribution at a point defined by the
amplitude and the phase of the probe field. By changing these
two parameters of the probe field, we may scan the complete
phase space, and obtain the full representation of the quantum
state of the measured light mode. We demonstrate an experimental
realisation of this scheme, and present measurements of the
Wigner function for several quantum states of light. The
experimental part of this thesis has been performed in
Division of Optics, Institute of Experimental Physics, Warsaw
University, in collaboration with Prof. Czes{\l}aw Radzewicz.

We shall study here in detail various aspects of the direct scheme for
measuring quasidistribution functions. On the practical side, there is a
question about the role of typical experimental imperfections. We shall
analyse how the result of the measurement is affected by such factors as
non-unit detection efficiency and imperfect interference visibility.
We shall also provide estimates for statistical error, which are
necessary to design an accurate experiment, and to specify confidence of
the experimental outcome. These theoretical results will be an important
tool for quantitative analysis of the performed experiment.  In addition,
our discussion of practical aspects has interesting consequences in
the recently disputed problem of compensating for detector losses
in photodetection measurements. Over past several years, there were
conflicting claims concerning the possibility of removing deleterious
effects of imperfect detection by appropriate numerical processing
of experimental data \cite{KissHerzPRA95,DAriMaccPRA98,KissHerzPRA98}.
Our measurement scheme provides a testing ground for this problem. We will
show that in general no compensation for detection losses is possible,
unless some {\em a priori} knowledge about the measured quantum state
is given. Discussion of this problem reveals the fundamental role of
statistical uncertainty in realistic quantum measurements, which results
from the fact that in a laboratory we always deal with finite ensembles.

An attractive feature of the presented approach to measuring
quasidistribution functions of a single light mode is the direct link
between the measured observable and the quantum state representation. One
may wonder whether this approach can be applied in other situations. We
shall describe here a generalization of the measurement scheme to
multimode radiation. We shall demonstrate that multimode quasidistribution
functions are also directly related to the photocount statistics, and that
they can be determined in an equally simple way.  Although our interest
in this thesis will be confined to detection of optical radiation,
it should be noted that the idea underlying our measurement scheme has
proven to be fruitful in the measurement of the vibrational state of
a trapped ion \cite{LeibMeekPRL96}. It has also motivated measurement
schemes for a cavity mode \cite{LuttDaviPRL97} and a diatomic molecule
\cite{DaviOrszPRA98}.

This thesis is organized as follows. In Chap.~\ref{Chap:Representations}
we review phase space representations of the quantum state. Starting from
the definition of the Wigner function, we show how this distribution
can be unified with other phase space representations, and we discuss
properties of generalized $s$-ordered quasidistribution functions.
In Chap.~\ref{Chap:Homodyne} we review briefly previous work on measuring
the quantum state of light. We present two techniques which have been
realized in experiments: optical homodyne tomography and balanced
homodyne detection. Next, in Chap.~\ref{Chap:Direct}, we introduce the
direct method for measuring quasidistribution functions of light. We
present the phase space picture of the measurement, and we develop the
multimode theory of the scheme. Various practical aspects of the proposed
measurement are discussed in Chap.~\ref{Chap:Practical}, including the
effects of imperfect detection, and the possibility of compensation for
detector losses.  In Chap.~\ref{Chap:Experiment} we present experimental
realization of the proposed scheme, and demonstrate the direct measurement
of the Wigner function of a single light mode. The issue of statistical
uncertainty in photodetection measurements is discussed from a more
general point of view in Chap.~\ref{Chap:Statistical}.  We show that
the statistical noise sometimes limits available information on the quantum
state. Finally, Chap.~\ref{Chap:Conclusions} concludes the thesis.
Major part of original results presented in this thesis has been
published in the following articles:
\begin{itemize}
\item
K. Banaszek and K. W\'{o}dkiewicz,\\
{\em Direct probing of quantum phase space by photon counting},\\
Phys.\ Rev.\ Lett.\ {\bf 76}, 4344 (1996).

\item
K. Banaszek and K. W\'odkiewicz\\
{\em Accuracy of sampling quantum phase space in a photon counting
experiment},\\
J. Mod.\ Opt.\ {\bf 44}, 2441 (1997).

\item
K. Banaszek, C. Radzewicz, K. W\'{o}dkiewicz, and J. S. Krasi\'{n}ski,\\
{\em Direct measurement of the Wigner function by photon counting},\\
Phys.\ Rev.\ A {\bf 60}, 674 (1999).

\item
K. Banaszek,\\
{\em Statistical uncertainty in
quantum-optical photodetection measurements},\\
J. Mod.\ Opt.\ {\bf 46}, 675 (1999).
\end{itemize}

\vspace*{\fill}

\section*{Acknowledgements}

First and foremost, I thank my supervisor, Prof.\
\foreignlanguage{polish}{Krzysztof W"odkiewicz}, for guiding me
throughout my PhD studies, and for providing numerous comments and
suggestions on drafts of this thesis. It has been both a pleasure and a
privilege to share his enthusiasm for scientific research. I am also
indebted to Prof.\ \foreignlanguage{polish}{Czes"law Radzewicz} and Prof.\
\foreignlanguage{polish}{Jerzy S. Krasi"nski}, with whom I collaborated
on the experiment, for teaching me how things {\em really} work.
My special thanks go to
Prof.\ \foreignlanguage{polish}{Kazimierz Rz"a"rewski}
and
Prof.\ \foreignlanguage{polish}{Jan Mostowski}.
I owe them my first encounters with quantum optics.

Some of the results presented in this thesis were obtained during my
stay in Laser Optics and Spectroscopy Group at Imperial College,
London. I would like to thank the Head of the Group, Prof.\ Peter
L. Knight FRS, as well as all its members, for making my stay so fruitful
and enjoyable.  In understanding foundations of quantum state measurement,
I have benefited a lot from the visit at Universit\`{a} di Pavia, and
collaboration with Prof.\ G. Mauro D'Ariano, Dr.\ Matteo Paris, and Dr.\
Massimiliano Sacchi.

Finally, I would like to thank the Foundation for Polish Science for
the Domestic Grant for Young Scholars.

\chapter{Phase space representations of quantum state}

\label{Chap:Representations}

\markright{CHAPTER \thechapter . \uppercase{Phase space representations
of quantum state}}

In the standard formulation of quantum mechanics, the quantum state is
characterized by a vector from the Hilbert space describing the physical
system. The state vector is related to measurable quantities by evaluating
expectation values with operators which represent observables.  This
formalism is very far from a classical, intuitive picture of statistical
fluctuations. Nevertheless, there is a possibility to transform the
standard quantum mechanical formalism into the form which resembles
a classical statistical theory. Such a representation is particularly
useful in investigating the classical limit of quantum mechanics. The
fundamental role in this approach is played by quasidistribution
functions, which can be regarded as quantum analogs of a phase space
probability distribution. However, due to noncommutativity of quantum
observables, the phase space representation of the quantum state is
not unique, and it is not possible to have in quantum mechanics a phase
space distribution that has all the properties of the classical one.

\section{Wigner function}

In 1932, Eugene Wigner \cite{WignPR32} introduced a quantum analog of
the classical phase space probability distribution. For a particle
travelling along one dimension, the Wigner function
is related to the wave function
$\psi(x)$ through the formula:
\begin{equation}
\label{Eq:WignerDef}
W(q,p) = \frac{1}{2\pi\hbar} \int \prtext{d}x \, \psi^{\ast}
(q+x/2) \, e^{ipx/\hbar} \, \psi(q-x/2),
\end{equation}
and it completely characterizes the quantum state.
The integral of $W(q,p)$ over $q$ and $p$ is one, which follows
from the normalization of the wave function.
Expectation values
of quantum observables can be obtained from the Wigner function by
integrating it with appropriate Wigner-Weyl expressions representing
these observables \cite{HillOConPRep84}. Furthermore,
marginals of the Wigner function yield quantum mechanical distributions
for the position and the momentum. However, the Wigner function has
one property which manifests that quantum mechanics is distinct from
a classical statistical theory: the Wigner function can take negative
values. We shall see later that this property is closely related to quantum
interference phenomena.

Difficulties with defining the quantum phase space distribution have
their origin in the noncommutativity of quantum observables. As the
position and momentum operators do not commute, we cannot introduce a
joint distribution of these two observables. This problem is closely
related to the issue of the ordering of observables, which appears
when passing from classical to quantum mechanics.  For example,
the classical expression $qp$ has the following nonequivalent
quantum counterparts: $\hat{q}\hat{p}$, $\hat{p}\hat{q}$, or
$\frac{1}{2}(\hat{q}\hat{p}+\hat{p}\hat{q})$.  The Wigner function
corresponds to a specific, symmetric ordering of the position and
momentum operators, called the Weyl ordering
\cite{WeylZfP27}.  We will now transform
Eq.~(\ref{Eq:WignerDef}) to the form which shows explicitly relation
between the Wigner function and the symmetric ordering of the position
and momentum operators. For this purpose, let us introduce an additional
delta function and represent it in an integral form:
\begin{eqnarray}
W(q,p) & = & \frac{1}{2\pi\hbar}
\int \prtext{d}x \int \prtext{d}y \,  e^{ipx/\hbar} \delta(y-q-x/2)
\psi^\ast(y) \psi(y-x) \nonumber \\
& = & \frac{1}{(2\pi\hbar)^2} \int \prtext{d}x \int \prtext{d}y 
\int \prtext{d}k \, e^{ipx/\hbar} e^{i(y-q-x/2)k/\hbar}
\psi^\ast(y) \psi(y-x) \nonumber \\
\label{Eq:WignerOriToWW}
& = & \frac{1}{(2\pi\hbar)^2} \int \prtext{d}x \int \prtext{d}k \,
e^{i(px-kq)/\hbar} \int \prtext{d}y \, e^{ikx/2\hbar}
\psi^{\ast}(y) e^{ik(y-x)/\hbar} \psi(y-x).
\end{eqnarray}
In the last expression, the integral over $y$ can be written as the
quantum expectation value:
\begin{equation}
\label{Eq:qpWW}
\int \prtext{d}y \, e^{ikx/2\hbar}
\psi^{\ast}(y) e^{ik(y-x)/\hbar} \psi(y-x)
 = \langle\psi| e^{i(k\hat{q}-\hat{p}x)/\hbar}
|\psi\rangle.
\end{equation}
This quantity is a function of two real parameters $k$ and
$x$. By differentiating over $k$ and $x$ we may obtain moments
of the position and momentum operators. These moments are ordered
symmetrically in $\hat{q}$ and $\hat{p}$, which follows from the
form of the exponent in Eq.~(\ref{Eq:qpWW}). The function $\langle\psi|
e^{i(k\hat{q}-\hat{p}x)/\hbar} |\psi\rangle$ is called the Wigner-Weyl
ordered characteristic function for the position and the momentum.
Coming back to Eq.~(\ref{Eq:WignerOriToWW}), we finally
arrive at the formula
\begin{equation}
\label{Eq:WqpSymOrd}
W(q,p)  = \frac{1}{(2\pi\hbar)^2} \int \prtext{d}x \int \prtext{d}k
\,
e^{i(px-kq)/\hbar} \langle\psi| e^{i(k\hat{q}-\hat{p}x)/\hbar}
|\psi\rangle
\end{equation}
which shows that the Wigner function is the Fourier transform of
the symmetrically ordered characteristic function for the position
and the momentum. Eq.~(\ref{Eq:WqpSymOrd}) can be used to evaluate
the Wigner function corresponding to a mixed state described by
the density matrix $\hat{\varrho}$. In such a case, we have to
replace $\langle\psi| e^{i(k\hat{q}-\hat{p}x)/\hbar} |\psi\rangle$ by
$\prtext{Tr}(\hat{\varrho}e^{i(k\hat{q}-\hat{p}x)/\hbar})$.  Equivalently,
the Wigner function of a mixed state can be obtained from a weighted sum
of the Wigner functions describing the pure components of the mixed state.

For a harmonic oscillator, it is convenient to introduce a pair of
dimensionless annihilation and creation operators, defined by the equations:
\begin{equation}
\hat{a} = \frac{1}{\sqrt{2}}(\lambda^{-1} \hat{q} + i \lambda \hbar^{-1}
\hat{p}), \hspace{1cm}
\hat{a}^\dagger = \frac{1}{\sqrt{2}}(\lambda^{-1} \hat{q} 
- i \lambda \hbar^{-1}
\hat{p})
\end{equation}
where $\lambda$ is a natural length scale defined by the mass and the
frequency of the oscillator. Analogously, the two real parameters of
the Wigner function can be combined into a single complex argument
$\alpha = (\lambda^{-1} q + i \lambda \hbar^{-1} p) /\sqrt{2}$.
The Wigner function in this parameterization is given by
\begin{equation}
\label{Eq:WigSymOrdered}
W(\alpha) = \frac{1}{\pi^2} \int \prtext{d}^2\zeta \,
e^{\zeta^\ast \alpha - \zeta \alpha^\ast} \langle
e^{\zeta \hat{a}^\dagger - \zeta^\ast \hat{a}} \rangle
\end{equation}
where the integration is performed over the whole complex plane
and the angular brackets $\langle \ldots \rangle$ denote the
quantum expectation value. Let us note, that the normalization
constant in Eq.~(\ref{Eq:WigSymOrdered}) has changed compared to
Eq.~(\ref{Eq:WqpSymOrd}). This is because the integration measure
over the phase space is now equal to
$\prtext{d}^2\alpha = \prtext{d}q \, \prtext{d}p / 2\hbar$.

The physical system which we shall describe in the phase space
representation, is optical radiation. In the standard procedure of
quantization, the electromagnetic field is decomposed into a set of
independent modes. Each of these modes is characterized by a pair of
creation and annihilation operators, which satisfy bosonic commutation
relations for a harmonic oscillator. When only one of the modes is
excited, we may describe its quantum state using the Wigner function
defined in Eq.~(\ref{Eq:WigSymOrdered}).  In the classical limit, the
parameter $\alpha$ characterizes the complex amplitude of the 
field, expressed in dimensionless units. We may also use the single-mode
description if our measuring apparatus is sensitive only to a selected
mode of the detected radiation.

\section{Quasidistribution functions}

As it is clearly seen from Eq.~(\ref{Eq:WigSymOrdered}), the
Wigner function corresponds to the characteristic function with the
symmetric ordering of the creation and annihilation operators. In
principle, we could consider also other orderings, for example
normal or antinormal.  In the normal ordering all creation operators
are placed before annihilation operators, and vice versa for the
antinormal ordering. Thus, we could think of replacing the quantum
expectation value in Eq.~(\ref{Eq:WigSymOrdered}) by the normally
ordered characteristic function $\langle e^{\zeta \hat{a}^\dagger}
e^{-\zeta^\ast \hat{a}} \rangle$, or by the antinormally ordered one
$\langle e^{-\zeta^\ast \hat{a}} e^{\zeta \hat{a}^\dagger} \rangle$.
These and other possibilities can be written jointly in a very elegant
way by introducing an exponential factor, which defines the ordering of
the creation and annihilation operators.  This idea leads to the concept
of more general  $s$-parameterized quasiprobability distributions.
The one-parameter family of quasidistribution functions
is given by the following formula \cite{CahiGlauPR69b}:
\begin{equation}
\label{Eq:QuasiDistDef}
W(\alpha ; s) = \frac{1}{\pi^2} 
\int \prtext{d}^2 \zeta \, 
e^{s |\zeta|^2 /2 + \zeta^\ast \alpha - \zeta \alpha^\ast}
\left\langle e^{\zeta \hat{a}^\dagger - \zeta^\ast \hat{a}}
\right\rangle .
 \end{equation}
The real parameter $s$ is associated with the ordering of the field
bosonic  operators through the exponential factor $\exp(s |\zeta|^2
/2)$. In particular, it can which can be easily checked using the
Baker-Campbell-Haussdorf formula that three values $s=1 ,0,$ and $-1$
generate the normal, symmetric and antinormal ordering, respectively.

The definition given by Eq.~(\ref{Eq:QuasiDistDef}) unifies the Wigner
function with other, independently developed quantum analogs of a
phase space distribution. For instance, normal ordering corresponds 
to the so-called $P$ function, introduced by Glauber 
\cite{GlauPRL63} and Sudarshan \cite{SudaPRL63}.
This function serves as a weight function in the diagonal coherent state
representation for the density matrix $\hat{\varrho}$:
\begin{equation}
\hat{\varrho} = \int \prtext{d}^2\alpha \, P(\alpha) \,
|\alpha\rangle\langle \alpha |.
\end{equation}
On the other hand, antinormal ordering yields the distribution
known as the Husimi \cite{Husimi} or $Q$ function
\cite{KanoJMP65,MehtSudaPR65},
which is given by the diagonal elements of the density matrix in the
coherent state basis:
\begin{equation}
\label{Eq:Qdiag}
Q(\alpha) = \frac{1}{\pi} \langle \alpha | \hat{\varrho} | \alpha \rangle.
\end{equation}

Properties of various $s$-parameterized quasidistribution functions are
quite different. This can be seen using the three examples of the $P$
function, the Wigner function, and the $Q$ function. The $P$ function
is highly singular for nonclassical states of light. For example, it is
given by derivatives of the delta function for eigenstates of the photon
number operator $\hat{a}^\dagger \hat{a}$.  The Wigner function is well
behaved for all states, but it may take negative values.  Finally the
$Q$ function is always positive definite, which follows directly from
Eq.~(\ref{Eq:Qdiag}).  The fact that quasidistribution functions with
lower ordering are more regular reflects a general relation linking
any two differently ordered quasidistributions via convolution with a
Gaussian function in the complex phase space:
\begin{equation}
\label{Eq:QDistConvolution}
W(\alpha;s') = \frac{2}{\pi(s-s')} \int \prtext{d}^2 \beta \,
\exp\left( - \frac{2|\alpha-\beta |^2}{s-s'} \right)
W(\beta;s),
\end{equation}
where $s > s'$.
Thus the lower the ordering, the smoother the quasidistribution is, and
fine details of the function are not easily visible. The Gaussian exponent
appearing in the above equation can be formally regarded as a propagator
for the diffusion equation, with the ordering parameter playing the role
of the time. Following this analogy, we may write a differential equation
for quasidistribution functions corresponding to a given quantum state:
\begin{equation}
\label{Eq:Wdiffusion}
\frac{\partial}{\partial s} W(\alpha ;s ) = - \frac{1}{2}
\frac{\partial^2}{\partial \alpha \partial \alpha^\ast}
W(\alpha ;s).
\end{equation}
Let us note that the above equation differs from the standard diffusion
equation by the minus sign. This difference originates from the fact,
that the ``diffusion'' of quasidistributions follows in the direction
of decreasing $s$.

In our calculations a normally ordered representation
of the quasidistribution functions will be very useful. 
Introducing normal ordering of the creation and annihilation
operators in Eq.~(\ref{Eq:QuasiDistDef})
allows to perform the integral explicitly, which yields:
\begin{equation}
\label{Eq:WalphasNormOrd}
W(\alpha; s)
 =  \frac{2}{\pi(1-s)} \left\langle : \exp \left( -
\frac{2}{1-s} (\hat{a}^\dagger - \alpha^\ast)(\hat{a} - \alpha)
\right) : \right\rangle.
\end{equation}
Thus the $s$-ordered quasidistribution function at a complex
phase space point $\alpha$ is given by the expectation value
of the operator
\begin{equation}
\label{Eq:Whatalphas}
\hat{W}(\alpha; s)
 =  \frac{2}{\pi(1-s)}
: \exp \left( -
\frac{2}{1-s} (\hat{a}^\dagger - \alpha^\ast)(\hat{a} - \alpha)
\right) :
\end{equation}
Using the operator identity \cite{LouisellOperatorAlgebra}
\begin{equation}
\label{Eq:ExpNormOrd}
 : \exp[(e^{i\zeta} -1) \hat{v}^\dagger
\hat{v}]: \, = \exp(i\zeta  \hat{v}^{\dagger} \hat{v})
\end{equation}
valid for an arbitrary bosonic annihilation operator $\hat{v}$, we may
transform Eq.~(\ref{Eq:Whatalphas}) to the following expression:
\begin{equation}
\label{WhatalphasSymOrd}
\hat{W}(\alpha;s)
= 
\frac{2}{\pi(1-s)} \left( 
\frac{s+1}{s-1} \right)^{(\hat{a}^\dagger - \alpha^{\ast})
(\hat{a} - \alpha)} .
\end{equation}
The operator appearing in the exponent is the displaced photon
number operator $\hat{n} = \hat{a}^\dagger \hat{a}$. Using the standard
displacement operator $\hat{D}(\alpha) = \exp(\alpha \hat{a}^\dagger 
- \alpha^\ast \hat{a})$, we may write $\hat{W}(\alpha ;s)$ as
\cite{EnglJPA89,MoyaKnigPRA93}:
\begin{eqnarray}
\hat{W}(\alpha;s)
& = &
\frac{2}{\pi(1-s)}
\hat{D}(\alpha)
\left(             
\frac{s+1}{s-1} \right)^{\hat{n}}
\hat{D}^\dagger(\alpha)
\nonumber \\
& = &
\label{Eq:WDisplacement}
\frac{2}{\pi(1-s)}
\sum_{n=0}^{\infty}
\left(
\frac{s+1}{s-1} \right)^{n}
\hat{D}(\alpha)
|n \rangle \langle n |
\hat{D}^\dagger(\alpha).
\end{eqnarray}
The last form is simply the spectral decomposition of $\hat{W}(\alpha;s)$.
The eigenvectors are displaced Fock states $\hat{D}(\alpha) |n \rangle$,
and the corresponding eigenvalues are $[(s+1)/(s-1)]^n$ times the
front normalization factor $2/\pi(1-s)$. It is instructive to see,
how the properties of the quasidistributions are reflected by the spectrum
of $\hat{W}(\alpha;s)$. First, let us note that for $s\rightarrow 1$,
the factor $(s+1)/(s-1)$ is divergent; this corresponds to the singular
character of the $P$ representation. For $0 < s < 1$ the set of
eigenvalues is unbounded; therefore, the corresponding quasidistributions
also may exhibit singular behaviour. The operator $\hat{W}(\alpha;s)$
becomes bounded for $s \le 0$, and the highest value, i.e.\ $s=0$,
corresponds to the Wigner function. Even when $\hat{W}(\alpha;s)$ is
bounded, its eigenvalues corresponding to odd $n$s can be negative. The
highest value of $s$ for which all the eigenvalues are nonnegative is
$s=-1$, which corresponds to the $Q$ function.

\section{Quantum interference in phase space}

We will now discuss, using a simple example, how quantum interference
phenomena are visualised in the phase space representation. A quantum
analog of a classical field with well defined amplitude and phase is
the coherent state $|\alpha_0\rangle$, defined as an eigenstate of the
annihilation operator $\hat{a} |\alpha_0\rangle = \alpha_0
|\alpha_0\rangle$. This equivalence originates from the fact, that
full quantum theory of photodetection gives for coherent states the same
predictions as semiclassical theory with quantized detector and classical
electromagnetic fields \cite{MandJOSA77}.
Coherent states are represented in the phase space by Gaussians
\begin{equation}
\hat{W}^{|\alpha_0\rangle}(\alpha; s)
 =  \frac{2}{\pi(1-s)}
 \exp \left( -
\frac{2}{1-s} |\alpha - \alpha_0|^2
\right).
\end{equation}

Quantum mechanics allows one to combine two such classical-like state,
for example $|\alpha_0\rangle$ and $|-\alpha_0\rangle$ into a coherent
superposition
\begin{equation}
\label{Eq:SchroedingerCatDef}
|\psi\rangle = \frac{1}{\sqrt{2(1+e^{-2|\alpha_0|^2})}} 
(|\alpha_0\rangle + |- \alpha_0 \rangle). 
\end{equation}
States of this type illustrate quantum coherence and 
interference between classical--like components, and are often
called  
quantum optical Schr\"{o}dinger cats 
\cite{SchlPernPRA91}. In contrast to coherent states,
they exhibit a variety of nonclassical properties \cite{BuzeKnigPIO95}.
The quasidistribution function of the superposition 
$|\psi\rangle$ is given by the formula
\begin{eqnarray}
W^{|\psi\rangle}(\alpha; s) & = & 
\frac{1}{\pi(1-s)(1+e^{-2|\alpha_0|^2})}
\left[
\exp\left(-\frac{2}{1-s}|\alpha - \alpha_0|^2\right)
\vphantom{\left(\frac{4 \prtext{Im}(\alpha_0 \alpha^{\ast})}{1-s} 
\right)}
\right.
\nonumber \\
& &
 +
\exp\left(-\frac{2}{1-s}|\alpha + \alpha_0|^2\right) 
\nonumber \\
& & 
\label{Eq:SchroedingerCat}
\left.
+ 2 
\exp\left(\frac{2s}{1-s}|\alpha_0|^2\right)
\exp\left( - \frac{2}{1-s}|\alpha|^2
\right) 
\cos \left(\frac{4 \prtext{Im}(\alpha_0 \alpha^{\ast})}{1-s} \right)
\right].
\end{eqnarray}
The first two terms in the square brackets describe the two coherent
components. The last term results from quantum interference between these
components. It contains an oscillating factor $\cos[4 \prtext{Im}(\alpha_0
\alpha^{\ast})/(1-s)]$. It is seen that the frequency of the oscillations
grows with the distance between the coherent components. The envelope
of this oscillating term is defined by the Gaussian
$\exp[-2|\alpha|^2/(1-s)]$, which is centered exactly half way between
the interfering states.

\begin{figure}
\begin{center}
\epsfig{file=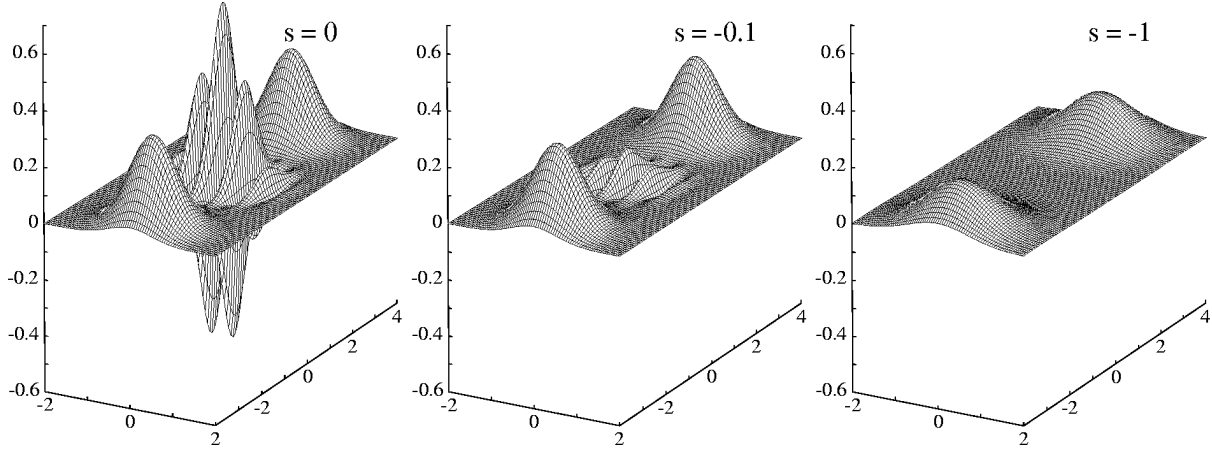}
\end{center}
\caption{Quasidistributions representing
the Schr\"{o}dinger cat state for
 $\alpha_0 = 3i$, depicted
for the ordering parameters $s=0, -0.1,$ and
$-1$.\label{Fig:CatQuasiDist}}
\end{figure}

Fig.~\ref{Fig:CatQuasiDist} shows quasidistributions plotted for three
different values of the ordering parameter $s$.  The Wigner function
contains an oscillating component originating from the interference
between the coherent states. This component is much smaller for $s=-0.1$
and it is completely smeared out in the $Q$ function, which can hardly
be distinguished from that of a statistical mixture of two coherent
states. This is because the whole interference term is multiplied by the
factor $\exp[2s|\alpha_0|^2/(1-s)]$, which quickly tends to zero with
decreasing $s$. Let us note that the larger is the distance between the
components, the faster this factor vanishes. The decay of the interference
component can be formally viewed as a result of diffusion, described 
by Eq.~(\ref{Eq:Wdiffusion}). Decreasing the ordering parameter $s$
makes the whole quasidistribution blurred, and this effect is particularly
deleterious to the quickly oscillating pattern.

\section{Deconvolution}
\label{Sec:Deconvolution}

We have seen, using the example of the Schr\"{o}dinger cat state,
that signatures of quantum interference can be visible better in
quasidistribution functions with higher ordering. Thus, what is
interesting, is the inversion of Eq.~(\ref{Eq:QDistConvolution}), i.e.\
deconvolution of a lower-ordered quasidistribution function.
This task is quite difficult.
Let us first note that in general the integral in
Eq.~(\ref{Eq:QDistConvolution}) fails to converge if we take $s<s'$.
Instead, we may use the Fourier transforms of
the quasidistribution functions
\begin{equation}
\tilde{W}(\zeta;s) = \int \prtext{d}^2\beta \, e^{\zeta \beta^\ast
- \zeta^\ast \beta} W(\beta;s) =
e^{s |\zeta|^2 /2}
\langle e^{\zeta \hat{a}^\dagger - \zeta^\ast \hat{a}} \rangle.
\end{equation}
Transition to a higher ordered quasidistribution consists now simply
in multiplication by an exponent:
\begin{equation}
\tilde{W}(\zeta;s') = e^{(s'-s)|\zeta|^2/2} \tilde{W}(\zeta; s),
\end{equation}
and evaluation of the inverse Fourier transform. The complete expression
of $W(\alpha;s')$ in terms of a lower ordered quasidistribution has
the form:
\begin{equation}
\label{Eq:Deconvolution}
W(\alpha ; s') = \frac{1}{\pi^2} 
\int \prtext{d}^2 \zeta \, 
e^{(s'-s) |\zeta|^2 /2 + \zeta^\ast \alpha - \zeta \alpha^\ast}
\int \prtext{d}^2\beta \, e^{\zeta \beta^\ast
- \zeta^\ast \beta} W(\beta;s).
\end{equation}
Anticipating for a moment the connection of the quasidistributions with
experiment, let us suppose that we are given an experimentally
determined quasidistribution $W(\beta;s)$, and that we are
trying to apply the deconvolution procedure described by
Eq.~(\ref{Eq:Deconvolution}). Usually, values of $W(\beta;s)$ will be
affected by errors originating from statistical uncertainty and various
experimental imperfections. These errors make the deconvolution a very
delicate matter. The crucial problem is that the Fourier transform
$\tilde{W}(\zeta; s)$ has to be multiplied by an {\em exploding}
factor $e^{(s'-s)|\zeta|^2/2}$. Experimental errors of $W(\beta;s)$ can
generate long, slowly decaying high-frequency components in its Fourier
transform. Multiplication by an exploding exponent enormously amplifies
contribution of these fluctuations, which leads to huge errors of the
reconstructed $W(\alpha;s)$.  Therefore, deconvolution of experimentally
determined quasidistributions according to Eq.~(\ref{Eq:Deconvolution})
is practically impossible.

\section{Multimode quasidistributions}

The concept of quasidistribution functions can be generalized
in a straightforward manner to multimode radiation. In analogy to
Eq.~(\ref{Eq:QuasiDistDef}), we need to take the symmetrically ordered
multimode characteristic function, and to evaluate its Fourier transform
with an appropriately chosen Gaussian factor which defines the ordering:
\begin{eqnarray}
\lefteqn{W(\alpha_1,\ldots, \alpha_M ; s)}
& & \nonumber \\
& = &
\frac{1}{\pi^{2M}} \int \prtext{d}\zeta_1 \ldots \prtext{d}\zeta_M \,
\exp \left( \sum_{i=1}^{M} \frac{s}{2} |\zeta_i|^2
+ \zeta_i^\ast \alpha_i - \zeta_i \alpha^\ast_i \right)
\left\langle
\exp\left(\sum_{i=1}^{M} \zeta_i \hat{a}^\dagger_i
- \zeta^\ast_i \hat{a}_i \right) \right\rangle.
\nonumber \\
& &
\end{eqnarray}
Introducing normal ordering allows one to perform the integrals, which
yields an explicit normally ordered representation:
\begin{equation}
\label{Eq:MultiQDFNormOrd}
W(\alpha_1,\ldots, \alpha_M ; s)
 =  \left( \frac{2}{\pi(1-s)} \right)^{M}
\left\langle : \exp \left( -
\frac{2}{1-s} \sum_{i=1}^{M}
(\hat{a}_i^\dagger - \alpha^\ast_i)(\hat{a}_i - \alpha_i)
\right) : \right\rangle.
\end{equation}
Using Eq.~(\ref{Eq:ExpNormOrd}), we may represent the quasidistribution
functions as:
\begin{equation}
W(\alpha_1,\ldots, \alpha_M ;s)
= 
\left( \frac{2}{\pi(1-s)} \right)^{M}
\left\langle \left( 
\frac{s+1}{s-1} \right)^{\sum_{i=1}^{M}
(\hat{a}_i^\dagger - \alpha^{\ast}_i)
(\hat{a}_i - \alpha_i)} \right\rangle.
\end{equation}
The expression $\sum_{i=1}^{M} (\hat{a}_i^\dagger - \alpha^{\ast}_i)
(\hat{a}_i - \alpha_i)$ appearing in the exponent is simply the phase
space displaced operator of the {\em total} number of photons. 
In analogy to the single-mode case, we may write the multimode
quasidistributions as an expectation value of the operator involving
the multimode displacement operator
\begin{equation}
\hat{D}(\{\alpha_i\}) = 
\exp \left( \sum_{i=1}^{M} \alpha_i \hat{a}^\dagger_i
- \alpha^\ast_i \hat{a}_i \right),
\end{equation}
and the total photon number operator, defined as
\begin{equation}
\hat{N} = \sum_{i=1}^{M} \hat{n}_i
\end{equation}
where $\hat{n}_i = \hat{a}^\dagger_i \hat{a}_i$. The explicit
expressions are:
\begin{eqnarray}
W(\alpha_1,\ldots, \alpha_M ; s)
& = & \left( \frac{2}{\pi(1-s)} \right)^{M}
\left\langle 
\hat{D}(\{\alpha_i\})
: \exp \left( -
\frac{2\hat{N}}{1-s} 
\right) : 
\hat{D}^{\dagger}(\{\alpha_i\})
\right\rangle
\nonumber 
\\
\label{Eq:WmultiND}
& = &
\left( \frac{2}{\pi(1-s)} \right)^{M}
\left\langle \hat{D}(\{\alpha_i\}) \left( 
\frac{s+1}{s-1} \right)^{\hat{N}}
\hat{D}^{\dagger}(\{\alpha_i\}) \right\rangle.
\end{eqnarray}

\section{Quasidistribution functionals}

The representation given in Eq.~(\ref{Eq:WmultiND}) suggests
generalization of the multimode quasidistribution functions to the form
independent of the specific decomposition into modes. Such generalized
quasidistributions are functionals of the electromagnetic field.
Instead of using a finite set of annihilation and creation operators,
we will now deal with the full description of the electromagnetic field,
involving the operator fields $\hat{\bf E}({\bf r},t)$ and $\hat{\bf
H}({\bf r},t)$.  In order to simplify the notation, we shall fix
the time $t$, and omit it in the subsequent formulae. The definition
of quasidistribution functionals involves two operators: the coherent
displacement operator $\hat{\cal D}$, and the operator of the total number
of photons $\hat{\cal N}$. The action of the displacement operator is
straightforward: it adds a classical amplitude to the field operators
according to the formula
\begin{eqnarray}
\hat{\cal D}[{\bf E}({\bf r}),{\bf H}({\bf r})]
\hat{\bf E}({\bf r})
\hat{\cal D}^\dagger [{\bf E}({\bf r}),{\bf H}({\bf r})]
& = & \hat{\bf E}({\bf r}) - {\bf E}({\bf r})
\nonumber \\
\hat{\cal D}[{\bf E}({\bf r}),{\bf H}({\bf r})]
\hat{\bf H}({\bf r})
\hat{\cal D}^\dagger [{\bf E}({\bf r}),{\bf H}({\bf r})]
& = & \hat{\bf H}({\bf r}) - {\bf H}({\bf r}).
\end{eqnarray}
In order to find an explicit formula for quasidistribution functionals,
we need to express the total photon number operator $\hat{\cal N}$
in terms of the electric and magnetic field. We shall start from the
standard decomposition of the electromagnetic field into plane waves
with periodic boundary conditions in a box of the volume $V$:
\begin{eqnarray}
\hat{\bf E}({\bf r}) & =  & i \sum_{l\sigma} \sqrt{
\frac{\hbar \omega_l}{2\epsilon_0 V}} {\bf e}_{l\sigma}
(\hat{a}_{l\sigma} e^{i{\bf k}_l {\bf r}} 
- \hat{a}^\dagger_{l\sigma} e^{- i{\bf k}_l {\bf r}})
\\
\hat{\bf H}({\bf r}) & = & - \frac{i}{c\mu_0} 
\sum_{l\sigma} \sqrt{
\frac{\hbar \omega_l}{2\epsilon_0 V}}
{\bf e}_{l\sigma} \times \frac{{\bf k}_l}{|{\bf k}_{l}|}
(\hat{a}_{l\sigma} e^{i{\bf k}_l {\bf r}} 
- \hat{a}^\dagger_{l\sigma} e^{- i{\bf k}_l {\bf r}}).
\end{eqnarray}
Here the indices $l$ and $\sigma$ label respectively the wave vectors
${\bf k}_l$ and the polarizations ${\bf e}_{l\sigma}$, and $\omega_l =
c|{\bf k}_l|$ is the frequency of an $l$th mode. 
Our goal is to represent the sum
\begin{equation}
\hat{\cal N} = \sum_{l\sigma} \hat{a}^\dagger_{l\sigma}
\hat{a}_{l\sigma}
\end{equation}
using $\hat{\bf E}({\bf r})$ and $\hat{\bf H}({\bf r})$. For this purpose
we shall take
Fourier transforms of these fields:
\begin{eqnarray}
\int \prtext{d}^3 {\bf r} \, \hat{\bf E}({\bf r}) e^{-i{\bf k}_l {\bf r}}
& = & 
i \sqrt{\frac{\hbar \omega_l V}{2\epsilon_0 }}
\sum_{\sigma} ( {\bf e}_{l\sigma} \hat{a}_{l\sigma}
- {\bf e}_{-l\sigma} \hat{a}^\dagger_{-l\sigma})
\\
\int \prtext{d}^3 {\bf r} \,
\hat{\bf H}({\bf r}) e^{-i{\bf k}_l {\bf r}}
& = & 
- i \sqrt{\frac{\hbar \omega_l V}{2\mu_0}}
\sum_\sigma
\left(
{\bf e}_{l\sigma} \times \frac{{\bf k}_l}{|{\bf k}_l|}
\hat{a}_{l\sigma}
-
{\bf e}_{-l\sigma} \times \frac{{\bf k}_{-l}}{|{\bf k}_{-l}|}
\hat{a}_{-l\sigma}^\dagger
\right).
\end{eqnarray}
Here on the right-hand sides we have used the fact that ${\bf k}_{-l}
= - {\bf k}_l$. The product of the Fourier transforms taken for
${\bf k}_l$ and ${\bf k}_{-l}$ can be expressed as:
\begin{equation}
\int \prtext{d}^3 {\bf r} \int \prtext{d}^3 {\bf r}' \,
\hat{\bf E}({\bf r}) \hat{\bf E}({\bf r}') e^{- i {\bf k}_l ({\bf r}
- {\bf r}')}
= 
\frac{\hbar \omega_l V}{2\epsilon_0}
\sum_{\sigma\sigma'}
({\bf e}_{l\sigma} \hat{a}_{l\sigma} - {\bf e}_{-l\sigma}
\hat{a}^{\dagger}_{-l\sigma})
({\bf e}_{l\sigma'} \hat{a}_{l\sigma'}^{\dagger} - {\bf e}_{-l\sigma'}
\hat{a}_{-l\sigma'})
\end{equation}
and
\begin{eqnarray}
\lefteqn{\int \prtext{d}^3 {\bf r} \int \prtext{d}^3 {\bf r}' \,
\hat{\bf H}({\bf r}) \hat{\bf H}({\bf r}') e^{- i {\bf k}_l ({\bf r}
- {\bf r}')}}
& & \nonumber \\
 & = & 
\frac{\hbar \omega_l V}{2\mu_0}
\sum_{\sigma\sigma'}
\left(
{\bf e}_{l\sigma} \times \frac{{\bf k}_l}{|{\bf k}_l|}
\hat{a}_{l\sigma}
-
{\bf e}_{-l\sigma} \times \frac{{\bf k}_{-l}}{|{\bf k}_{-l}|}
\hat{a}_{-l\sigma}^\dagger
\right)
\left(
{\bf e}_{l\sigma'} \times \frac{{\bf k}_l}{|{\bf k}_l|}
\hat{a}_{l\sigma'}^\dagger
-
{\bf e}_{-l\sigma'} \times \frac{{\bf k}_{-l}}{|{\bf k}_{-l}|}
\hat{a}_{-l\sigma'}
\right)
\nonumber \\
& = &
\frac{\hbar \omega_l V}{2\mu_0}
\sum_{\sigma\sigma'} 
(\delta_{\sigma\sigma'} \hat{a}_{l\sigma} \hat{a}_{l\sigma'}^\dagger
 + {\bf e}_{l\sigma} {\bf e}_{-l\sigma'} \hat{a}_{l\sigma}
\hat{a}_{-l\sigma'} + {\bf e}_{-l\sigma} {\bf e}_{l\sigma'}
\hat{a}^\dagger_{-l\sigma} \hat{a}^\dagger_{l\sigma'}
+
\delta_{\sigma\sigma'} \hat{a}_{-l\sigma}^\dagger
\hat{a}_{-l\sigma'}).
\end{eqnarray}
We shall now add the expressions for the electric and magnetic fields
multiplied by the factors $\epsilon_0/2\hbar\omega_l V$ and
$\mu_0/2\hbar\omega_l V$ respectively. This yields:
\begin{equation}
\int \prtext{d}^3 {\bf r} \int \prtext{d}^3 {\bf r}' \,
\left(
\frac{\epsilon_0}{2}
\hat{\bf E}({\bf r}) \hat{\bf E}({\bf r}')
+
\frac{\mu_0}{2}
\hat{\bf H}({\bf r}) \hat{\bf H}({\bf r}')
\right)
\frac{ e^{- i {\bf k}_l ({\bf r} - {\bf r}')}}{\hbar\omega_l V}
=
\sum_{\sigma} {\textstyle \frac{1}{2}}
(\hat{a}_{l\sigma} \hat{a}^\dagger_{l\sigma}
+ \hat{a}^\dagger_{-l\sigma} \hat{a}_{-l\sigma}
).
\end{equation}
This formula is close to the standard expression for the energy of the
electromagnetic field. Indeed, we could obtain it via multiplication
of both the sides by $\hbar\omega$, and summation over $l$. However,
we are now interested in a different quantity, namely the total number
of photons, and we need to perform the summation with the factor
$\hbar\omega$ in the denominator of the left hand side. In this way
we obtain:
\begin{equation}
\label{Eq:Sumlsigma}
\sum_{l\sigma} {\textstyle \frac{1}{2}}
(\hat{a}_{l\sigma} \hat{a}^\dagger_{l\sigma}
+ \hat{a}^\dagger_{l\sigma} \hat{a}_{l\sigma}
)
= 
\int \prtext{d}^3 {\bf r} \int \prtext{d}^3 {\bf r}' \,
\left(
\frac{\epsilon_0}{2}
\hat{\bf E}({\bf r}) \hat{\bf E}({\bf r}')
+
\frac{\mu_0}{2}
\hat{\bf H}({\bf r}) \hat{\bf H}({\bf r}')
\right)
K({\bf r}-{\bf r}').
\end{equation}
In the second term of the left-hand side we have changed the summation index
$-l \rightarrow l$. The integral kernel $K({\bf r}-{\bf r}')$ appearing
on the right-hand side is given by:
\begin{equation}
\label{Eq:Kr}
K({\bf r}) = \sum_{l} 
\frac{e^{- i {\bf k}_l {\bf r}}}{\hbar\omega_l V}.
\end{equation}
We shall evaluate it in the continuous limit, when the sum over $l$ can
be replaced by a three-dimensional integral over the wave vector ${\bf k}$.
In this limit, there occurs a singularity at  ${\bf r} =0$, which is
a result of the slowly decaying integrand with large ${\bf k}$.
We shall regularize the integral by introducing the upper cut-off
$k_{\prtext{\scriptsize max}}$ for the wave number. 
Physically, this means that we do not take into account photons with
energy larger than $\hbar c k_{\prtext{\scriptsize max}}$.
The regularized kernel can be evaluated in a straightforward manner:
\begin{eqnarray}
K({\bf r}) & = &
\frac{1}{(2\pi)^3\hbar c} \int \prtext{d}^3 {\bf k} \, 
\frac{e^{-i{\bf k}{\bf r}}}{|{\bf k}|}
=
\frac{1}{(2\pi)^2\hbar c}
\int_0^{k_{\prtext{\tiny max}}} \prtext{d}k \, k
\int_{0}^{\pi} \prtext{d}\vartheta \, \sin\vartheta
\, e^{-ik|{\bf r}|\cos\vartheta}
\nonumber \\
& = &
\frac{1}{2\pi^2 \hbar c}
\frac{1 - \cos k_{\prtext{\scriptsize max}} |{\bf r}|}{{\bf r}^2}.
\end{eqnarray}
It is easily seen that truncation of the wave vector magnitude has
removed singularity of the kernel $K({\bf r})$ occurring at ${\bf r}=0$.

On the left-hand side of Eq.~(\ref{Eq:Sumlsigma}), we have a symmetrically
ordered product of the creation and annihilation operators
$\frac{1}{2}
(\hat{a}_{l\sigma} \hat{a}^\dagger_{l\sigma}
+ \hat{a}^\dagger_{l\sigma} \hat{a}_{l\sigma})$.
In order to obtain the total photon number operator, we need to introduce
the normal ordering of the right-hand side of Eq.~(\ref{Eq:Sumlsigma}).
Using this expression, we can easily define quasidistribution functionals
of the electromagnetic field. There is a small difficulty arising
from the fact that we now deal with the infinite number of degrees of
freedom. In Eq.~(\ref{Eq:MultiQDFNormOrd}), we cannot pass to infinity
with the number of modes in the normalization prefactor $[2/\pi(1-s)]^M$.
We shall solve this difficulty by absorbing the
normalization prefactor into the functional integration measure over the
fields ${\bf E}$ and ${\bf H}$. Thus, we define the quasidistribution
functional as:
\begin{equation}
{\cal W}[{\bf E}({\bf r}), {\bf H}({\bf r}) ; s]
=
\left\langle
\hat{\cal D}[{\bf E}({\bf r}),{\bf H}({\bf r})]
: \exp \left( -\frac{2\hat{\cal N}}{1-s} \right) :
\hat{\cal D}^{\dagger} [{\bf E}({\bf r}),{\bf H}({\bf r})]
\right\rangle.
\end{equation}
This definition can be written explicitly using the electromagnetic
field operators with the help of the derived expression for the total
photon number operator as:
\begin{eqnarray}
\lefteqn{{\cal W}[{\bf E}({\bf r}), {\bf H}({\bf r}) ; s]} & &
\nonumber \\
& = &
\left\langle : \exp\left[
- \frac{2}{1-s}
\int \prtext{d}^3 {\bf r} \int \prtext{d}^3 {\bf r}' \,
K({\bf r}-{\bf r}')
\left(
\frac{\epsilon_0}{2}
[\hat{\bf E}({\bf r})-{\bf E}({\bf r})]
[\hat{\bf E}({\bf r}')-{\bf E}({\bf r}')]
\right. \right. \right.
\nonumber \\
 & & 
+
\left. \left.
\vphantom{\frac{2}{1-s} \int}
\left.
\frac{\mu_0}{2}
[\hat{\bf H}({\bf r})-{\bf H}({\bf r})]
[\hat{\bf H}({\bf r}')-{\bf H}({\bf r}')]
\right)
\right]
: \right\rangle.
\end{eqnarray}
The integral kernel appearing in the above formula is given by
Eq.~(\ref{Eq:Kr}).

\chapter{Homodyne techniques for quantum state measurement}

\label{Chap:Homodyne}

\markright{CHAPTER \thechapter . \uppercase{Homodyne techniques\ldots}}

Over the last decade, the domain of quantum state measurement has
passed a long way from first theoretical proposals to well understood
experimental realizations. Complete presentation of the current state of
this field would require a separate book, encompassing a wide range of
experimental techniques and concepts of data analysis. In this chapter
we shall set the scene for further parts of the thesis by describing
briefly earlier works on measuring the quantum state of light. We shall
restrict our attention to detection of optical radiation, and describe
two techniques which have been successfully realized in experiments:
double homodyne detection and optical homodyne tomography.

Double homodyne detection allows one to measure the $Q$ function
of a light mode. It was demonstrated in 1986 by Walker and Caroll
\cite{WalkCaroOQE86}. Their experiment had as a main purpose
the demonstration
of a homodyne measurement near the quantum noise limit, and it later
attracted attention as a complete characterization of the quantum
state. As we discussed in the previous chapter, the $Q$ function is a
positive definite distribution, and it exhibits only faint traces of
quantum interference.  Optical homodyne tomography was realized first
by Smithey {\em et al.} in 1993 \cite{SmitBeckPRL93}.  This technique
is capable of measuring the Wigner function. Apparently, this fact
added extra excitement to the development of homodyne tomography, as
the Wigner function is a nonclassical distribution function which may
take negative values resulting from quantum interference.

Both these techniques are based on the same experimental apparatus,
namely the balanced homodyne detector. This device provides information on
phase-sensitive properties of light. In the quantum mechanical formalism,
it performs the measurement of a family of observables called
{\em quadratures}. We shall start this chapter with a description of the
balanced homodyne detector in Sec.~\ref{Sec:BalancedHomodyne}.  The double
homodyne detection scheme is discussed in Sec.~\ref{Sec:DoubleHomodyne}.
We show that this scheme can be used to measure two noncommuting
observables at the cost of introducing extra noise to the measurement.
Sec.~\ref{Sec:OHT} is devoted to optical homodyne tomography. It
describes the physical principle of the method, as well as mathematical
transformations involved in the processing of experimental data.

\section{Balanced homodyne detector}
\label{Sec:BalancedHomodyne}

Standard photodetection is insensitive to phase properties of optical
radiation. This is because the observed signal depends only on the
operator of the number of photons $\hat{n} = \hat{a}^{\dagger} \hat{a}$.
Nevertheless, we may use a photodetector to measure phase-dependent
quantities by superposing the measured beam with an auxiliary coherent
field using a beam splitter. The auxiliary field has the name
of the {\em local oscillator}.  When measuring such a superposition,
the signal from the photodetector will be described by an expression
involving terms linear in $\hat{a}$ and $\hat{a}^{\dagger}$. Thus, it
carries information on the phase properties of the measured field. This
is the basic idea of homodyne detection.

In quantum optics, homodyne detection has played an important
role in investigating the squeezed states of light. These states
exhibit interesting noise properties in certain phase-de\-pen\-dent
observables. More specifically, for a single light mode we may introduce
a family of quadrature observables dependent on the phase $\theta$:
\begin{equation}
\hat{x}_{\theta}  = \frac{e^{i\theta} \hat{a}^\dagger +
e^{- i \theta}  \hat{a}}{\sqrt{2}}.
\end{equation}
It is easy to check that the commutator of two quadratures corresponding
to phases which differ by $\pi/2$ is $[\hat{x}_{\theta}, \hat{x}_{\theta
+ \pi/2}] = i$. Consequently, variances of these two observables
satisfy the uncertainty relation in the form $\Delta x_\theta \Delta
x_{\theta+\pi/2} \ge 1/2$. For coherent states, this variance is evenly
distributed over all the quadratures, and it equals to $\Delta x_\theta =
1/\sqrt{2}$. Squeezed states are such states of the electromagnetic field,
which for a certain phase $\theta$ have the variance smaller than the
coherent state level. These states cannot be described within classical
theory of radiation, and the squeezing is clearly a non-classical
property. Squeezed states can find application in very precise
interferometric measurements \cite{WallsMilburnGravitation}.

\begin{figure}
\begin{center}
\psset{unit=8mm}
\begin{pspicture}(10,10)
\rput{45}(4.5,4){%
\psframe[fillstyle=solid,fillcolor=lightgray](-2,-.1)(2,.1)%
} 
\psset{linewidth=.5mm}
\psline{->}(1,4)(4,4)
\psline{->}(5,4)(8,4)
\psline{->}(4.5,.5)(4.5,3.5)
\psline{->}(4.5,4.5)(4.5,7.5)
\pswedge(8.4,4){.6}{270}{90}
\pswedge(4.5,7.9){.6}{0}{180}
\pscircle(9.5,9){0.4}
\psline{-}(9,4)(9.5,4)(9.5,8.6)
\psline{-}(4.5,8.5)(4.5,9)(9.1,9)
\rput(9.5,9){\large $-$}
\rput(2.3,2){\large 50:50}
\rput(1.5,4.4){\large $\hat{\rho}$}
\rput(5.5,0.9){\large $|\beta\rangle_{LO}$}
\end{pspicture}
\end{center}
\caption{The balanced homodyne setup.
The signal field, described by a density matrix $\hat{\rho}$,
is combined with a coherent local oscillator $|\beta\rangle_{LO}$.
The two outgoing fields are measured using photodetectors.
The difference of their counts is the statistical data recorded
in the experiment.
\label{Fig:HomodyneSetup}}
\end{figure}
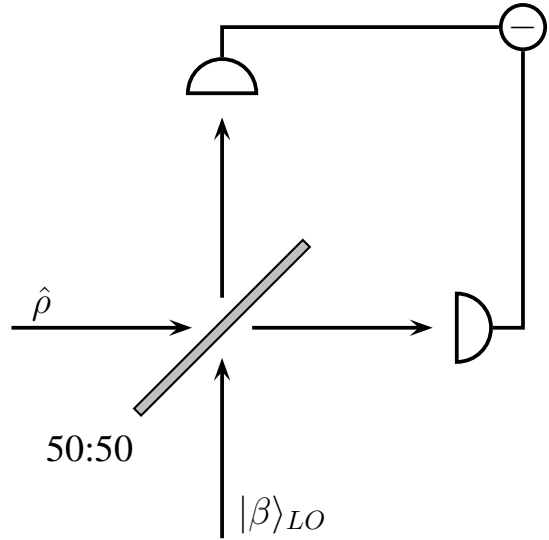

Of course, it is obvious from the definition of quadratures that their
measurement requires a phase-sensitive technique,
such as homodyne detection. In practice, we need to take into account
various experimental imperfections. One of them is the excess noise
of the local oscillator field. This noise adds to the observed level
of fluctuations, and it may mask subtle quantum effects related to
squeezing. Fortunately, there is a possibility to subtract the local
oscillator noise by using the so-called balanced scheme.  In the balanced
homodyne detection scheme, depicted in Fig.~\ref{Fig:HomodyneSetup},
the signal field is superposed using a 50:50 beam splitter with the
local oscillator. Two photodetectors monitor the output ports of the
beam splitter, and the recorded signal is the difference of the detector
photocurrents. In this way, we cancel the effect of local oscillator
noise when measuring the variance of the difference photocurrent.

Moreover, theoretical description of the balanced homodyne detector shows
that this setup, in an idealized limit, is the optical realization of
the quantum measurement of the quadrature operator
$\hat{x}_\theta$. The idealization is
based  on two assumptions: the unit efficiency of the photodetectors and
the classical limit of the local oscillator. The latter condition can be
easily satisfied in an experiment. Also, efficiency of photodetectors
used in a homodyne setup can be close to 100\%. If we rely on these
two assumptions, theoretical analysis of the balanced scheme becomes 
quite compact. More detailed studies can be found in
Ref.~\cite{BrauPRA90,VogeGrabPRA93,BanaWodkPRA97}.

Let us describe the signal field with the annihilation operator
$\hat{a}$, and the local oscillator field with $\hat{b}$.  These two
fields, superposed on a 50:50 beam splitter, yield two outgoing modes. In
general, combination of the modes at a beam splitter is given by an SU(2)
transformation \cite{VogelWelschBeamSplitter}. For a 50:50 beam splitter,
we may simplify this transformation to the matrix
\begin{equation}
\label{Eq:50:50}
\left(
\begin{array}{c} \hat{c} \\ \hat{d} \end{array}
\right)
=
\frac{1}{\sqrt{2}}
\left(
\begin{array}{rr} 1 & 1 \\ 1 & -1 \end{array}
\right)
\left(
\begin{array}{c} \hat{a} \\ \hat{b} \end{array}
\right),
\end{equation}
where $\hat{c}$ and $\hat{d}$ are the annihilation operators of the
outgoing fields. We assume that the local oscillator is in a coherent
state $|\beta\rangle_{LO}$, and the quantum state of the mode $\hat{a}$
is given by the density matrix $\hat{\varrho}$.

The quantity we are interested in is the difference of photocurrents
generated by the detectors monitoring the modes $\hat{c}$ and $\hat{d}$.
On the microscopic level, these photocurrents consist of a discrete number
of electrons $n_1$ and $n_2$. In a real experiment, this discreteness
is not observed due to the large average number of the generated
photoelectrons. The observable measured in balanced homodyne detection
is the difference of the electron number $\Delta N = n_1 - n_2$.
The probability $p (\Delta N)$ of obtaining a specific
value for $\Delta N$ can be easily derived using the standard theory
of photoelectric detection. It is given by 
the expression:
\begin{equation}
p (\Delta N) = \sum_{n_1 - n_2 = \Delta N}
\prtext{Tr} \{ \hat{\varrho} \otimes |\beta\rangle\langle\beta|_{LO}
: e^{ -
\hat{c}^\dagger \hat{c}} \frac{( \hat{c}^\dagger \hat{c}
)^{n_1}}{n_1!} \, e^{ - \hat{d}^\dagger \hat{d}} \frac{(
\hat{d}^\dagger \hat{d})^{n_2}}{n_2!} : \}.
\end{equation}
In further calculations, it is more convenient to deal with the generating
function for the probability distribution $p (\Delta N)$. The generating
function is obtained by evaluating the Fourier transform:
\begin{eqnarray}
Z (\xi) & = & \sum_{\Delta N = - \infty}^{\infty}
e^{i\xi \Delta N} p ( \Delta N)
\nonumber \\
& = &
\prtext{Tr} \{ \hat{\varrho} \otimes |\beta\rangle\langle\beta|_{LO}
 : \exp [ (e^{i\xi} - 1 )
\hat{c}^\dagger \hat{c} + (e^{-i\xi} -1) \hat{d}^\dagger \hat{d}
] : \, \}.
\end{eqnarray}
In this way, we managed to get rid of the troublesome constrained sum with
the condition $n_1 - n_2 = \Delta N$. We can now remove the normal ordering
symbol by making use of the
operator identity given in Eq.~(\ref{Eq:ExpNormOrd}). This yields:
\begin{equation}
Z(\xi) = \prtext{Tr} \{ \hat{\varrho} \otimes |\beta\rangle\langle\beta|_{LO}
\,
e^{i\xi(\hat{c}^\dagger \hat{c} - \hat{d}^\dagger \hat{d})} \}
= \prtext{Tr} \{ \hat{\varrho} \otimes |\beta\rangle\langle\beta|_{LO}
\,
e^{i\xi(\hat{a}^\dagger \hat{b} + \hat{a} \hat{b}^\dagger)} \} .
\end{equation}
When the local oscillator is in a strong coherent state, the
bosonic operators $\hat{b}, \hat{b}^\dagger$ in the exponent
$e^{i\xi(\hat{a}^\dagger \hat{b} + \hat{a} \hat{b}^\dagger)}$ can be
replaced by $c$-numbers $\beta, \beta^\ast$. In this regime, it is also
convenient to rescale the difference photocurrent $\Delta N$, which grows
as the first power of the local oscillator amplitude. Dividing
$\Delta N$ by $|\beta|$, we obtain a quantity which is independent
of the magnitude $|\beta|$ in the regime of the classical local
oscillator. We shall introduce an extra factor of $1/\sqrt{2}$,
and define the homodyne variable as $x = \Delta N/\sqrt{2}|\beta|$.
This variable can be treated as a continuous one, as the local
oscillator amplitude is very large. The rescaling of the homodyne
variable corresponds to changing the parameter of the generating
function according to
$\lambda=\xi\sqrt{2}|\beta|$. In the new parameterization, the
generating function takes the form:
\begin{equation}
\label{Eq:Zthetalambda}
Z_\theta(\lambda) =
\left\langle \exp \left( \frac{i\lambda}{\sqrt{2}}
(e^{i\theta} \hat{a}^\dagger +
e^{- i \theta}  \hat{a})
\right)
\right\rangle
= \langle e^{i\lambda \hat{x}_\theta} \rangle,
\end{equation}
where $\theta$ is the phase of the local
oscillator: $\beta = |\beta|e^{i\theta}$.  We have added here a subscript
$\theta$ to the generating function $Z_\theta(\lambda)$ to stress
that the measured observable depends on the local oscillator phase.
Using the last form of the $Z_\theta(\lambda)$, we may easily obtain
the probability distribution $p_\theta(x)$ for the homodyne
variable $x$ by evaluating the inverse
Fourier transform. Let us note that we should now integrate over
all real values $\lambda$ because of the introduced rescaling. The
inverse Fourier transform yields:
\begin{equation}
p_\theta(x) = \frac{1}{2\pi} \int_{-\infty}^{\infty} \prtext{d}\lambda
\, e^{-i\lambda x} \langle e^{i\lambda \hat{x}_\theta} \rangle
= \langle \delta(x-\hat{x}_{\theta}) \rangle
= \left\langle |x\rangle_{\theta\, \theta} \langle x| \right\rangle.
\end{equation}
This expression clearly shows, that balanced homodyne detection is the
measurement of the quadrature operator $\hat{x}_\theta$. The probability of
obtaining the result $x$ is given by the projection on the corresponding
eigenstate of the quadrature operator, defined as $\hat{x}_{\theta}
|x\rangle_{\theta} = x |x\rangle_{\theta}$.

\section{Double homodyne detection}
\label{Sec:DoubleHomodyne}

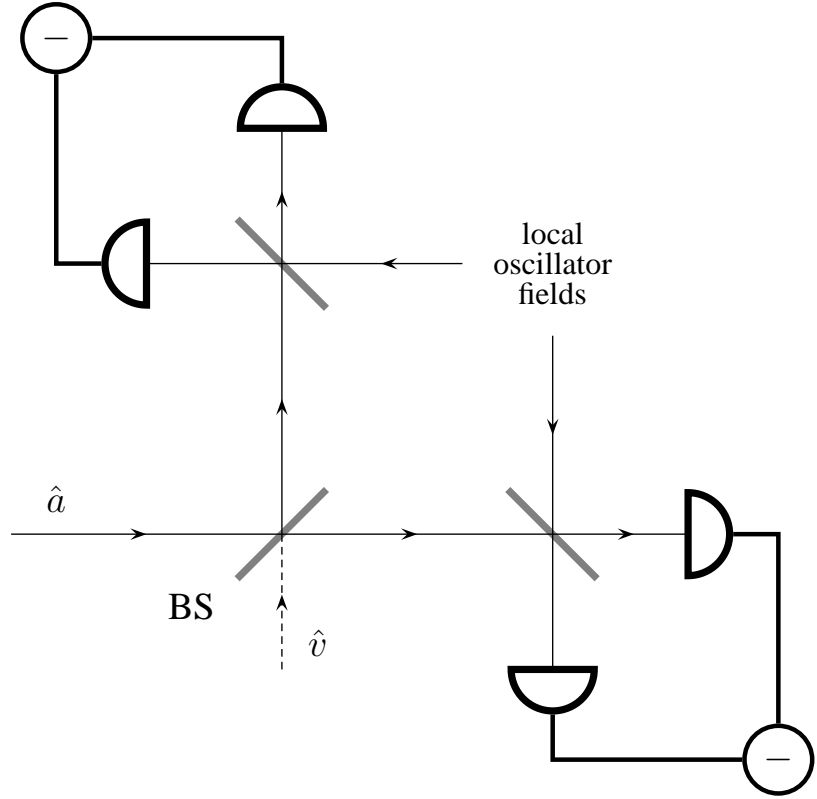
\begin{figure}[t]
\begin{center}
\psset{unit=1.2cm}
\begin{pspicture}(-1,-1)(9,8)
\rput{45}(3,2){\psline[linewidth=0.08,linecolor=gray](-0.7,0)(0.7,0)}
\rput(2,1.2){\large BS}
\rput{135}(6,2){\psline[linewidth=0.08,linecolor=gray](-0.7,0)(0.7,0)}
\rput{135}(3,5){\psline[linewidth=0.08,linecolor=gray](-0.7,0)(0.7,0)}
\pswedge[linewidth=0.08](7.5,2){.5}{270}{90}
\pswedge[linewidth=0.08](6,0.5){.5}{180}{360}
\pswedge[linewidth=0.08](3,6.5){.5}{0}{180}
\pswedge[linewidth=0.08](1.5,5){.5}{90}{270}
\psline[linewidth=0.015](0,2)(7.5,2)
\psline[linewidth=0.015](3,2)(3,6.5)
\rput(6,5){\shortstack{local\\oscillator\\fields}}
\psline[linewidth=0.015](6,4.2)(6,0.5)
\psline[linewidth=0.015](1.5,5)(5,5)
\psline{->}(1.4,2)(1.5,2)
\psline{->}(4.4,2)(4.5,2)
\psline{->}(6.8,2)(6.9,2)
\psline{->}(6,3.2)(6,3.1)
\psline{->}(4.2,5)(4.1,5)
\psline{->}(3,3.4)(3,3.5)
\psline{->}(3,5.7)(3,5.8)
\psline{->}(3,1.28)(3,1.33)
\psline[linewidth=0.015,linestyle=dashed,dash=3pt 2pt](3,0.5)(3,2)
\rput(0.5,2.4){\large $\hat{a}$}
\rput(3.4,0.8){\large $\hat{v}$}
\pscircle[linewidth=0.05](0.5,7.5){0.4}
\psline[linewidth=0.05](1,5)(0.5,5)(0.5,7.1)
\psline[linewidth=0.05](3,7)(3,7.5)(0.9,7.5)
\rput(0.5,7.5){\large $-$}
\pscircle[linewidth=0.05](8.5,-0.5){0.4}
\psline[linewidth=0.05](6,0)(6,-0.5)(8.1,-0.5)
\psline[linewidth=0.05](8,2)(8.5,2)(8.5,-0.1)
\rput(8.5,-0.5){\large $-$}
\end{pspicture}
\end{center}
\caption{Double homodyne detection setup. The signal field, denoted
by the annihilation operator $\hat{a}$, is divided using a 50:50 beam
splitter BS. The two outgoing fields fall onto balanced
homodyne detectors. The local oscillator phases are adjusted such that
two conjugate quadratures are measured. In the quantum mechanical
description of the setup, one has to take into account the vacuum field
$\hat{v}$ entering through the unused input port of the beam splitter BS.}
\label{Fig:DoubleHomodyneSetup}
\end{figure}

In homodyne detection, we have to select the phase of the local
oscillator, which defines the measured quadrature. Simultaneous
measurement of different quadratures is not possible, because they
correspond to noncommuting observables: it is easy to check that
for example $[x_\theta, x_{\theta+\pi/2}] = i$.  However, we may try
to circumvent this difficulty by splitting first the input beam on
a 50:50 beam splitter and performing {\em two} homodyne measurements
on the outgoing fields. This is the idea of double homodyne detection.
The corresponding setup is shown in Fig.~\ref{Fig:DoubleHomodyneSetup}.
The signal field, described by an annihilation operator $\hat{a}$,
is divided using the 50:50 beam splitter BS. The two outgoing beams
are measured with two separate balanced homodyne detectors. The phases
of local oscillators can be independently adjusted in each of the arms
of the setup, which allows one to measure two arbitrary quadratures of
the fields leaving the beam splitter BS.  For simplicity, let us choose
the two local oscillator phases to be $0$ and $\pi/2$, and to denote
the corresponding quadratures by $\hat{q} = \hat{x}_0$ and $\hat{p} =
\hat{x}_{\pi/2}$.  These two quantities commute to the imaginary unit
$[\hat{q}, \hat{p}] = i$, and they are optical analogs of the position
and the momentum operators for a particle.

In the quantum description of the setup, we need to take into account
the vacuum field entering through the unused input port of the beam
splitter BS dividing the signal field. This vacuum field is denoted with the
annihilation operator $\hat{v}$ in Fig.~\ref{Fig:DoubleHomodyneSetup}.
The quadratures measured at the two homodyne detectors are given by
the combinations
\begin{equation}
\label{Eq:q1p2}
\hat{q}_1 = \frac{1}{\sqrt{2}}(\hat{q}_a + \hat{q}_v), \hspace{1cm}
\hat{p}_2 = \frac{1}{\sqrt{2}}(\hat{p}_a - \hat{p}_v)
\end{equation}
where the indices $a$ and $v$ denote quadrature operators corresponding
the signal and the vacuum mode respectively. The form of these
combinations follows directly from Eq.~(\ref{Eq:50:50}), describing
transformation of the field operators at a 50:50 beam splitter.

The probability distribution $p(q_1,p_2)$ for the outcomes of the
measurement is now defined on the two-dimensional space spanned by the
variables $q_1$ and $p_2$. Analogously to the previous section, it will
be more convenient to use the generating function
$Z(\lambda_1, \lambda_2)$, which depends now on two parameters $\lambda_1$
and $\lambda_2$:
\begin{equation}
Z(\lambda_1, \lambda_2)
=
\int \prtext{d}\lambda_1
\int \prtext{d}\lambda_2
\,
e^{i\lambda_1 q_1 + i \lambda_2 q_2} p(q_1,p_2).
\end{equation}
The generating function describing the joint measurement of the
quadratures $\hat{q}_1$ and $\hat{p}_2$ is given by a straightforward
generalization of Eq.~(\ref{Eq:Zthetalambda}):
\begin{eqnarray}
Z(\lambda_1, \lambda_2) & = & \langle \exp(i\lambda_1 \hat{q}_1 + i
\lambda_2 \hat{p}_2) \rangle_{a,v} 
\nonumber \\
& = & \exp\left(-\frac{1}{8}(\lambda_1^2 + \lambda_2^2)\right)
\left\langle \exp\left(\frac{i}{\sqrt{2}}
( \lambda_1 \hat{q}_a + \lambda_2 \hat{p}_a ) \right)
\right\rangle_a .
\end{eqnarray}
In the second line we have evaluated explicitly the quantum
expectation value over the vacuum mode.
The joint probability distribution $p(q_1,p_2)$
can be obtained from the double inverse Fourier
transform of the generating function. We shall rearrange this expression
 to the form:
\begin{eqnarray}
p(q_1,p_2) & = & \frac{1}{(2\pi)^2} \int \prtext{d}\lambda_1
\int \prtext{d}\lambda_2 \, e^{-i\lambda_1 q_1 - i \lambda_2 p_2}
Z(\lambda_1,\lambda_2) 
\nonumber \\
& = & \frac{1}{\pi^2}
\int \prtext{d}^2 \zeta \, e^{-|\zeta|^2/2 + \zeta^\ast (q_1 + ip_2)
- \zeta (q_1 - i p_2)}
\langle e^{\zeta \hat{a}^\dagger - \zeta^\ast \hat{a}} \rangle_a
\end{eqnarray}
where we have substituted $\zeta = (i\lambda_1 - \lambda_2)/2$. The last
expression can be directly related to the definition of quasidistribution
functions in Eq.~(\ref{Eq:QuasiDistDef}), with $\alpha = q_1 + ip_2$,
and $s=-1$. Thus, the joint probability distribution of homodyne
events measured in double homodyne detection is equal to the $Q$ function
of the mode $\hat{a}$:
\begin{equation}
p(q_1,p_2)  = Q_a(q_1 + ip_2) .
\end{equation}

One may wonder how this formula changes when we inject an arbitrary state
in the second input port of the beam splitter BS dividing the signal field.
In this case $Z(\lambda_1, \lambda_2)$ can be factorized to the product
of the symmetrically ordered characteristic functions for the position and
the momentum:
\begin{equation}
Z(\lambda_1, \lambda_2)  = 
\left\langle \exp\left(\frac{i}{\sqrt{2}}
( \lambda_1 \hat{q}_a + \lambda_2 \hat{p}_a ) \right)
\right\rangle_a
\left\langle \exp\left(\frac{i}{\sqrt{2}}
( \lambda_1 \hat{q}_v + \lambda_2 \hat{p}_v ) \right)
\right\rangle_v .
\end{equation}
The inverse Fourier transform maps the product of the symmetrically ordered
characteristic functions onto a convolution of the corresponding Wigner
functions. After a simple calculation,  we obtain:
\begin{equation}
p(q_1,p_2) = 2 \int \prtext{d}q \int \prtext{d}p \,
W_a(q,p) W_v(\sqrt{2}q_1-q,\sqrt{2} p_2 -p),
\end{equation}
where $W_a(q,p)$ and $W_v(q,p)$ are the Wigner functions describing the
quantum state of the fields incident on the beam splitter BS.

The above results illustrates the operational approach to the joint
measurement of the position and the momentum \cite{WodkPRL84,StenAnP92}.
These two observables do not commute and they cannot be measured
simultaneously. Nevertheless, we may introduce an auxiliary system,
called the ``quantum ruler'', and measure two commuting combinations of
positions and momenta. Such a pair of combinations has been defined in
Eq.~(\ref{Eq:q1p2}). These two operational observables can be detected
simultaneously, and their measurement yields a joint two-dimensional
probability distribution of two variables which can be related to
the position and the momentum. The resulting operational phase space
distribution is given by a convolution of the Wigner functions of the
measured system and the ruler. Double homodyne detection is an optical
realisation of this approach, with the role of the quantum ruler played by
the vacuum field. The vacuum field is described by the gaussian Wigner
function, and the double homodyne detection yields a smeared Wigner
function of the signal field, which coincides with the $Q$ function.

Double homodyne detection has been realized experimentally by Walker and
Caroll \cite{WalkCaroOQE86}. A thorough discussion of this technique
can be found in the article by Walker \cite{WalkJMO87}. The same
experimental scheme has been applied in the operational measurement of
the quantum phase \cite{NohFougPRL91}, and it was shown later that the
phase distribution measured in this scheme corresponds to the radially
integrated $Q$ function \cite{FreySchlPRA93,LeonPaulPRA93a}.

\section{Optical homodyne tomography}
\label{Sec:OHT}

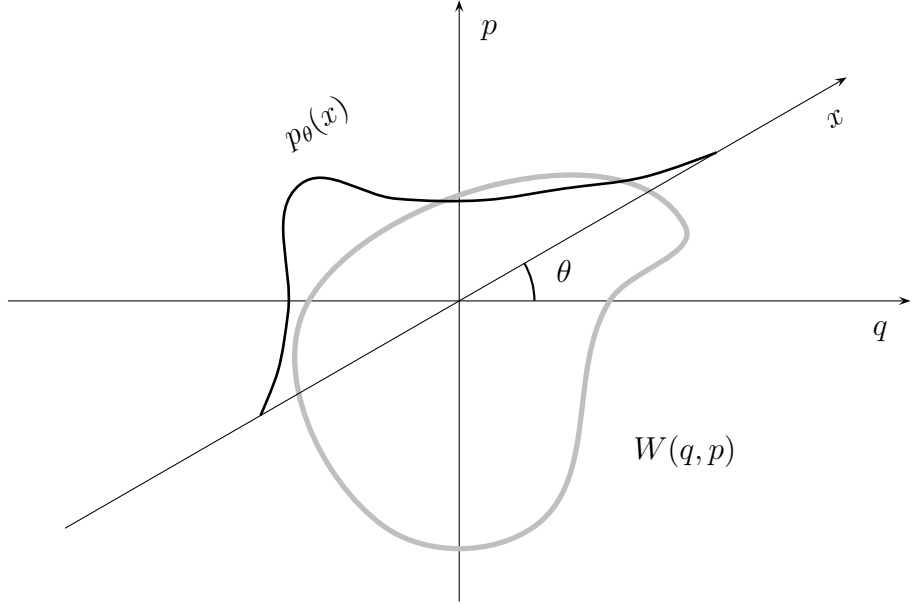
\begin{figure}
\begin{center}
\psset{unit=2cm}
\begin{pspicture}(-3,-2)(3,2)
\psline[linewidth=.4pt]{->}(0,-2)(0,2)
\rput(.2,1.8){$p$}
\psline[linewidth=.4pt]{->}(-3,0)(3,0)
\rput(2.8,-.2){$q$}
\psccurve[linewidth=2pt,linecolor=lightgray]%
(-0.5,-1.5)(-1,0)(1.5,0.5)(1,0)(0.5,-1.5)
\rput(1.5,-1){$W(q,p)$}
\psarc(0,0){.5}{0}{30}
\rput(0.7,0.2){$\theta$}
\rput{30}(0,0){\begin{pspicture}(-3,-1)(3,1)
\psline[linewidth=.4pt]{->}(-3,0)(3,0)
\rput(2.8,-.2){$x$}
\pscurve[linewidth=1pt]%
(-1.5,0)(-1.25,.2)(-1,.5)(-.5,1.2)(0,.8)(.5,.5)(1,.3)(1.5,.1)(2,0)
\rput(-.2,1.5){$p_{\theta}(x)$}
\end{pspicture}
} 
\end{pspicture}
\end{center}
\caption{Quadrature distributions $p_\theta(x)$ are one-dimensional
projections of the Wigner function $W(q,p)$, obtained by integrating
it along parallel stripes in the phase space.}
\label{Fig:WignerProjections}
\end{figure}

In contrast to the $Q$ function, the Wigner function does not have the
operational meaning of a probability distribution, simply because it
may take negative values. Therefore, one cannot design an experiment,
in which the joint statistics of two real variables would be described
by the Wigner function.  Nevertheless, one-di\-men\-sional projections of
the Wigner function are positive definite. Furthermore, these projections
describe quadrature distributions according to the formula:
\begin{equation}
\label{Eq:QuadDistandWigner}
p_\theta (x) = \int \mbox{\normalsize d}y \,
W(x \cos\theta - y \sin\theta, x \sin\theta + y \cos\theta) .
\end{equation}
This equation is a generalization of the marginal properties of the
Wigner function for the position and the momentum.  As we have seen
in Sec.~\ref{Sec:BalancedHomodyne}, quadrature distributions can be
measured by means of a balanced homodyne detector. Thus we may obtain
experimentally a family of one-dimensional projections of the Wigner
function, depicted schematically in Fig.~\ref{Fig:WignerProjections}. What
we would like to do, is to reconstruct from these ``shadows'' the
two-dimensional Wigner function.

A very similar problem is encountered in medical tomography
\cite{BarrPIO84}. By measuring
absorption of radiation across the body, we can obtain density of the
tissue integrated along the direction of the measurement. These data
are subsequently processed to reconstruct the full density distribution
using numerical back-projections algorithms. In analogy to this technique,
the method of reconstructing the Wigner function from homodyne statistics
has been called {\em optical homodyne tomography}.

The relation between $p_\theta(x)$ and $W(q,p)$ defined
in Eq.~(\ref{Eq:QuadDistandWigner}) can be viewed as a
transformation between two-dimensional functions according to the formula:
\begin{equation}
p_\theta(x) = \int \prtext{d}q \int \prtext{d}p \,
\delta(x-q\cos\theta-p\sin\theta) W(q,p),
\end{equation}
which is known in the field of image processing as the {\em Radon
transform}. A function $W(q,p)$ of two real variables $q$ and $p$
is transformed into another function $p_\theta(x)$ which depends on
the angular variable $\theta$ and a real variable $x$.

Inversion of the relation between the Wigner function and
quadrature distributions becomes quite obvious, if we rewrite
Eq.~(\ref{Eq:QuadDistandWigner}) in terms of the Fourier transforms
of both the sides. The generating function for the quadrature distribution
can be expressed using the Wigner function as:
\begin{eqnarray}
Z_\theta(\lambda) & = & \int \prtext{d}x \, e^{i\lambda x} p_\theta(x)
\nonumber \\
& = & \int \prtext{d}x \int \prtext{d}y \, e^{i\lambda x}
W(x \cos\theta - y \sin\theta, x \sin\theta + y \cos\theta)
\nonumber \\
& = & \int \prtext{d}q \int \prtext{d}p \, e^{i\lambda
(q\cos\theta+p\sin\theta)} W(q,p) .
\end{eqnarray}
The last expression is simply the Fourier transform of the Wigner
function taken at the point $(q\cos\theta,p\sin\theta)$. Thus, the projection
relation expressed in terms of the Fourier transforms consists in the change
of the coordinate system, from the Cartesian one (Wigner function) to the
polar one (quadrature distributions).

With this observation in hand, the way to invert
Eq.~(\ref{Eq:QuadDistandWigner}) is straightforward: we need to write
the Wigner function as the inverse Fourier transform, and to change the
integration variables from Cartesian to polar. This allows us to insert
the generating function for quadrature distributions:
\begin{equation}
W(q,p) = \frac{1}{(2\pi)^2} \int_{-\infty}^{\infty} |\lambda| \prtext{d}
\lambda \int_{0}^{\pi} \prtext{d}\theta \,
e^{-iq\lambda\cos\theta - i p \lambda \sin \theta} Z_\theta (\lambda) .
\end{equation}
Expressing $Z_\theta (\lambda)$ in terms of quadrature distributions and
performing the integral over $\lambda$ yields:
\begin{equation}
W(q,p) = \frac{1}{2\pi^2} \int_{-\infty}^{\infty}
\prtext{d}x \int_{0}^{\pi} \prtext{d}\theta \, p_\theta(x) 
\frac{\prtext{d}}{\prtext{d}x} P \frac{1}{x-q\cos\theta-p\sin\theta}
\end{equation}
where $P$ denotes the principal value. The above formula is known
as {\em the inverse Radon transform}. It is clearly seen that this
transformation is singular. Therefore, its numerical implementation
is quite complicated. When processing experimental distributions
$p_\theta(x)$, which are affected by statistical noise, one has to apply
a regularization scheme.

The close link between the quadrature distributions and the
Wigner function could be noted already during the discussion of the
balanced homodyne detector. The first expression for $Z_\theta(\lambda)$
in Eq.~(\ref{Eq:Zthetalambda}) is exactly the symmetrically ordered
characteristic function that appears in Eq.~(\ref{Eq:WigSymOrdered}),
with $\zeta = i\lambda e^{i\theta}/\sqrt{2}$.

The first experimental realization of optical homodyne tomography has
been demonstrated by Smithey {\em et al.} \cite{SmitBeckPRL93}. This
seminal work has been followed by extensive theoretical and experimental
research. The effects of imperfect detection were analysed
\cite{LeonPaulPRA93}, and
it was shown that in such a case the inverse Radon transform yields
a generalized quasidistribution function with the ordering parameter equal
to $-(1-\eta)/\eta$, where $\eta$
is the efficiency of the photodetectors. Thus, detector losses result
in blurring of the measured Wigner function. A more fundamental problem
related to optical homodyne tomography was the determination of other
quantum state representations from homodyne statistics. In principle,
once we have the Wigner function, we can evaluate the expectation
value of any quantum observable $\hat{O}$, and obtain for example
the density matrix in the Fock basis.  However, it would be appealing
to reconstruct the observables directly from the homodyne statistics,
in order to avoid the detour via the singular inverse Radon transform.
The formula needed for this purpose is of the form:
\begin{equation}
\langle \hat{O} \rangle
=
\int_{-\infty}^{\infty} \prtext{d}x \int_{0}^{\pi} \prtext{d}\theta
\, f_{\hat{O}}(x,\theta) p_\theta(x),
\end{equation} 
where $f_{\hat{O}}(x,\theta)$ is called the {\em pattern
function} related to the observable $\hat{O}$. The problem
of deriving pattern functions for the elements of the density
matrix in the Fock basis was studied first by D'Ariano {\em et al.}
\cite{DAriMaccPRA94}.  It was later generalized to a more fundamental form
\cite{DAriLeonPRA95,LeonPaulPRA95}.  The statistical error of optical
homodyne tomography has been thoroughly studied in a series of papers
\cite{DAriMaccPLA94,DAriMaccQSO97,LeonMunrOpC96,DAriPariPLA97}.  On the
experimental side, optical homodyne tomography has been demonstrated for
cw fields \cite{BreiMullJOSAB95}, and a beautiful gallery of squeezed
states of light has been presented \cite{BreiSchiNAT97}. An analogous
tomographic method has been used to characterize transversal degrees
of freedom of a laser beam \cite{EppichPreprint}.

\section{Random phase homodyne detection}
\label{Sec:RandomHomodyne}

In the context of optical homodyne tomography, a new technique for
measuring light has been developed. This
technique is balanced homodyne detection
with the phase $\theta$ made a uniformly distributed random variable
\cite{MunrBoggPRA95}.
The distribution of events $p_{\cal R}(x)$
observed in such a case is described by phase-averaged homodyne
statistics:
\begin{equation}
\label{Eq:pRx}
p_{\cal R}(x) = \frac{1}{2\pi} \int_{0}^{2\pi} \prtext{d}\theta \,
p_\theta(x) =
\frac{1}{2\pi} \int_{0}^{2\pi} \prtext{d}\theta \,
\left\langle |x\rangle_{\theta\, \theta} \langle x| \right\rangle.
\end{equation}
In this regime, the phase sensitivity of homodyne detection is
completely lost, and the phase-averaged homodyne statistics $p_{\cal
R}(x)$ contains information only on phase-independent properties of the
measured light. Nevertheless, random phase homodyne detection has some
advantages compared to direct photodetection. First, ultrafast sampling
time can be achieved by using the local oscillator field in the form of
a short pulse. Second, information on the photon distribution is carried
by two rather intense fields, which can be detected with substantially
higher efficiency than the signal field itself. This feature has enabled
an experimental demonstration of even-odd oscillations in the photon
distribution of the squeezed vacuum state \cite{SchiBreiPRL96}.

Let us now see, how the phase-averaged homodyne statistics depends
on the photon distribution. We shall use the fact that eigenvectors
of the quadrature operator $\hat{x}_\theta$ can be obtained from
the position eigenvectors $|x\rangle$ by the unitary transformation
$|x\rangle_\theta = e^{i\theta\hat{a}^{\dagger} \hat{a}} |x\rangle$.
This unitary transformation is diagonal in the Fock basis, and
$e^{i\theta\hat{a}^{\dagger} \hat{a}} |n \rangle = e^{in\theta}
|n \rangle$.
Introducing two decompositions of the identity operator in the Fock basis,
we have:
\begin{eqnarray}
p_{\cal R}(x) & = &
\frac{1}{2\pi} \int_{0}^{2\pi} \prtext{d}\theta 
\sum_{m,n=0}^{\infty} 
 \left\langle |m\rangle \langle m |x\rangle_{\theta\, \theta} \langle x| 
n \rangle \langle n| \right\rangle
\nonumber \\
& = &
\sum_{m,n=0}^{\infty} 
\langle m | x \rangle \langle x | n \rangle
\int_{0}^{2\pi} \frac{\prtext{d}\theta}{2\pi}
e^{i(n-m)\theta}
\langle | m \rangle \langle n | \rangle
\nonumber \\
& = &
\label{pRx=sum}
\sum_{m=0}^{\infty} | \langle m | x \rangle |^2
\langle | m \rangle \langle m | \rangle
\end{eqnarray}
Thus, occupations of the Fock states given by the expectation
values $\langle | m \rangle \langle m | \rangle$ contribute
to the phase-averaged homodyne statistics with the coefficients
\begin{equation}
| \langle m | x \rangle |^2 = \frac{1}{\sqrt{\pi}2^m m!}
H^2_m(x) e^{-x^2},
\end{equation}
where $H_m(x)$ denote Hermite polynomials. These coefficients
correspond to the position distributions for the eigenstates
of the harmonic oscillator.

Integrating $p_{\cal R}(x)$ with appropriate pattern functions, we may
reconstruct the photon statistics of the measured field, as well as
other phase independent observables. An alternative method of processing
the phase-averaged homodyne statistics, based on maximum-likelihood
estimation, has been described in Ref.~\cite{BanaPRA98}.

\chapter{Direct probing of quantum phase space}
\label{Chap:Direct}

\markright{CHAPTER \thechapter . \uppercase{Direct probing of quantum phase space}}

Quasidistribution functions contain complete characterization of the
quantum state. An interesting and nontrivial problem is how to determine
quasidistributions from quantities which can be detected in a
feasible experimental scheme. In the previous chapter, we have discussed
two experimental techniques for measuring the quantum state of a light
mode: double homodyne detection, and optical homodyne tomography. These
techniques are based on detection of quadratures, which are continuous
variables. The two-dimensional probability distribution observed in double
homodyne detection yields directly the $Q$ function of a light mode.
In optical homodyne tomography, a family of one-dimensional projections
of the Wigner function is measured and then processed numerically using
the back-projection algorithm.

In this chapter, we shall present a different approach to
measuring quasidistributions of a light mode. We have seen in
Chap.~\ref{Chap:Representations} that quasidistributions at a specific
point of the phase space are given by expectation values of certain
Hermitian operators. We shall demonstrate that this definition leads to a
novel optical scheme for measuring quasidistribution functions of light.
This scheme is based on photon counting. In contrast to the homodyne
techniques discussed in the previous chapter, it is essential in our
approach that the signal obtained from the detector is discrete, and
that it is described by an integer variable characterizing the number
of absorbed photons.

Our starting point in Sec.~\ref{Sec:Wignerandphoton} will be a simple
relation between the Wigner function and the photon statistics. We show
that this relation can be implemented using a simple optical setup,
which allows one to determine quasidistributions from photon statistics.
In Sec.~\ref{Sec:PhaseSpacePicture} we discuss the proposed scheme using
the phase space picture. This picture explains in an intuitive way how
physical parameters of the setup determine the measured quantity. In
Sec.~\ref{Sec:Generalization} we generalize the relation linking the
quasidistributions and the photon statistics. We also discuss effects
of non-unit detector efficiency. The full multimode theory of the
proposed setup is developed in Sec.~\ref{Sec:MultimodeApproach}. We
show there that the direct method for measuring quasidistributions
can be easily extended to multimode radiation. Finally, in
Sec.~\ref{Sec:ExamplesOfCountStatistics} we discuss theoretically
several examples of photon statistics which would be obtained from the
photodetector when measuring quasidistribution functions.

\section{Wigner function and photon statistics}
\label{Sec:Wignerandphoton}

We will start from deriving a simple relation between the Wigner function at
the origin of the complex phase space $W(0)$ and the photon statistics.
Let us take the operator (\ref{Eq:Whatalphas}) for $\alpha=0$ and $s=0$,
and expand it into a power series:
\begin{eqnarray} 
\label{Eq:W(0,0)} 
{\hat W}(0;0) &  = & \frac{2}{\pi}  : \exp \left( - 2 
\hat{a}^\dagger \hat{a} \right) : \nonumber \\ 
& = & \frac{2}{\pi} \sum_{n=0}^{\infty} (-1)^n  : 
e^{- \hat{a}^\dagger \hat{a}} \frac{(\hat{a}^\dagger 
\hat{a})^n}{n!} :\nonumber \\ 
& = & \frac{2}{\pi}\sum_{n=0}^{\infty} (-1)^n | n \rangle 
\langle n|  .
\end{eqnarray} 
In the last step, we have used the normally ordered operator
representation of the $n$ photon number projection operator. The last
expression contains a sum which assigns $+1$ to even Fock states, and
$-1$ to odd Fock states.  Therefore, the whole sum is simply the parity
operator, and the Wigner function at the phase space origin is given, up
to the front factor $2/\pi$, by its expectation value \cite{RoyePRA77}.
Taking the quantum average of Eq.~(\ref{Eq:W(0,0)}) gives:
\begin{equation} 
\label{Eq:W(0)=sum} 
W(0) = \langle {\hat W}(0;0) \rangle = \frac{2}{\pi} 
\sum_{n=0}^{\infty} 
(-1)^n p_{n} 
\end{equation} 
where the value $p_{n}$ appearing 
in this expansion is just the probability of counting $n$ 
photons  by an ideal photodetector. 
Thus the photon statistics allows one to evaluate the Wigner 
function at the origin of the phase space.

It would be appealing to generalize this relation to an arbitrary point
of the phase space. In principle, the only thing we have to do is to shift
the system or equivalently the frame of reference in the phase
space. The problem is, how to realize this in practice for optical
fields. We will show that this goal can be achieved using a very simple
optical arrangement, and that in a certain limit the displacement
transformation is realized.

\begin{figure}
\begin{center}
\psset{unit=.3375in}
\begin{pspicture}(10,6)
\rput{45}(4.5,4){%
\psframe[fillstyle=solid,fillcolor=lightgray](-2,-.1)(2,.1)%
} 
\psset{linewidth=.5mm}
\psline{->}(1,4)(4,4)
\psline{->}(5,4)(8,4)
\psline{->}(4.5,.5)(4.5,3.5)
\pswedge(8.4,4){.6}{270}{90}
\rput(2.5,2){\large BS}
\rput(8.6,2.8){\large PD}
\rput(1.5,4.4){\large $\hat{a}_S$}
\rput(7.4,4.4){\large $\hat{a}_{\mbox{\small\rm out}}$}
\rput(5.0,0.9){\large $\hat{a}_P$}
\end{pspicture}
\end{center}
\caption{Experimental setup for measuring directly quasidistribution
functions of a single light mode. BS denotes the
beam splitter, PD is the photodetector, and the annihilation
operators of the modes are indicated.\label{Fig:Setup}}
\end{figure}
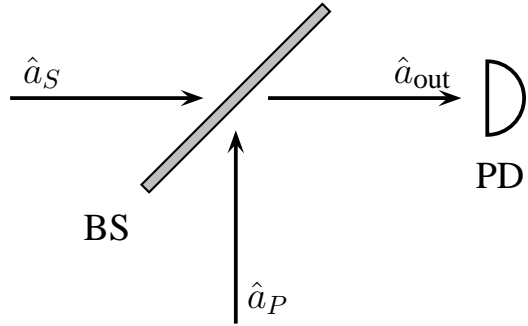

Let us consider the setup presented in Fig.~\ref{Fig:Setup}.  We take
the detected field to be a superposition of two single-mode fields, which
we shall call the signal and the probe. The corresponding annihilation
operators are denoted by $\hat{a}_S$ and $\hat{a}_P$ respectively. The
superposition is realized by means of a beam splitter BS with the power
transmission characterized by the parameter $T$.  In general, the action
of the beam splitter is described by an $\prtext{SU}(2)$ transformation
between the annihilation operators of the incoming and outgoing modes
\cite{VogelWelschBeamSplitter}. As the phase shifts appearing in this
transformation can be eliminated by appropriate redefinition of the modes,
the annihilation operator $\hat{a}_{\prtext{\scriptsize out}}$ of the
outgoing mode falling onto the detector surface can be assumed to be
a combination
\begin{equation} 
\hat{a}_{\prtext{\scriptsize out}} 
= \sqrt{T} \hat{a}_S - \sqrt{1-T} \hat{a}_P. 
\end{equation} 

The photon statistics of the field
$\hat{a}_{\prtext{\scriptsize out}}$ is used to evaluate
the alternating series according
to Eq.~(\ref{Eq:W(0)=sum}).
In this way we obtain
the Wigner function of the outgoing mode at the origin of the phase
space. This quantity depends on the quantum state of both the modes
$\hat{a}_S$ and $\hat{a}_P$.  The Wigner function of the outgoing mode
at the phase space origin is given in terms of the incoming modes by
the expectation value of
\begin{eqnarray} 
{\hat W}_{\prtext{\scriptsize out}}(0;0) & = & \frac{2}{\pi} \, : 
\exp \left( - 2 
\hat{a}^\dagger_{\prtext{\scriptsize out}} \hat{a}_{\prtext{\scriptsize
out}} \right) : 
\nonumber \\ 
\label{Eq:W=SP} 
& = & \frac{2}{\pi} \, : \exp \left( - 2 T 
(\hat{a}^{\dagger}_{S}- \sqrt{(1-T)/T}\hat{a}^{\dagger}_{P}) 
 ( \hat{a}_{S} 
- \sqrt{(1-T)/T} \hat{a}_{P} ) \right) :  \;  .
\end{eqnarray} 
This simple relation provides an interesting link between the detected
quantity and the $S$ mode. Let us consider the case when the probe
field is a coherent state $\hat{a}_P | \alpha \rangle = \alpha | \alpha
\rangle$ uncorrelated with the signal mode. Performing the quantum average
over the $P$ mode in Eq.\ (\ref{Eq:W=SP}) is straightforward due to the
normal ordering of the operators. Taking the expectation value over the
signal and recalling the normally ordered definition of quasidistributions
(\ref{Eq:WalphasNormOrd}), we can recognize a quasidistribution of
the signal field. Comparing its parameters, we obtain that  the Wigner
function (\ref{Eq:W(0)=sum}) for the outgoing mode is proportional to
an $s=1-1/T$ ordered  quasidistribution function of the $S$ mode:
\begin{equation} 
\label{Eq:Wout(0)=W_S} 
W_{\prtext{\scriptsize out}}(0)= \frac{1}{T} W_S \left( \sqrt{\frac{1-T}{T}} 
\alpha ; - \frac{1-T}{T} \right),
\end{equation}
taken at the point $\sqrt{(1-T)/T} \alpha$. Thus our setup delivers
directly the value of the signal quasidistribution function at the phase
space point dependent on the amplitude and the phase of the probe coherent
state. Since both these parameters can be controlled experimentally
without difficulties, we may simply scan the phase space by changing
the amplitude and the phase of the probe field and thus determine
the complete quasidistribution function. Eq.\ (\ref{Eq:Wout(0)=W_S})
shows that its ordering depends on the beam splitter transmission. For
$T$ near one the ordering is close to zero, which means that the measured
quasidistribution approaches the Wigner function of the signal field.
In contrast to optical homodyne
tomography the Wigner function is measured directly and
no sophisticated computer processing of the
experimental data is necessary. The quantity measured in the 
experiment is proportional to the quasiprobability distribution 
at the phase space point depending only on the amplitude and phase of the
probe state.

\section{Phase space picture}
\label{Sec:PhaseSpacePicture}

The quantity measured in the setup discussed in the previous section has an
interesting phase space interpretation for arbitrary states of the $S$
and $P$ modes. To show this, we shall disentangle the two-mode operator
defined in Eq.~(\ref{Eq:W=SP}) using the following Gaussian integral of
normally ordered operators for the $S$ and $P$ modes:
\begin{eqnarray} 
\hat{W}_{\prtext{\scriptsize out}}(0;0)
& = & \frac{4}{\pi^2}  \int \prtext{d}^2 \beta \, 
: \exp \left( - 2 (\sqrt{T} \beta^{*} - \hat{a}^{\dagger}_{P} ) 
(\sqrt{T} \beta - \hat{a}_{P} ) \right): 
\nonumber \\ 
\label{U=S+P} 
&  &  \times : \exp \left( - 2 
(\sqrt{1-T}\beta^{*} -\hat{a}^{\dagger}_{S} ) 
(\sqrt{1-T}\beta - \hat{a}_{S} ) \right) : \; . 
\end{eqnarray} 
Under the assumption that the $S$ and $P$ modes are uncorrelated,
we can evaluate separately expectation values over the signal
and the probe modes. It is easily seen that these expectation values
yield the values of Wigner functions of the signal and the probe modes
$W_S(\sqrt{1-T}\beta)$ and $W_P (\sqrt{T} \beta)$. Thus we obtain
the following expression for the 
quantity detected by our setup: 
\begin{eqnarray} 
\label{Eq:W(0)=int} 
W_{\prtext{\scriptsize out}}(0)
& = & \int \prtext{d}^2 \beta \, W_S(\sqrt{1-T} 
\beta) W_P (\sqrt{T} \beta) \nonumber \\ 
& = & \frac{1}{1-T} \int \prtext{d}^2\beta \, W_S(\beta) W_P 
(\sqrt{T/(1-T)} \beta). 
\end{eqnarray} 
This formula establishes a connection between the photon number parity of
the outgoing mode and the Wigner functions of the $S$ and $P$ modes.
The object of interest in the above formula is the Wigner function
of the signal field $W_S(\beta)$. The quantity we obtain from the
measurement is $W_{\prtext{\scriptsize out}}(0)$, which is given
by the integral of $W_S(\beta)$ with the function $ W_P(\sqrt{T/(1-T)}
\beta)$. Let us now see, what information on the signal state can be
obtained from this integral.

In the case when the beam splitter splits the light equally, i.e.\
the power transmission is $T=50\%$, we have $\sqrt{T/(1-T)} = 1$ and
$W_{\prtext{\scriptsize out}}(0)$ is simply a doubled overlap of the
signal and probe Wigner functions. If we take the probe field to be
a coherent state with variable amplitude and phase, we obtain the $Q$
function of the signal field. Analogously to double homodyne detection
discussed in Sec.~\ref{Sec:DoubleHomodyne}, the probe Wigner function
can be considered as a ``quantum ruler'', which smoothes the signal
Wigner function to a positive definite phase space distribution.

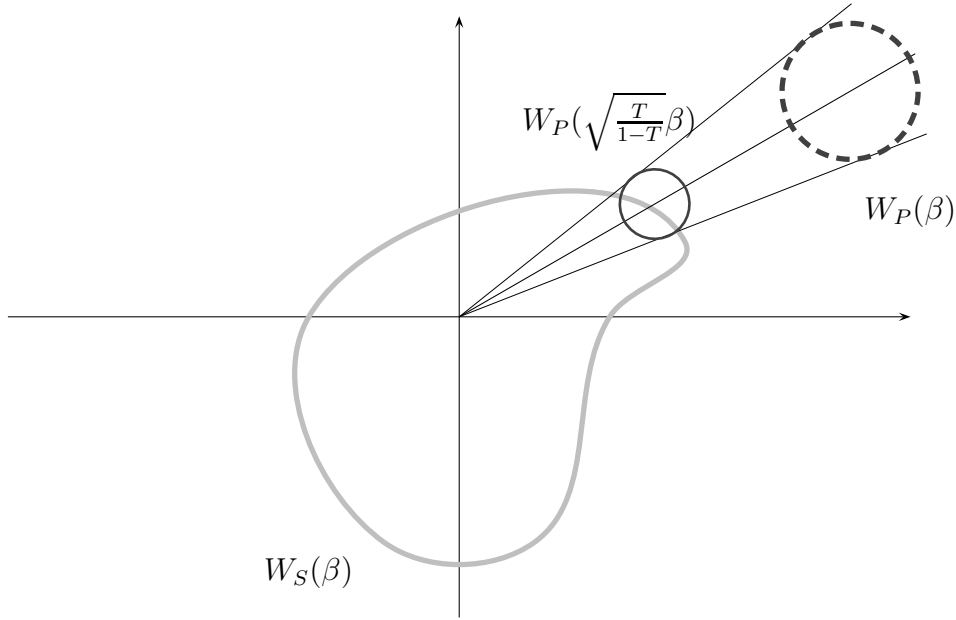
\begin{figure}[t]
\begin{center}
\psset{unit=2cm}
\begin{pspicture}(-3,-2)(3,2.2)
\psline[linewidth=.4pt]{->}(0,-2)(0,2)
\psline[linewidth=.4pt]{->}(-3,0)(3,0)
\psccurve[linewidth=2pt,linecolor=lightgray]%
(-0.5,-1.5)(-1,0)(1.5,0.5)(1,0)(0.5,-1.5)
\rput(-1,-1.7){\normalsize $W_S (\beta)$}
\rput{30}(0,0){%
\psline[linewidth=.2pt](0,0)(3.3,-.5)
\psline[linewidth=.2pt](0,0)(3.3,.5)
\psline[linewidth=.4pt](0,0)(3.5,0)
\pscircle[linewidth=2pt,linecolor=darkgray,linestyle=dashed](3,0){.47}
\pscircle[linewidth=1pt,linecolor=darkgray](1.5,0){.24}
}
\rput(3,.7){\normalsize $W_P (\beta)$}
\rput(1,1.3){\normalsize $W_P( \sqrt{\frac{T}{1-T}} \beta)$}
\end{pspicture}
\end{center}
\caption{Phase space interpretation of the observable measured
in the discussed setup. The Wigner function of the signal field
$W_S(\beta)$ is integrated with the rescaled Wigner function of the
probe field $W_P( \sqrt{\frac{T}{1-T}} \beta)$. If the rescaling
factor $\sqrt{\frac{T}{1-T}}$ is larger than one, the probe Wigner
function becomes effectively contracted in all directions. Of course,
the rescaled Wigner function does not describe physical fields at any
point of the setup.}
\label{Fig:Rescaling}
\end{figure}

In a general case, the phase space parameterization of the probe
Wigner function is rescaled by the factor $\sqrt{T/(1-T)}$. It is easy
to see that this factor can take an arbitrary positive value depending
on the beam splitter transmission. The most interesting region is for
$T>1/2$, where the scaling factor $\sqrt{T/(1-T)}$ is greater than one.
This situation is shown pictorially in Fig.~\ref{Fig:Rescaling}.
The scaling factor effectively ``contracts'' the probe Wigner function
in all the directions simultaneously. Consequently, the area occupied
by the rescaled probe Wigner function becomes smaller, and the integral
in Eq.~(\ref{Eq:W(0)=int}) provides more ``local'' information on the
behaviour of the signal Wigner function $W_S(\beta)$.
It is seen from the form of the scaling factor, that this effect of
``contraction'' grows unlimitedly with the beam splitter transmission
tending to one.

In a particular case of the
coherent probe field $|\alpha\rangle$, the $P$
mode Wigner function is of the form
\begin{equation}
\label{WPcoh}
W_P(\beta) = \frac{2}{\pi} \exp \left ( -2 | \beta - \alpha | ^2
\right).
\end{equation}
When $T$ tends to one, the rescaled probe Wigner function
$W_P(\sqrt{T/(1-T)}\beta)$ approaches the form of the delta function,
and the effect of smoothing becomes negligible.  Thus, in the limiting
case of $T\rightarrow 1$ the integral (\ref{Eq:W(0)=int}) yields the
value the signal Wigner function at a single phase space point. However,
it it seen from Eq.~(\ref{Eq:Wout(0)=W_S}) that the rescaling of the
probe Wigner function in the integral (\ref{Eq:W(0)=int}) has another
consequence.  When $T$ tends to one, the factor multiplying the probe
amplitude $\alpha$ becomes very small. This effect is seen also in
Fig.~\ref{Fig:Rescaling}.  Therefore, in order to scan the interesting
region of the signal phase space we need to use a probe field of large
intensity.  From the mathematical point of view, the discussed limiting
case involves two transitions: with $T\rightarrow 1$ and $|\alpha|
\rightarrow \infty$, such that the product $\sqrt{(1-T)/T}\alpha$ is
fixed. This product determines the point of the phase space at which
the signal Wigner function is measured.

Let us stress that the rescaled Wigner function, not obeying the
Heisenberg uncertainty principle, does not describe any fields appearing
physically in the setup; it is a purely abstract object introduced in
the phase space interpretation of our measurement scheme.

The integral representation derived in Eq.~(\ref{Eq:W(0)=int})
shows a connection of our setup with the model scheme of 
a phase space measurement, discussed in Sec.~\ref{Sec:DoubleHomodyne}.
In this model scheme, a filter device---a ``quantum ruler''---is
introduced in addition to the system
the measured phase space probability distribution 
is the  convolution of the system and filter Wigner functions. Our 
scheme is more general, since the Wigner function of the 
filter can be rescaled by an arbitrary factor. Consequently the 
rescaled probe Wigner function does not have to obey the 
Heisenberg uncertainty principle and it may even 
approach the shape of a delta function, which leads to the direct 
measurement of the Wigner function. 

\section{Generalization}
\label{Sec:Generalization}

In the remaining discussion we will introduce two generalizations. 
First we will make our considerations more realistic by taking 
into account the imperfection of the photodetector. When the 
detector efficiency is $\eta$, the probability of counting $n$ 
photons is given by the expectation value:
\begin{equation}
p_n = 
\left\langle :
e^{- \eta\hat{a}_{\prtext{\tiny out}}^\dagger
\hat{a}_{\prtext{\tiny out}}}      
\frac{(\eta\hat{a}_{\prtext{\scriptsize out}}^\dagger
\hat{a}_{\prtext{\scriptsize out}})^{n}}{n!} :
\right\rangle.
\end{equation}
The second extension is the substitution of the factor 
$(-1)^n$ in Eq.\ (\ref{Eq:W(0)=sum}) by $-(s+1)^n/(s-1)^{n+1}$, 
where $s$ is a real parameter. The origin and the role of the 
parameters $\eta$ and $s$ is different: $\eta$ describes 
experimental limitations, while $s$ is an artificial number 
introduced in the numerical processing of the measured data. 
With these two parameters we obtain the following simple 
generalization of the formula (\ref{Eq:W=SP}), when expressed in 
terms of the $S$ and $P$ modes 
\begin{eqnarray} 
{\hat W}_{\prtext{\scriptsize out}}^{(\eta)}(0;s)
&=& \frac{2}{\pi (1-s)} \sum_{n=0}^{\infty} \left( 
\frac{s+1}{s-1} \right)^{n} : 
e^{- \eta\hat{a}_{\prtext{\tiny out}}^\dagger
\hat{a}_{\prtext{\tiny out}}} 
\frac{(\eta\hat{a}_{\prtext{\scriptsize out}}^\dagger 
\hat{a}_{\prtext{\scriptsize out}})^{n}}{n!} : 
\nonumber \\ 
& = & \frac{2}{\pi (1-s)} : 
\exp \left( - \frac{2\eta}{1-s} 
\hat{a}_{\prtext{\scriptsize out}}^\dagger \hat{a}_{\prtext{\scriptsize
out}}\right) : 
\nonumber \\ 
& = & \frac{2}{\pi (1-s)} \  : \exp \left( - \frac{2\eta T}{1-s} 
(\hat{a}^{\dagger}_{S}-  \sqrt{(1-T)/T}\hat{a}^{\dagger}_{P} ) 
( \hat{a}_{S} - \sqrt{(1-T)/T} \hat{a}_{P} ) \right) : .
\nonumber \\
& &
\label{Eq:W(eta)} 
\end{eqnarray} 
The third line of this equation suggests that the parameter $s$ 
can be used to compensate the imperfectness of the 
photodetector. Indeed if we selected $s =1 - \eta$, we would 
determine the expectation value of $: \exp ( - 2 
\hat{a}_{\prtext{\scriptsize
out}}^\dagger \hat{a}_{\prtext{\scriptsize out}}):$ regardless 
of the detector efficiency. But in this case the factor multiplying the
probability of counting $n$ photons is $(1-2/\eta)^n$ and its magnitude
diverges to infinity with $n \rightarrow \infty$. Therefore we may expect
problems with the convergence of the series. Even when the series is
convergent, some singularities can be encountered in the processing
of the experimentally measured photon statistics, which is affected by
statistical fluctuations. Statistical noise may be a source of problems,
as the increasing factor in Eq.~(\ref{Eq:W(eta)}) causes that an
important contribution comes from the ``tail'' of the experimental
counts distribution, which usually has a very poor statistics.
The diverging factor $[(s+1)/(s-1)]^n$ amplifies fluctuations in this
tail, and consequently the final result has a huge statistical error.
The simplest way to avoid all these problems is to assume that the factors
multiplying the counts statistics are bounded, which is equivalent to
the condition $s \leq 0 $.  We shall discuss thoroughly the effects
of statistical noise in the next chapter. This discussion will fully
confirm the present conclusion drawn from qualitative arguments that in
principle it is not possible to compensate for detector losses, and that
in general we should restrict the range of $s$ to nonpositive values.

As before, let us consider the case when
the coherent state $|\alpha\rangle$
is employed as a probe. The expectation value of the generalized 
operator $\hat{W}_{\prtext{\scriptsize
out}}^{(\eta)}(0;s)$ is again given by the 
quasidistribution function of the signal mode: 
\begin{equation}
\label{Eq:W(eta)=WS} 
 \langle \hat{W}^{(\eta)}_{\prtext{\scriptsize
out}} (0;s) \rangle 
= \frac{1}{\eta T} W_S \left( 
\sqrt{\frac{1-T}{T}} \alpha ; - \frac{1-s-\eta T}{\eta T} 
\right). 
\end{equation} 
Let us now analyze the ordering of this function. As we discussed
in Sec.~\ref{Sec:Deconvolution},
although from a theoretical point of view an arbitrarily ordered 
distribution contains the complete characterization of the quantum 
state, experimental errors 
make it difficult to compute higher ordered distributions from the
measured one.
Thus what is interesting is the highest ordering 
achievable in our scheme. 
Analysis of the role of the parameter $s$ is the simplest, 
since the greater its value, the higher is the ordering obtained. 
Because we have restricted its range to nonpositive values, it
should be consequently set to zero. Thus we are left with the
product of the parameters $\eta$ and $T$.  For fixed $\eta$ the
highest ordering is achieved when $T\rightarrow 1$, and its limit
value is now $-(1-\eta)/\eta$. Under the assumption that $\eta$ and
$T$ are close to one, the ordering of the measured distribution is
effectively equal to this limiting value if the difference $1-T$ is
much smaller than $1-\eta$. This is a realistic condition for currently
existing photodetectors, which have the maximum efficiency about 80\%
\cite{KwiaSteiPRA93}.  Thus the highest ordering achievable in our
scheme is effectively determined by the photodetector efficiency and
is equal to $-(1-\eta)/\eta$. It is noteworthy that this is exactly
equal to the ordering of the distribution reconstructed tomographically
from data measured in the homodyne detection with imperfect detectors
\cite{LeonPaulPRA93}.

\section{Multimode approach}
\label{Sec:MultimodeApproach}

So far, we have discussed the experimental scheme assuming that the signal
and the probe beams are single-mode fields.  We shall now present the
full multimode theory of the direct scheme for measuring the quantum
optical quasidistribution functions. In the single-mode description
it was sufficient to use a pair of annihilation operators $\hat{a}_S$
and $\hat{a}_P$. The spatio-temporal characteristics of these two modes
was not important in this approach, and it was implicitly assumed that 
the modes are matched perfectly at the beam splitter. We shall free
our further analysis from these simplifying assumptions.

Let us denote by $\hat{\bf E}_{\prtext{\scriptsize out}}^{(+)}({\bf r},t)$
the positive-frequency part of the electric field operator at the surface
of the detector. This field is a superposition of the signal and the probe
fields combined at the beam splitter BS. Mathematical representation of
this combination is a slightly delicate matter. If we wanted to express
$\hat{\bf E}_{\prtext{\scriptsize out}}^{(+)}({\bf r},t)$ in terms of
the signal and probe field operators before the beam splitter, we would
have to introduce appropriate propagators. This would obscure the
physical picture of the measurement. Therefore we shall choose another
notation for the signal and the probe fields, which will make the discussion
much more transparent.
We shall denote by
$\hat{\bf E}_S^{(+)}({\bf r}, t)$ the electric field operator of the
signal beam that would fall onto the detector surface {\em in the absence
of the beam splitter BS}. Analogously, let $\hat{\bf E}_P^{(+)}({\bf r},
t)$ be the probe field at the detector surface, assuming that the
beam splitter BS {\em was replaced by a perfectly reflecting mirror}. 
With these definitions,
the field $\hat{\bf
E}^{(+)}_{\prtext{\scriptsize out}}({\bf r}, t)$ resulting from the
interference of the signal and the probe beams is given simply by
\begin{equation}
\hat{\bf E}^{(+)}_{\prtext{\scriptsize out}}({\bf r}, t)
=
\sqrt{T}
\hat{\bf E}^{(+)}_S({\bf r}, t)
- \sqrt{1-T}
\hat{\bf E}^{(+)}_P({\bf r}, t).
\end{equation}
We have assumed here that the characteristics of the beam splitter
is constant over the spectral and polarization range of the considered
fields. 

Further, we shall assume that the detected fields are quasi-monochromatic
with the central frequency $\omega_0$. This will allow us to relate easily
the number of photons to the energy of the field absorbed by the detector.
Assuming that the direction of propagation of the field
$\hat{\bf E}^{(+)}_{\prtext{\scriptsize out}}({\bf r}, t)$ 
is perpendicular
to the detector, the operator of the
photon flux through the detector surface is given by
\begin{equation}
\hat{\cal J}_{\prtext{\scriptsize out}} = 
\frac{2\epsilon_0 c}{\hbar\omega_0}
\int_{\Delta t} \prtext{d}t 
\int_{D} \prtext{d}^2 {\bf r} \,
\hat{\bf E}^{(-)}_{\prtext{\scriptsize out}}({\bf r}, t)
\hat{\bf E}^{(+)}_{\prtext{\scriptsize out}}({\bf r}, t)
\end{equation}
where $\hat{\bf E}^{(-)}_{\prtext{\scriptsize out}}({\bf r}, t)
=[\hat{\bf E}^{(+)}_{\prtext{\scriptsize out}}({\bf r}, t)]^\dagger$,
and
the temporal and the spatial integrals are performed respectively
over the detector opening time $\Delta t$ and its active surface $D$.
The probability of registering $n$ photons is given by
\begin{equation}
\label{Eq:MultimodeCountStatistics}
p_n = \left\langle : e^{-\eta \hat{\cal J}_{\prtext{\scriptsize out}}}
\frac{(\eta \hat{\cal J}_{\prtext{\scriptsize out}})^n}{n!} :
\right\rangle_{S,P},
\end{equation}
where $\eta$ is the detector quantum efficiency.

As before, we will use the count statistics to calculate the average
parity of the registered photons. It can be expressed in terms
of the photon flux operator as:
\begin{equation}
\sum_{n=0}^{\infty} (-1)^{n} p_n = \langle \, : \! \exp(-2 \eta
\hat{\cal J}_{\prtext{\scriptsize out}} ) \! : \, \rangle_{S,P} .
\end{equation}
If a coherent field is used as the probe, we may immediately evaluate
the quantum expectation value over the $P$ mode and obtain
\begin{eqnarray}
\sum_{n=0}^{\infty} (-1)^{n} p_n
& = & \left\langle  : \exp \left( -
\frac{4\eta \epsilon_0 c}{\hbar\omega_0}
\int_{\Delta t} \prtext{d}t 
\int_{D} \prtext{d}^2 {\bf r} \,
[\sqrt{T}
\hat{\bf E}_S^{(-)}({\bf r}, t)
- \sqrt{1-T}
{\bf E}_P^\ast({\bf r}, t)]
\right. \right.
\nonumber \\
\label{Eq:Sum=ExpInt}
& & \times
\left. \left.
\vphantom{\int_{D}}
[\sqrt{T}
\hat{\bf E}^{(+)}_S({\bf r}, t)
- \sqrt{1-T}
{\bf E}_P({\bf r}, t)]
\right)
 :  \right\rangle,
\end{eqnarray}
where ${\bf E}_P({\bf r}, t)
= \langle \hat{\bf E}^{(+)}_P({\bf r}, t) \rangle_P$
is the amplitude of the coherent probe field.

We will now consider the signal field
$\hat{\bf E}^{(+)}_S({\bf r}, t)$ in which a finite number of $M$ modes
is possibly excited. We shall denote the corresponding annihilation
operators by $\hat{a}_i$, and the mode functions by ${\bf u}_i
({\bf r}, t)$, where $i=1,2,\ldots M$. Our goal will be to relate the photon
statistics $p_n$ to the multimode
quasidistribution characterizing these modes.
Thus we decompose the signal field $\hat{\bf E}^{(+)}_S({\bf r}, t)$
in the form
\begin{equation}
\hat{\bf E}^{(+)}_S({\bf r}, t)
= \sum_{i=1}^{M} \hat{a}_i {\bf u}_i ({\bf r}, t)
+ \hat{\bf V}({\bf r}, t),
\end{equation}
where the operator $\hat{\bf V}({\bf r}, t)$ is a sum of all the other
modes remaining in the vacuum state. This part of the field does not
contribute to the detector counts in the normally ordered expression
given in Eq.~(\ref{Eq:Sum=ExpInt}), because its normally ordered moments
are zero.

Further, we shall assume
that virtually all the excited part of the signal field is absorbed
by the detector within the gate opening time.  This allows us to write
orthonormality relations for the mode functions ${\bf u}_i({\bf r}, t)$
in the form
\begin{equation}
\label{Eq:uOrthonormality}
\frac{2\epsilon_0 c}{\hbar\omega_0}
\int_{\Delta t} \prtext{d}t 
\int_{D} \prtext{d}^2 {\bf r} \,
{\bf u}_i^\ast ({\bf r}, t){\bf u}_j ({\bf r}, t) = \delta_{ij}
\end{equation}
where the integrals are restricted to the domain defined by the
detection process. With these assumptions, we may simplify the exponent of
Eq.~(\ref{Eq:Sum=ExpInt}). It is convenient to introduce dimensionless
amplitudes $\alpha_i$, which are projections of the probe field onto
the mode functions:
\begin{equation}
\alpha_i = 
\frac{2\epsilon_0 c}{\hbar\omega_0}
\int_{\Delta t} \prtext{d}t 
\int_{D} \prtext{d}^2 {\bf r} \,
{\bf u}_i^\ast ({\bf r}, t)
{\bf E}_P({\bf r}, t).
\end{equation}
Using these amplitudes, we may write the measured quantity as:
\begin{eqnarray}
\sum_{n=0}^{\infty} (-1)^{n} p_n
& = & 
\left\langle 
: \exp\left(-2\eta T \sum_{i=1}^{M} (\hat{a}_i^\dagger -
\sqrt{(1-T)/T}
\alpha_i^\ast )(\hat{a} - 
\sqrt{(1-T)/T}
\alpha_i) \right) :  \right\rangle
\nonumber \\
& & \times \exp \left[
 - 2\eta(1-T) \left( \frac{2 \epsilon_0 c}{\hbar\omega_0}
\int_{\Delta t} \prtext{d}t 
\int_{D} \prtext{d}^2 {\bf r} \,
|{\bf E}_P({\bf r}, t)|^2
- \sum_{i=1}^{M} |\alpha_i|^2
\right) \right] .
\nonumber \\
\label{Eq:MultimodeGeneral}
& &
\end{eqnarray}
The exponent appearing in the second line  of the above expression results
from the part of the probe field that is orthogonal (in the sense of
Eq.~(\ref{Eq:uOrthonormality})) to the mode functions describing the
excited component of the signal field. This exponent is equal to one
if the probe field matches the $M$ signal modes of interest. This condition
can be written as:
\begin{equation}
{\bf E}_P({\bf r}, t) = \sum_{i=1}^{M} \alpha_i {\bf u}_i ({\bf r}, t).
\end{equation}
In this case, we can easily recognize in the quantum expectation value
in Eq.~(\ref{Eq:MultimodeGeneral}) the multimode quasidistribution
function defined in Eq.~(\ref{Eq:MultiQDFNormOrd}), and write:
\begin{equation}
\sum_{n=0}^{\infty} (-1)^{n} p_n
=
\left(
\frac{\pi}{2\eta T} \right) ^M
W_S \left(
\sqrt{\frac{1-T}{T}}\alpha_1, \ldots,
\sqrt{\frac{1-T}{T}}\alpha_M
;
- \frac{1-\eta T}{\eta T}
\right).
\end{equation}
Thus, the direct scheme allows one to measure multimode quasidistribution
functions, even if the modes cannot be spatially separated. What one
needs to do, is to combine the multimode signal field with appropriately
chosen probe field, and to measure the count statistics of the resulting
superposition. Let us note, that it is not necessary to resolve
contributions to the count statistics from each of the modes; the only
observable we need to reconstruct the multimode quasidistribution
is the parity of the {\em total} number of photocounts.

The direct scheme for measuring multimode quasidistributions can be
also used if some of the modes are spatially separated. In this case,
each mode (or group of spatially overlapping modes) has to be displaced
in the phase space by combining at a beam splitter with a coherent probe
field, and then measured using a photon counting detector. The parity
of the number of photocounts obtained on all the detectors yields the
value of the quasidistribution function at a point defined by the values
of coherent displacements.

\section{Examples of photocount statistics}
\label{Sec:ExamplesOfCountStatistics}

We will close this chapter by presenting
several examples of the photocount statistics for different quantum
states of the signal probed by a coherent source of light. 
The  most straightforward case is when a coherent state
 $|\alpha_0\rangle$ enters through the signal
port of the beam splitter. Then the statistics of the registered
counts is given by the Poisson distribution:
\begin{equation}
\label{Eq:pncoh}
p_n^{|\alpha_0\rangle} = \frac{[J(\alpha_0)]^n}{n!} e^{-J(\alpha_0)},
\end{equation}
where $J(\alpha_0) = \eta T|\beta - \alpha_0|^2$ 
is the average number of registered photons, and $\beta
= \sqrt{(1-T)/T}\alpha$ denotes the point of the phase space at which
the value of the quasidistribution is measured. When the
measurement is performed at the point 
where the quasidistribution of the signal field is
centered, i.e., $\beta = \alpha_0$, the fields
interfere destructively and no photons are detected.
In general, for an arbitrary phase space point, the average
number of registered photons is proportional to the squared 
distance from $\alpha_0$. Averaging Eq.~(\ref{Eq:pncoh}) over
an appropriate $P$ representation yields
the photocount statistics for a thermal signal
state characterized by an average photon number $\bar{n}$:
\begin{equation}
\label{pnThermal}
p_n^{\prtext{\scriptsize
th}} = \frac{(\eta T \bar{n})^n}{(1+\eta T \bar{n})^{n+1}} 
L_n \left( - \frac{|\beta|^2}{\bar{n}(1+\eta T \bar{n})} 
\right)
\exp \left( - \frac{\eta T |\beta|^2}{1 + \eta T \bar{n}} \right),
\end{equation}
where $L_n$ denotes the $n$th Laguerre polynomial. 

A more interesting case is when the signal field is in a nonclassical
state, which cannot described by a positive definite $P$ function. Then
the interference between the signal and the probe fields cannot be
described within the classical theory of radiation. We will consider two
nonclassical states: the one photon Fock state and the Schr\"{o}dinger cat
state.  The most straightforward way to calculate the photocount statistics
is to evaluate explicitly the general quasidistribution function, and to
substitute as its parameters $\sqrt{(1-T)/T}\alpha$ and
$-(1-s-\eta T)/\eta T$ according to Eq.~(\ref{Eq:W(eta)=WS}).  Expanding
this expression into the powers of $(s+1)/(s-1)$ yields the photocount
statistics, which follows from the first line of Eq.~(\ref{Eq:W(eta)}).

\begin{figure}
\begin{center}
\epsfig{file=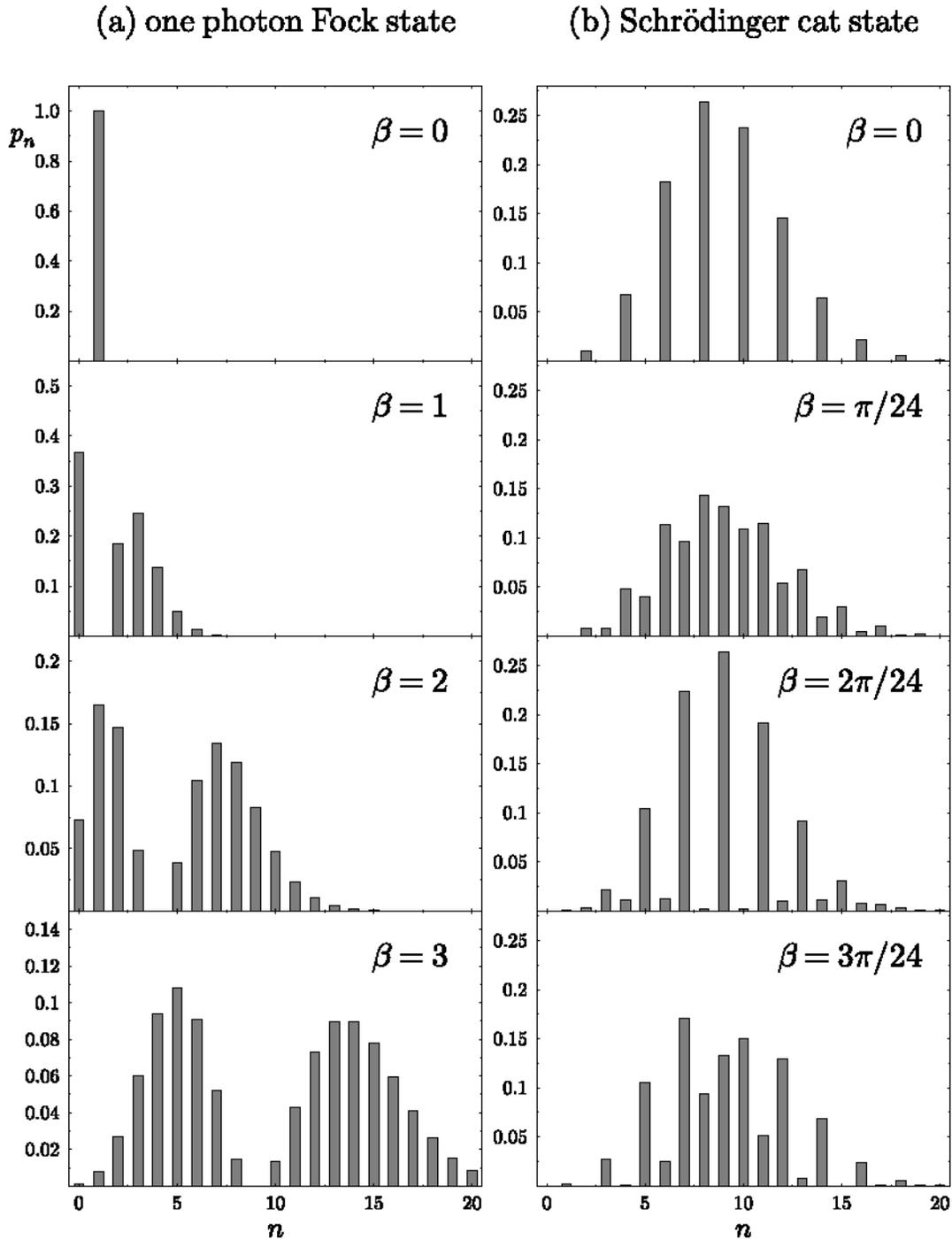}
\end{center}
\caption{The photocount statistics of (a) the one photon Fock state
and (b) the Schr\"{o}dinger cat state for $\alpha_0 = 3i$,
shown for several values of the rescaled probe field amplitude
$\beta = \protect\sqrt{(1-T)/T}\alpha$ in the limit $\eta T =1$.}
\label{Fig:PhotonStatistics}
\end{figure}

The photocount distribution for the one photon Fock 
state $|1\rangle$ can be written as an 
average of two terms with the weights $\eta T$ and $1- \eta T$:
\begin{equation}
p^{|1\rangle}_n =  \eta T [n-J(0)]^2
\frac{[J(0)]^{n-1}}{n!} e^{-J(0)}
+ (1- \eta T)
\frac{[J(0)]^{n}}{n!} e^{-J(0)} \; .
\end{equation}
The second term corresponds to the  detection of the vacuum signal
field. Its presence is a result of the detector imperfection and the
leakage of the signal field through the unused output port of the beam
splitter. This term vanishes in the limit of $\eta T \rightarrow 1$, 
where the Wigner function is measured in the setup. The first term
describes the detection of the one photon Fock state. In
Fig.~\ref{Fig:PhotonStatistics}(a) we show the statistics generated
by this term for different values 
of $\beta$. If the amplitude of the probe field is
zero, we detect the undisturbed signal field and the statistics is
nonzero only for $n=1$. The distribution becomes flatter with
increasing $\beta$. Its characteristic feature is that it vanishes
around $n \approx J(0)$. 

For the Schr\"{o}dinger cat state defined in
Eq.~(\ref{Eq:SchroedingerCatDef}) the photocount
statistics is a sum of three terms:
\begin{eqnarray}
p_n^{|\psi\rangle} & = &
\frac{1}{2(1+e^{-2|\alpha_0|^2})}
\left[
\frac{[J(\alpha_0)]^n}{n!} e^{-J(\alpha_0)} + 
\frac{[J(-\alpha_0)]^n}{n!} e^{-J(-\alpha_0)}
\right. \nonumber \\
& & \left. + 2 {\prtext{Re}} \left(
\frac{[\eta T (\beta^{\ast} - \alpha^{\ast}_0) 
(\beta + \alpha_0)]^n}{n!} 
e^{\eta T (\alpha_0^{\ast} \beta - \alpha_0 \beta^{\ast})}
\right) e ^{-(2-\eta T)|\alpha_0|^2 - \eta T |\beta|^2}
\right] \; .
\end{eqnarray}
The first two terms describe  the two coherent components of the cat
state, whereas the last one contributes to the quantum interference 
structure. In Fig.~\ref{Fig:PhotonStatistics}(b) we plot the photocount
statistics for different values
of $\beta$ probing this structure, in the 
limit $\eta T \rightarrow 1$. The four values of
$\beta$ correspond to the cosine function in
Eq.~(\ref{Eq:SchroedingerCat}) equal to $1, 0, -1,$ and $0$,
respectively, for $t=0$. 
It is seen that the form of the statistics changes very
rapidly with $\beta$. This behavior becomes clear if we recall that the
Wigner function is given by the expectation value of the displaced 
photon number parity operator.
Therefore, in order to obtain a large positive (negative) value of
the Wigner function, the photocount statistics has to be concentrated
in even (odd) values of $n$.

\chapter{Practical aspects}
\label{Chap:Practical}

\markright{CHAPTER \thechapter . \uppercase{Practical aspects}}

We will now discuss practical aspects of the direct scheme for measuring
quasidistributions of a single light mode. An important problem is the
role of the statistical noise. The proposed measurement scheme is based
on the relation between the quasidistributions and the photocount
statistics. In a real experiment the statistics of the detector
counts cannot be known with perfect accuracy, as it is determined
from a finite sample of $N$ measurements. This statistical uncertainty
affects the experimental value of the quasidistribution. Theoretical
analysis of the statistical error is important for two reasons. First,
we need an estimate for the number of the measurements required
to determine the quasidistribution with a given accuracy. Such an
estimate is needed when designing an experiment. The total number of
measurements is usually limited by various factors, such as temporal
stability of the optical setup. Estimation of the statistical error
tells us, how precise result can be expected in a realistic scheme.
Secondly, we have seen in Sec.~\ref{Sec:Generalization} that one may
attempt to compensate the imperfection of the detector and the non-unit
transmission of the beam splitter by appropriate numerical processing of
the measured statistics. However, a preliminary qualitative discussion
of the statistical noise strongly indicated that such a procedure may
amplify the statistical noise. Our present calculations will provide a
detailed, quantitative analysis of this problem. In Sec.~\ref{Sec:PCGF}
we define the observable whose statistical properties will be studied. In
Sec.~\ref{Sec:Error} we derive expressions for the mean value and
the statistical variance. The basic element of this derivation is the
multinomial distribution, which defines the probability of obtaining
a specific photocount statistics from a series of $N$ measurements.
The behaviour of the mean value and statistical variance is discussed
in Sec.~\ref{Sec:Compensation}. Particular attention is paid to the
possibility of loss compensation. We present a collection of pathological
cases, where the compensation leads to an explosion of the statistical
uncertainty. These cases show clearly that in a general case the
compensation of detector losses is not possible.

In Sec.~\ref{Sec:ModeMismatch} we discuss effects of the mode-mismatch
between the signal and the probe fields. In practice, the fields
superposed at a beam splitter never exhibit 100\% visibility of
interference. This fact has to be taken into account, when we analyse
the result of a real experiment. The effects of the mode-mismatch will
be discussed using the multimode theory developed in
Sec.~\ref{Sec:MultimodeApproach}.

\section{Photon count generating function}
\label{Sec:PCGF}

In order to make our discussion more transparent, we shall
redefine the quantity evaluated from the photocount statistics $p_n$
to the form:
\begin{equation}
\label{Eq:Pialphas}
\Pi (s)  =  \sum_{n=0}^{\infty} 
\left( \frac{s+1}{s-1} \right)^n p_n .
\end{equation}
Thus, $\Pi (s)$ corresponds to the observable defined in
Eq.~(\ref{Eq:W(eta)}) without the front normalization factor. We shall
call $\Pi (s)$ the {\em photon count generating function (PCGF)}, as the
full photon statistics can be retrieved from $\Pi (s)$ as an analytical
function of $s$. In our scheme, the PCGF is given by the expectation value
\begin{equation}
\Pi (s)  =  
\left\langle : \exp \left(
- \frac{2\eta \hat{\cal J}_{\prtext{\scriptsize
out}}}{1-s} \right) : \right\rangle,
\end{equation}
where  $\hat{\cal J}_{\prtext{\scriptsize
out}}$ is the operator of the time-integrated 
flux of the light incident onto the surface of the detector. 
This operator can be expressed in terms of the signal and probe fields as
\begin{equation}
\hat{\cal J}_{\prtext{\scriptsize
out}} = (\sqrt{T}\hat{a}_S^\dagger - 
\sqrt{1-T} \hat{a}_P^\dagger)(\sqrt{T} \hat{a}_S
- \sqrt{1-T} \hat{a}_P), 
\end{equation}
with $T$ being the beam splitter power transmission. 
When a coherent state $|\alpha\rangle_P$ is used as a probe,
we have
\begin{equation}
\label{Eq:PiandQDist}
\Pi (s) = \frac{\pi(1-s)}{2\eta T} W_S \left( 
\sqrt{\frac{1-T}{T}} \alpha ; - \frac{1 - s - \eta T}{\eta T}
\right),
\end{equation}
and the PCGF is proportional to the quasidistribution function of the
signal mode.

\section{Statistical error}
\label{Sec:Error}

In a real experiment, the photon statistics is obtained from a finite
series of $N$ measurements. The result of these measurements has
the form of a histogram $\{k_n\}$, where $k_n$ denotes the number of
measurements when $n$ photons have been detected.  Dividing $k_n$ by
the total number of measurements, we obtain an estimate for the photon
distribution $p_n$. This estimate is
subsequently used to evaluate the PCGF defined
in Eq.~(\ref{Eq:Pialphas}). In the analysis of the statistical properties
of the PCGF, we shall introduce a cut-off parameter $K$ for the maximum
photon number. In this way, the analysed quantity will depend on a finite
number of variables $k_n$, where $n=0,1,\ldots,K$. Thus, we consider an
experimental estimate for $\Pi(s)$ in the form:
\begin{equation}
\Pi_{\prtext{\scriptsize exp}} (s) = \frac{1}{N} \sum_{n=0}^{K} 
\left( \frac{s+1}{s-1} \right)^n k_n.
\end{equation}

Our goal is to see how well this estimate approximates the ideal quantity
$\Pi(s)$. 
Due to extreme simplicity of the relation between the count statistics
the quasidistributions, it is possible to perform a rigorous analysis of
the statistical error and to obtain an exact expression for the uncertainty
of the final result. The basic tool in our analysis is the probability
distribution for the histograms $\{k_n\}$ which can be obtained from
$N$ experimental runs. Because all the runs are statistically independent,
the set of $k_n$s obeys the multinomial distribution
\cite{EadieMultinomial}:
\begin{equation}
\label{Eq:Pk0k1}
{\cal P}(k_0,k_1, \ldots , k_K ) = \frac{N!}{k_0! k_1! \ldots 
\left(N - \sum_{n=0}^{K}k_n \right)!} 
p_0^{k_0} p_1^{k_1} \ldots p_K^{k_K}
\left( 1 - \sum_{n=0}^{K} p_n 
\right)^{N - \sum_{n=0}^{K} k_n}.
\end{equation}
In this formula, $p_n$ is the ideal, noise-free photocount
distribution that
would be obtained in the limit of the infinite number of measurements.

Using the distribution given in Eq.~(\ref{Eq:Pk0k1}), we may calculate
quantities characterizing statistical properties of the
experimental PCGF.  In order to see how well $\Pi_{\prtext{\scriptsize
exp}}(s)$ approximates the ideal quantity we will find its mean value and
its variance.  This task is quite easy, since the only expressions we
need in the calculations are the following moments:
\begin{eqnarray}
\overline{k_n} & = & N p_n, \nonumber \\
\overline{k_l k_n} & = & N (N-1) p_l p_n + \delta_{ln} N p_n.
\end{eqnarray}
We use the bar to denote the statistical average with respect to the
distribution ${\cal P}(k_0,\ldots k_K)$. Given this result, it is
straightforward to obtain:
\begin{equation}
\label{Eq:PiexpAv}
\overline{\Pi_{\prtext{\scriptsize exp}}(s)} = \sum_{n=0}^{K} 
\left(\frac{s+1}{s-1}\right)^{n} p_n \; ,
\end{equation}
and
\begin{eqnarray}
\delta\Pi_{\prtext{\scriptsize exp}}^{2}(s) & = &
\overline{\left(\Pi_{\prtext{\scriptsize exp}}(s) - 
\overline{\Pi_{\prtext{\scriptsize exp}} (s)}\right)^2}
\nonumber \\
& = & \frac{1}{N} \left[ \sum_{n=0}^{K} 
\left(\frac{s+1}{s-1}\right)^{2n} p_n 
- \left( \sum_{n=0}^{K} 
\left(\frac{s+1}{s-1}\right)^{n} p_n
\right)^2 
\right].
\label{Eq:PiexpDispersion}
\end{eqnarray}
The error introduced by the cut-off of the photocount statistics can
be estimated by
\begin{equation}
|\overline{\Pi_{\prtext{\scriptsize exp}}(s)} - \Pi(s)| 
=
\left| \sum_{n=K+1}^{\infty} \left( \frac{s+1}{s-1} 
\right)^n p_n \right|
\le 
\sum_{n=K+1}^{\infty} 
\left|\frac{s+1}{s-1}\right|^{n} p_n.
\end{equation}
The variance $\delta\Pi^{2}_{\prtext{\scriptsize exp}}$,
derived in Eq.~(\ref{Eq:PiexpDispersion}),
is a difference of two terms. 
The second one is simply the squared average of
$\Pi_{\prtext{\scriptsize exp}}$.
The first term is  a sum over the count statistics
multiplied by the powers of a {\it positive} factor 
$[(s+1)(s-1)]^2$. If $s>0$, this factor is greater than one and 
the sum may be arbitrarily large. In the case when the contribution
from the cut tail of the statistics is negligible, i.e., if 
$K \rightarrow \infty$, it can be estimated by the average number of
registered photons:
\begin{equation}
\sum_{n=0}^{\infty} \left(
\frac{s+1}{s-1} \right)^{2n} p_n \ge 1 + \frac{4 s}{(s-1)^2}
\langle \eta \hat{\cal J}_{\prtext{\scriptsize out}} \rangle.
\end{equation}
Thus, the variance grows unlimited as we probe phase space points
far from the area 
where the quasidistribution is localized. Several examples in the next
section will demonstrate that the variance usually explodes much more 
rapidly, exponentially rather than linearly. This makes the
compensation of the detector inefficiency a very subtle matter. It can
be successful only for very restricted regions of the phase space,
where the count statistics is concentrated for a small number of counts 
and vanishes sufficiently quickly for larger $n$'s. 

Therefore, in order to ensure that the statistical error remains bounded 
over the whole phase space, we have to impose the condition $s\le 0$.
Since we are interested in achieving the highest possible ordering 
of the measured
quasidistribution, we should consequently set $s=0$. For this particular
value the estimations for the uncertainty of $\Pi_{\prtext{\scriptsize
exp}}$
take a much simpler form. The error 
caused by the cut-off of the count distribution can be estimated 
by the ``lacking'' part of the probability:
\begin{equation}
\label{Eq:CutOffError}
|\Pi_{\prtext{\scriptsize exp}}(0) - \Pi (0)|
\le 1 - \sum_{n=0}^{K} p_n,
\end{equation}
which shows that the cut-off is unimportant as long as the probability
of registering more than $K$ photons is negligible. The variance of 
$\Pi_{\prtext{\scriptsize exp}}$ is given by
\begin{equation}
\delta \Pi_{\prtext{\scriptsize exp}}^{2}(0)
= \frac{1}{N}\left[
\sum_{n=0}^{K} p_n - 
\left(
\overline{\Pi_{\prtext{\scriptsize exp}}(0)}
\right)^2
\right]
\le
\frac{1}{N}\left[1 - \left(
\overline{\Pi_{\prtext{\scriptsize exp}}(0)}
\right)^2
\right]
\le 
\frac{1}{N}.
\end{equation}
Thus, the statistical uncertainty of the measured quasidistribution
can be simply estimated as $1/\sqrt{N}$
multiplied by the proportionality constant given 
in Eq.~(\ref{Eq:PiandQDist}). It is also seen that the uncertainty is
smaller for the phase space points where the 
magnitude of the quasidistribution is large.

\section{Compensation of detector losses}
\label{Sec:Compensation}

We will now consider several examples of the reconstruction of the
quasidistributions from the data collected in a photon counting
experiment. Our discussion will be based on Monte Carlo simulations
compared with the analytical results obtained in the previous
section. 

First, let us note that the huge statistical error is not the only
problem in compensating the detector inefficiency. If $s>0$, the sum
(\ref{Eq:PiexpAv})  does not even have to converge in the limit of
$K\rightarrow \infty$. An example of this pathological behaviour is
provided by a thermal state, which has been calculated in
Eq.~(\ref{pnThermal}). For the zero probe field we obtain 
\begin{equation}
\overline{\Pi_{\prtext{\scriptsize exp}}^{\prtext{ \scriptsize
th}}(s)} 
= \frac{1}{1+\eta T \bar{n}} 
\sum_{n=0}^{K} 
\left(
\frac{s+1}{s-1}
\right)^{n}
\left(
\frac{\eta T \bar{n}}{1 + \eta T \bar{n}}
\right)^{n},
\end{equation}
which shows that if $s>0$, then for a sufficiently intense thermal
state the magnitude of the summand is larger than one and consequently
the sum diverges, when $K \rightarrow \infty$. This behaviour is due
to the very slowly vanishing count distribution for large $n$, and it
does not appear for the other examples of the count statistics derived
in Sec.~\ref{Sec:ExamplesOfCountStatistics}.

\begin{figure}
\begin{center}
\epsfig{file=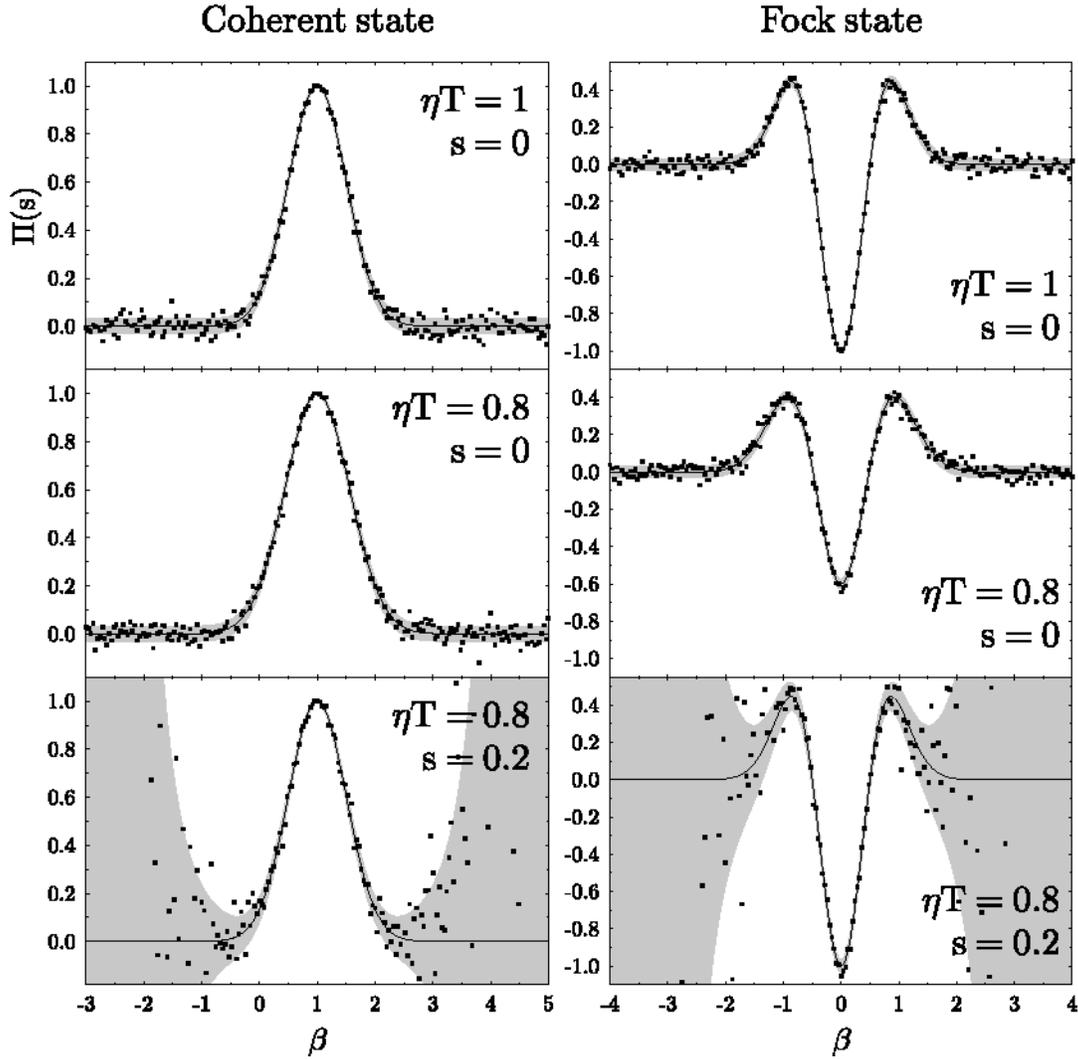}
\end{center}
\caption{Reconstruction of the quasiprobability distributions of the
coherent state $|\alpha_0=1\rangle$ (left) and the one photon Fock
state (right) from $N=1000$ events. The solid lines are the analytical
quasidistributions and the grey areas mark the statistical dispersion.
The plots are parameterized with the rescaled probe field amplitude
$\beta = \sqrt{(1-T)/T}\alpha$.}
\label{Fig:CohAndFockReconstruction}
\end{figure}

In Fig.~\ref{Fig:CohAndFockReconstruction} we plot the
reconstructed quasidistributions for the coherent state 
\mbox{$|\alpha_0 = 1\rangle$} 
and the one photon Fock state. Due to the symmetry of these
states, it is sufficient to discuss the behaviour of the reconstructed
quasidistribution on the real axis of the phase space. The cut-off
parameter is set high enough to make the contribution
from the cut tail of the statistics negligibly small.
The quasidistributions are
determined at each phase space point from the Monte Carlo simulations
of $N=1000$ events. The grey areas denote the statistical uncertainty
calculated 
according to Eq.~(\ref{Eq:PiexpDispersion}). 
The two top graphs show the reconstruction of the Wigner
function in the ideal case $\eta T = 1$. 
It is seen that the statistical error is
smaller, where the magnitude of the Wigner function is large. In the
outer regions it approaches its maximum value $1/\sqrt{N}$. The effect
of the nonunit $\eta T$ is shown in the center graphs. The measured
quasidistributions become wider and the negative dip in the case of
the Fock state is shallower. In the bottom graphs we depict the
result of compensating the nonunit value of $\eta T$ by setting
 $s=1-\eta T$. The compensation works quite well in the central region,
where the average number of detected photons is small, but
outside of this region the statistical error explodes exponentially. 
Of course, the statistical error can be in principle 
suppressed by increasing the
number of measurements. However, this is not a practical method, since
the statistical error decreases with the size of the sample only as 
 $1/\sqrt{N}$. 

\begin{figure}
\begin{center}
\epsfig{file=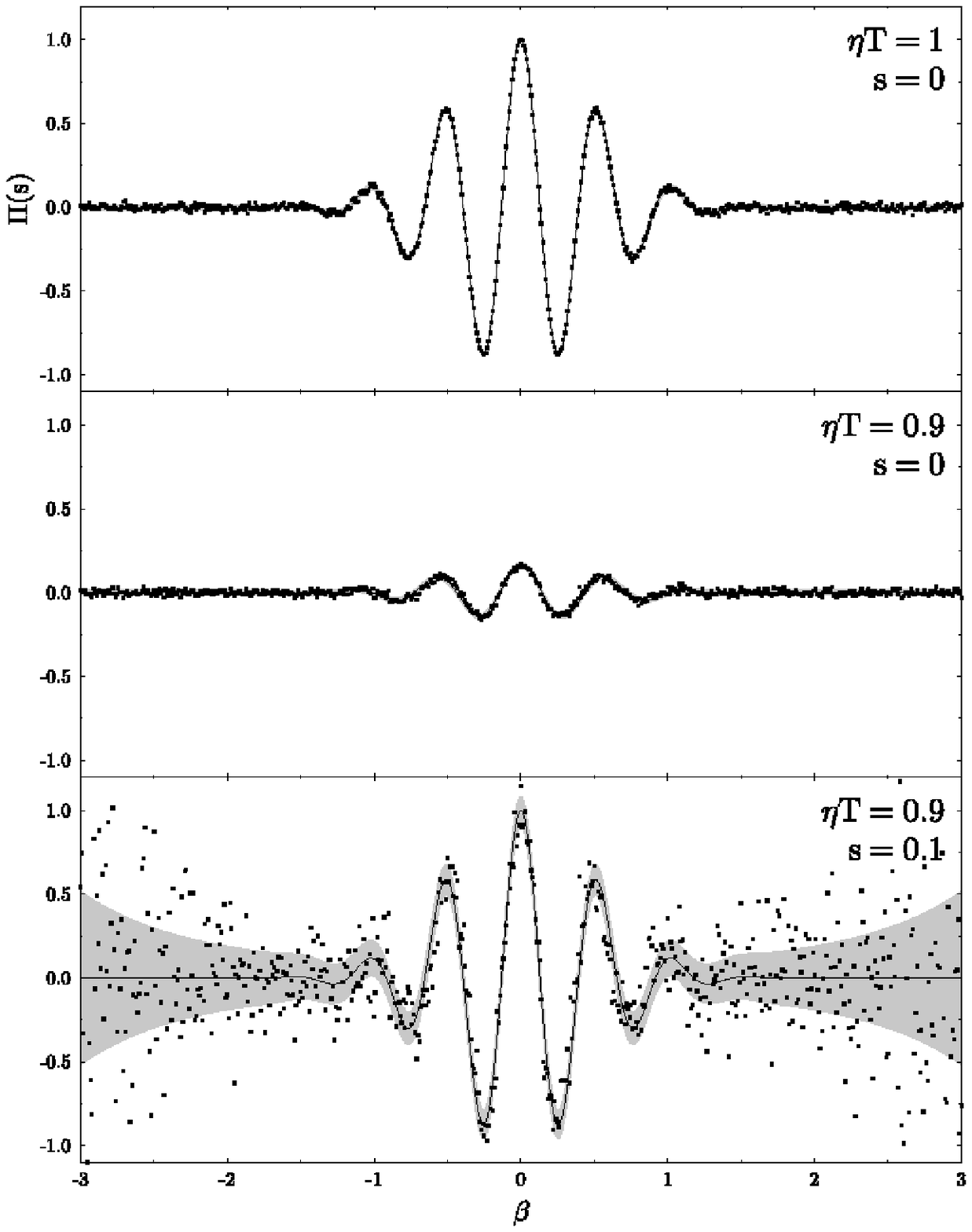}
\end{center}
\caption{Reconstruction of the interference structure of the
Schr\"{o}dinger cat state for $\alpha_0 = 3i$ from $N=5000$ events at
each point.}
\label{Fig:CatReconstruction}
\end{figure}

The reconstruction of the interference structure of the
Schr\"{o}dinger cat state is plotted in
Fig.~\ref{Fig:CatReconstruction}. 
We have used the state defined in Eq.~(\ref{Eq:SchroedingerCatDef})
with $\alpha_0 = 3i$. 
The interference structure is very fragile, and
its precise measurement requires a large sample of events. In the
case of the presented plot,  $N = 5000$ simulations were performed at
each phase space point. Comparison of the top and the center graphs
shows how even  
relatively small imperfection destroys the interference pattern. 
The data collected in a non-ideal setup can be processed to recover
the Wigner function, but at the cost of a significantly larger
statistical error, as it is shown in the bottom graph. Outside the
interference structure, we again observe the exponential explosion of
the dispersion due to the increasing intensity of the detected light. 

\begin{figure}
\vspace*{5mm}
\begin{center}
\epsfig{file=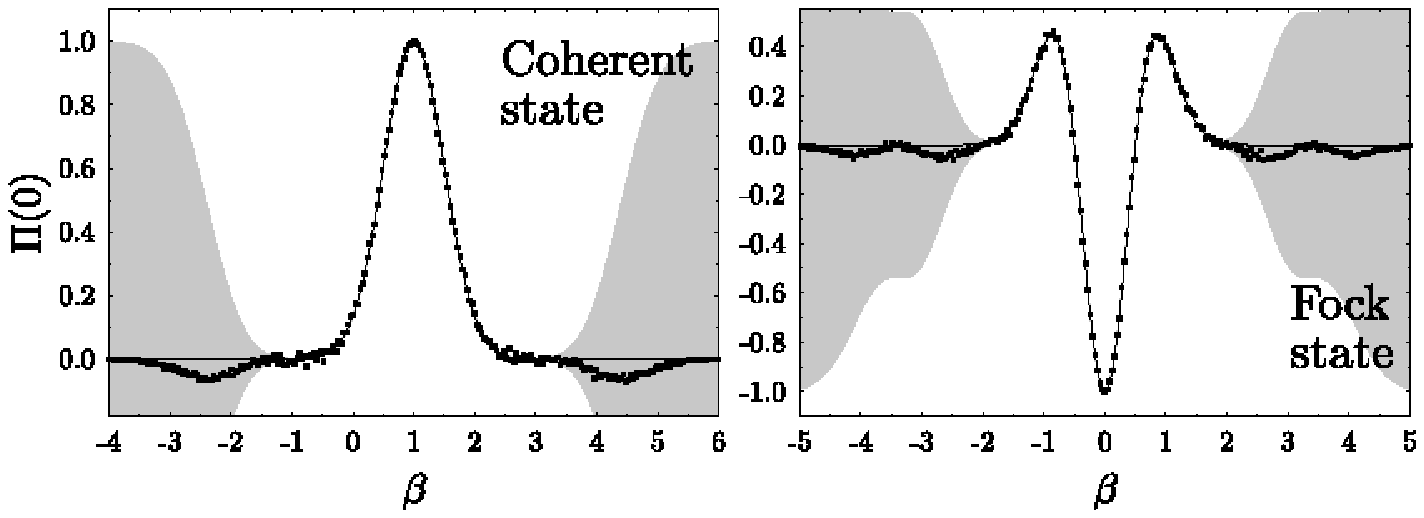}
\end{center}
\caption{Reconstruction of the Wigner function of the coherent state
and the one photon Fock state from the count statistics cut at
$K=11$, for $\eta T =1$ and $s=0$. The number of events is $N=10^4$.}
\label{Fig:CutOff}
\end{figure}

From the above examples we clearly see, that compensation of any
efficiency lower than 100\% is a delicate matter. This conclusion may
be quite puzzling, if we compare it with recently published results on
compensation of losses in photodetection \cite{KissHerzPRA95}.  It was
shown there that the true photon distribution can be reconstructed from
data measured by an imperfect detector, provided that its efficiency is
greater than 50\%. If this condition is fulfilled, the photon distribution
can be obtained from experimental data via the so-called inverse Bernoulli
transformation, and the statistical properties of such reconstruction have
been shown to behave regularly.
One could think of using this recipe in our scheme for measuring
quasidistribution functions: first, application of the inverse Bernoulli
transformation would yield the loss-free photon distribution, and then
evaluation of the alternating series would yield the unblurred Wigner
function. However, a simple calculation shows that such a two-step
method is completely equivalent to setting $s=1-\eta T$ in the PCGF,
and we end up with all the difficulties discussed above. Explanation
of this seeming contradiction is simple: quantities obtained from the
inverse Bernoulli transformation have strongly correlated statistical
errors, which accumulate when evaluating the alternating series. We
shall discuss this issue in detail in Sec.~\ref{Sec:Random}, using the
general theory of statistical uncertainty in photodetection measurements.

Finally, let us look at the effect of cutting the statistics at a
finite value. Fig.~\ref{Fig:CutOff} shows the Wigner functions for
the one photon coherent and Fock states reconstructed from the count
distributions cut at $K=11$. We performed a large number of $N=10^4$
simulations in order to get the pure effect of the cut-off that is not
spoiled by the statistical uncertainty. The grey areas show the cut-off
error, estimated using Eq.~(\ref{Eq:CutOffError}). The reconstruction
works well as long as the probability of detecting more than $K$ photons
is negligible. When the average number of incident photons starts to
be comparable with the cut-off parameter, ``ghost'' structures appear.
When we go even further, the Wigner function again tends to zero, but
this is a purely artificial effect due to the virtually vanishing count
distribution below $K$.  This kind of ``ghost'' structures resulting from
the cut-off were observed in the characterization of the motional state
of a trapped ion \cite{LeibMeekPRL96}. In this experiment, the analog of
photon statistics $p_n$ was reconstructed in an indirect way by monitoring
the fluorescence \cite{MeekMonrPRL96}, and it was necessary to truncate
the statistics $p_n$ in order to keep the error on a reasonably low level.
We have seen that in the regions of the phase space where a substantial
part of the statistics $p_n$ is lost by the truncation, we do not have
sufficient data to determine the value of the Wigner function.

\section{Mode mismatch}
\label{Sec:ModeMismatch}

Another experimental imperfection that occurs in a realistic setup is
the non-unit matching of the modes interfered at the high-transmission
beam splitter BS. In order to analyse consequences
of the mode mismatch, we will use the multimode
approach developed in Sec.~\ref{Sec:MultimodeApproach}. Let us take the
signal field and the coherent probe field to be of the form
\begin{eqnarray}
\hat{\bf E}^{(+)}_S({\bf r}, t) & = & \hat{a}_S {\bf u}_S ({\bf r}, t)
+ \hat{\bf V}({\bf r}, t),
\nonumber \\
{\bf E}_P({\bf r}, t) & = & \alpha {\bf u}_P({\bf r}, t)
\end{eqnarray}
where ${\bf u}_S ({\bf r}, t)$ and ${\bf u}_P({\bf r}, t)$ are the
corresponding normalized mode functions, and
the operator $\hat{\bf V}({\bf r}, t)$ is the sum of all other signal modes
remaining in the vacuum state. We will assume that the
support of these mode functions lies within the domain defined by the
detector surface and the gate opening time. This allows us to write the
normalization of the mode functions as
\begin{equation}
\int_{\Delta t} \prtext{d}t
\int_{D} \prtext{d}^2 {\bf r} \,
|{\bf u}_S ({\bf r}, t)|^2
=
\int_{\Delta t} \prtext{d}t
\int_{D} \prtext{d}^2 {\bf r} \,
|{\bf u}_P ({\bf r}, t)|^2
=
\frac{\hbar\omega_0}{2\epsilon_0 c}
\end{equation}
Further, we will assume that the overlap of the functions
${\bf u}_S ({\bf r}, t)$ and ${\bf u}_P({\bf r}, t)$ is real and positive.
This can be always achieved by multiplying ${\bf u}_P({\bf r}, t)$
and $\alpha$ by appropriate conjugated phase factors. We will denote
\begin{equation}
\frac{2 \epsilon_0 c}{\hbar \omega_0}
\int_{\Delta t} \prtext{d}t
\int_{D} \prtext{d}^2 {\bf r} \,
{\bf u}_S^\ast ({\bf r}, t)
{\bf u}_P ({\bf r}, t)
= \sqrt{\xi}.
\end{equation}

Under these assumptions, we may represent the PCGF calculated from
the count statistics given by Eq.~(\ref{Eq:MultimodeCountStatistics})
in the form:
\begin{equation}
\Pi(s) = \left\langle : \exp \left( -\frac{2\eta}{1-s}
[T\hat{a}_S^\dagger \hat{a}_S
- \sqrt{\xi T(1-T)}(\hat{a}_S^\dagger \alpha + \hat{a}_S \alpha^\ast)
+ (1-T)|\alpha|^2] \right) : \right\rangle
\end{equation}
Rearranging the terms in the exponent yields:
\begin{equation}
\label{Eq:WStimesGaussian}
\Pi (s) = \frac{\pi(1-s)}{2\eta T} W_S \left( 
\sqrt{\frac{\xi(1-T)}{T}} \alpha ; - \frac{1 - s - \eta T}{\eta T}
\right) \exp\left(-\frac{2\eta(1-T)(1-\xi)}{1-s}|\alpha|^2 \right).
\end{equation}
Thus, if the signal and the probe modes are not matched perfectly,
the PCGF is given by the quasidistribution function of the signal, but
multiplied by a Gaussian envelope $\exp[-2\eta (1-T) (1-\xi) |\alpha|^2
/(1-s)]$. This envelope is centered at the origin of the phase space,
and the faster it decays, the larger is the mode mismatch, characterized
by the difference $1-\xi$. Additionally, the parameterization of the
signal quasidistribution is rescaled by $\sqrt{\xi}$.

The effect of the mode-mismatch is more severe in outer regions of the
phase space. It is not important if the quasidistribution is localized
around the center of the phase space within the width of the Gaussian
envelope. In particular, for the vacuum signal state we obtain:
\begin{equation}
\Pi (s) = \exp \left(- \frac{2\eta(1-T)}{1-s}|\alpha|^2 \right).
\end{equation}
This result does not depend at all on the mode overlap parameter
$\xi$, simply because with the vacuum signal field no interference occurs
at the beam splitter BS and all the recorded photons come from the probe
beam. The above expression can be interpreted as the quasidistribution
function of the signal mode characterized by the mode function ${\bf
u}_P({\bf r}, t)$, which perfectly overlaps with the probe field.

In a general case, we may rewrite Eq.~(\ref{Eq:WStimesGaussian}) to the form:
\begin{eqnarray}
\Pi(s) & = &
\frac{\pi(1-s)}{2\eta T} W_S \left( 
\sqrt{\frac{\xi(1-T)}{T}} \alpha ; - \frac{1 - s - \eta T}{\eta T}
\right) 
\nonumber \\
 & &
\times
\frac{\pi(1-s)}{2\eta T}
W_{\prtext{\scriptsize vac}}
\left(
\sqrt{\frac{(1-\xi)(1-T)}{T}} \alpha ; - \frac{1 - s - \eta T}{\eta T}
\right),
\end{eqnarray}
where $W_{\prtext{\scriptsize vac}}(\beta;s )$ is the vacuum
quasidistribution function. This representation allows us to interpret
Eq.~(\ref{Eq:WStimesGaussian}) as a two-mode quasidistribution which
is a product of the signal mode quasidistribution and an additional
vacuum quasidistribution. These two modes are probed at the phase space
points proportional to $\sqrt{\xi}\alpha$ and $\sqrt{1-\xi}\alpha$. The
amplitude $\sqrt{\xi}\alpha$ describes the part of the probe field that
overlaps perfectly with the signal field, whereas $\sqrt{1-\xi}\alpha$
corresponds to the orthogonal remainder (orthogonality is understood
here in the sense of the scalar product between the mode functions).

The parameter $\xi$ characterizing the overlap of the signal and the
probe modes can be related to the visibility of the interference. This
expression will be useful in the discussion of the practical realization
of the scheme. Let as assume that the signal mode is in a coherent
state $|\alpha_0\rangle$. The intensity of the field measured by the
photodetector is given by
\begin{equation}
I = |\sqrt{1-T}\alpha - \sqrt{\xi T} \alpha_0|^2
+ (1 - \xi)  T |\alpha_0|^2,
\end{equation}
where $\alpha$ is the amplitude of the coherent probe mode. 
It is seen that for given $\alpha_0$
the minimum intensity is obtained for
$\sqrt{1-T}\alpha = \sqrt{\xi T} \alpha_0$, and it equals to:
\begin{equation}
I_{\prtext{\scriptsize min}} = (1 - \xi) T |\alpha_0|^2.
\end{equation}
When we now vary with the phase of $\alpha$, keeping its absolute
value fixed, the maximum intensity is achieved for
$\sqrt{1-T}\alpha = - \sqrt{\xi T} \alpha_0$, and its value is:
\begin{equation}
I_{\prtext{\scriptsize max}} = 
4 \xi T |\alpha_0|^2 + (1 - \xi) T |\alpha_0|^2.
\end{equation}
A simple calculation shows that the interference visibility $v$
can be expressed using the overlap parameter $\xi$ as:
\begin{equation}
v = \frac{I_{\prtext{\scriptsize max}} - I_{\prtext{\scriptsize min}}}%
{I_{\prtext{\scriptsize max}} + I_{\prtext{\scriptsize min}}}
= \frac{4\xi}{2 + 2\xi}.
\end{equation}
Inverting this relation, we obtain that $\xi = v/(2-v)$.

Finally, let us note that the effect of mode-mismatch in our scheme is
quite different from balanced homodyne detection, where
it can be simply included in the overall detection efficiency parameter.
In our scheme, it generates a Gaussian envelope multiplying the measured
quasidistribution function, and its importance depends on the probed
point of the phase space.

\chapter{Experiment}
\label{Chap:Experiment}

\markright{CHAPTER \thechapter . EXPERIMENT}

We shall now present direct measurement of the Wigner function by photon
counting, using the method presented in Chap.~\ref{Chap:Direct}. Previous
measurements of the Wigner function of light, performed at
the University of Oregon \cite{SmitBeckPRL93} and Universit\"{a}t Konstanz
\cite{BreiSchiNAT97}, were realizations of optical homodyne tomography. In
our experiment, the Wigner function is determined directly from the
statistics of photocounts. Apart from the different principle of the
measurement, we use a different technique for light detection. As we
discussed in Chap.~\ref{Chap:Homodyne}, optical homodyne tomography
is based on detection of the signal light superposed on a strong,
classical local oscillator. The light incident on photodetectors has
macroscopic intensity, and it is converted into an electronic current
with the help of {\em p-i-n} photodiodes. At this level of intensity,
it is not possible to resolve contributions from single photons, and
the electric current is practically a continuous variable. Information
on the quantum state of the measured light is contained in fluctuations
of the difference signal between two detectors. This signal is recorded
using an analog-to-digital converter.

In our scheme, the intensity of the detected light is comparable with
the intensity of the signal field itself. Therefore, we need to use a
detector that is sensitive to single optical photons. Currently, the
most efficient commercially available detectors on single-photon level
are avalanche photodiodes operated in the so-called Geiger mode. The
Geiger mode of operation consists in biasing the diode slightly above
breakdown. Absorption of a photon triggers the breakdown, which is
a macroscopic, recordable event. The diode is placed in a circuit
which quenches the breakdown by lowering the voltage, and after a while
restores the higher bias thus preparing the diode for the detection of a
next photon. The electronic signal obtained from the diode has the form
of pulses of uniform height and duration, which correspond to single
detection events. These pulses, after shaping, can be counted using
a standard digital logic device.

We open this chapter with a review of the principle of the measurement
in Sec.~\ref{Sec:Principle}. The experimental setup is described in
Sec.~\ref{Sec:Setup}, and the results of the measurements are reported
in Sec.~\ref{Sec:Results}. In Sec.~\ref{Sec:Discussions} we discuss the
effect of various experimental factors, and summarize the presentation
of the experiment.

\section{Principle}
\label{Sec:Principle}

In order to make the presentation of the experiment self-contained,
let us start with a brief review of the principle of the measurement.
The Wigner function at a given phase space point is itself a well defined
quantum observable. Furthermore, the measurement of this observable
can be implemented for optical fields using an arrangement employing an
auxiliary coherent probe beam. The amplitude and the phase of the probe
field define the point in the phase space at which the Wigner function
is measured. This allows one to scan the phase space point-by-point,
simply by changing the parameters of the probe field.

Our experiment is based on the representation of the Wigner function
at a complex phase space point denoted by $\alpha$ as the expectation
value of the following operator:
\begin{equation}
\label{Eq:Wdef}
\hat{W}(\alpha) = \frac{2}{\pi}
\sum_{n=0}^{\infty} (-1)^{n} \hat{D}(\alpha)|n\rangle \langle
n | \hat{D}^{\dagger}(\alpha),
\end{equation}
where $\hat{D}(\alpha)$ is the displacement operator and $|n\rangle$
denote Fock states, $\hat{n}|n\rangle = n|n\rangle$.
Thus, $\hat{W}(\alpha)$ has two eigenvalues:
$2/\pi$ and $-2/\pi$, corresponding to degenerate subspaces spanned
respectively by even and odd displaced Fock states.  Practical means to
translate this formula into an optical arrangement are quite simple.
The displacement transformation can be realized by superposing the
measured field at a low-reflection beam splitter with a strong coherent
probe beam. The value of the displacement $\alpha$ is equal in this
setup to the reflected amplitude of the probe field. Furthermore, the
projections on Fock states can be obtained by photon counting assuming
unit quantum efficiency. These two procedures, combined together, provide
a practical way to measure the Wigner function at an arbitrarily selected
phase space point $\alpha$.

\section{Setup}
\label{Sec:Setup}

\begin{figure}[t]
\begin{center}
\begin{pspicture}(2,1)(16.5,8)
\psline[linewidth=1.5pt](4.5,2.5)(12.5,2.5)(12.5,6.5)(15,6.5)(15,3.5)
\psline[linewidth=1.5pt](6.5,2.5)(6.5,6.5)(12.5,6.5)
\psline[linewidth=1.5pt]{->}(5.4,2.5)(5.5,2.5)
\psline[linewidth=1.5pt]{->}(8,6.5)(8.1,6.5)
\psline[linewidth=1.5pt]{->}(13.75,6.5)(13.8,6.5)
\psset{linecolor=black}
\psline[linewidth=4pt,linecolor=gray](6,5)(7,5)
\rput(5.5,5){\large ND}
\psframe(2.5,2)(4.5,3)
\rput(3.5,2.5){\large laser}
\rput{45}(6.5,2.5){\beamsplitter}
\rput(6.5,1.5){\large BS1}
\psframe(7.9,2)(8.1,3)
\rput(8,1.5){\large $\lambda/2$}
\psframe(9.6,2.1)(10.9,2.9)
\rput(10.25,1.5){\large EOM1}
\rput{45}(12.5,2.5){\mirror}
\psframe(12.1,4.1)(12.9,4.9)
\psline(12.1,4.1)(12.9,4.9)
\rput{45}(6.5,6.5){\mirror\psframe(-0.4,0)(0.4,0.3)}
\rput(6.5,7.5){\large PZT}
\psframe(9.6,6.1)(10.9,6.9)
\rput(10.25,7.5){\large EOM2}
\rput{45}(12.5,6.5){\beamsplitter}
\rput(12.5,7.5){\large BS2}
\rput{135}(15,6.5){\mirror}
\rput(15,5.5){\psline[linewidth=1.5pt](-0.5,0)(-0.05,0)%
\psline[linewidth=1.5pt](0.05,0)(0.5,0)}
\rput(15.9,5.5){\large A}
\rput(15,4.5){\psarc[linewidth=1pt](0,0.866){1}{240}{300}%
\psarc[linewidth=1pt](0,-0.866){1}{60}{120}}
\psframe(14.5,1.5)(15.5,3.5)
\rput{90}(15,2.5){\large SPCM}
\end{pspicture}
\end{center}
\caption{The experimental setup for measuring the Wigner function.
BS1 and BS2 are quartz plates serving as low-reflection beam splitters.
The quantum state is prepared using the neutral density filter ND and
a mirror mounted an a piezoelectric translator PZT. The
electrooptic modulators EOM1 and EOM2 control respectively the amplitude
and the phase of the point at which the Wigner function is measured. The
signal field, after removing spurious reflections using the aperture A,
is focused on a single photon counting module SPCM.}
\label{Fig:ExpSetup}
\end{figure}
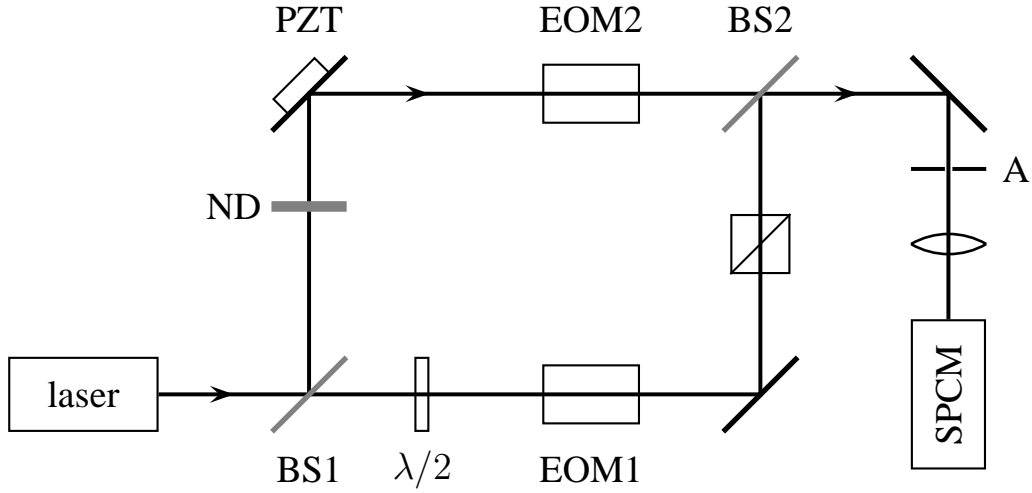

The experimental setup we used to measure the Wigner function is
shown schematically in Fig.~\ref{Fig:ExpSetup}.  In principle, it is a
Mach-Zender interferometric scheme with the beams in two arms
of the interferometer serving as the
signal and the probe fields.  An attenuated, linearly polarized 
(in the plane of Fig.~\ref{Fig:Setup}) 632.8~nm
beam from a frequency-stabilized single-mode He:Ne laser is divided by
a low-reflection beam splitter BS1. The weak reflected beam is used
to generate the signal field whose Wigner function will be measured.
The state preparation stage consists of a neutral density filter ND
and a mirror mounted on a piezoelectric translator PZT. With this
arrangement, we are able to create pure coherent states with variable
phase as well as their incoherent mixtures. Though these states do not
exhibit nonclassical properties, they constitute a nontrivial family
to demonstrate the principle of the method, which provides complete
characterization of both quantum and classical field fluctuations.

The strong beam leaving the beam splitter BS1 plays the role of the
probe field with which we perform the displacement transformation
$\hat{D}(\alpha)$. In order to scan the phase space one should be able to
set freely its amplitude and phase, which define respectively the radial
and angular coordinate in the phase space. The amplitude modulation is
achieved with a half-wave plate, a longitudinal Pockels cell EOM1, and a
polarizer oriented parallel to the initial direction of polarization. The
phase modulation is done with the help of an ADP crystal electrooptic
phase modulator EOM2 on the signal field.  This is completely equivalent
to modulating the probe field phase, but more convenient for technical
reasons: in this arrangement optical paths in both the arms of
the Mach-Zender interferometer are approximately the same, and better
overlap of the signal and the probe modes is achieved at the output
of the interferometer.

The signal and the probe fields are interfered at a nearly
completely transmitting beam splitter BS2 with the power transmission
$T=98.6\%$. In this regime, the transmitted signal field effectively
undergoes the required displacement transformation.  
Spurious reflections
that accompany the beam leaving the interferometer are removed using
the aperture A. Finally, the transmitted signal is focused on an EG\&G
photon counting module SPCM-AQ-CD2749, whose photosensitive element is a
silicon avalanche diode operated in the Geiger regime. The overall
quantum efficiency of the module specified by the manufacturer is
$\eta \ge 70\%$.  The count rate is kept low in the experiment, so that
the chance of two or more photons triggering a single avalanche signal
is very small, and the probability of another photon arriving during
the detector dead time can be neglected. Under these assumptions, each
pulse generated by the module corresponds to the detection of a single
photon\footnote{The
operating mode of the SPCM results in a certain amount of  extraneous
pulses originating from dark counts and afterpulsing. The dark count
rate of our module is less than 100~s$^{-1}$. This gives on average $<
3 \cdot 10^{-3}$ during a single counting interval, which is 30~$\mu$s
long. The typical afterpulsing probability is $0.2\%$.  Thus, both these
effects give a negligible contribution to the measured count statistics.}.
The pulses are acquired by a computer, which
also controls the voltages applied to the electrooptic modulators. The
interference visibility in our setup has been measured to be $v \ge
98.5\%$, and the phase difference between the two arms was stable up to
few percent over times of the order of ten minutes.

\section{Results}
\label{Sec:Results}

The voltages applied to electrooptic modulators were generated by
high-voltage power supplies controlled by analog output ports of a
multifunction I/O card (National Instruments PCI-MIO-16E-4).  A typical
scan of the phase space consisted of sampling a sequence of circles with
increasing radius. This was because changing the voltage applied to the
amplitude modulator required a settling time of the order of 1~s. For
a fixed amplitude the phase could be scanned much faster, as the phase
modulator driver had the bandwidth up to 10~kHz. For a selected point
of the phase space, the photon statistics was collected using a digital
counter on the same I/O card, operated in the buffered event counting
mode.

\begin{figure}[t]
\vspace*{5mm}
\epsfig{file=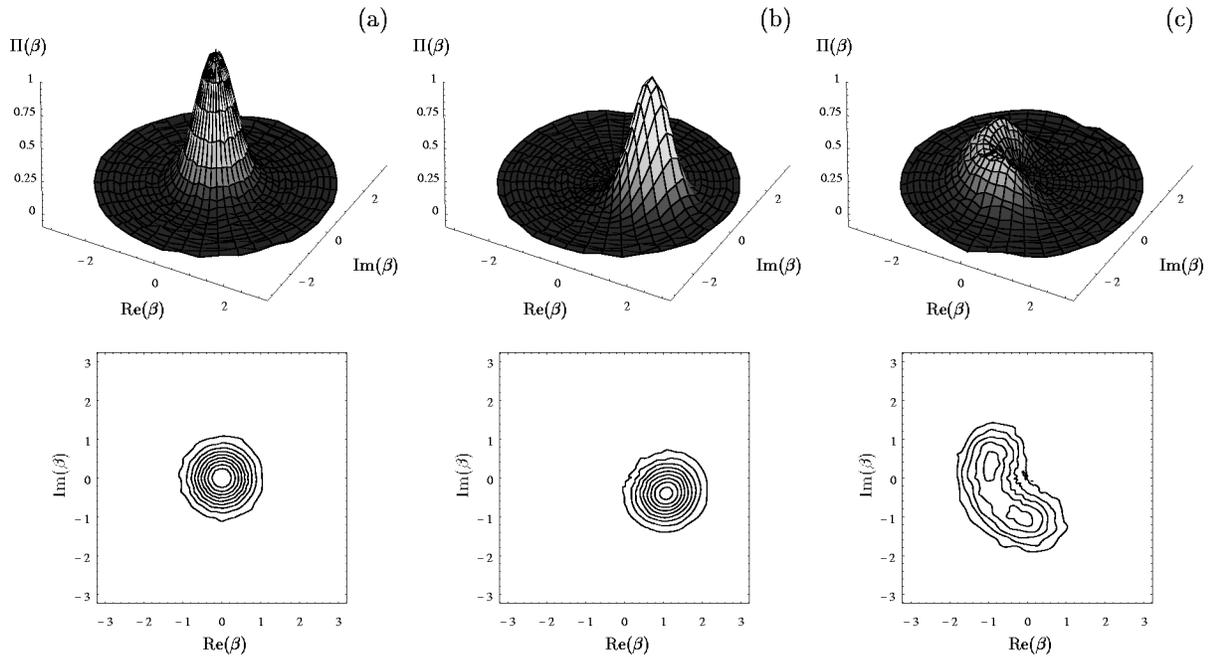}
\caption{The measured Wigner functions of (a) the vacuum, (b) a weak
coherent state, and (c) a phase diffused coherent state. The contour plots
depict interpolated heights given by mutliples of 0.1 for the
plots (a) and (b), and by $0.08, 0.14, 0.20, 0.26, 0.32$ for the plot
(c).}
\label{Fig:Wigner}
\end{figure}

In Fig.~\ref{Fig:Wigner} we depict the measured Wigner functions of the
vacuum, a weak coherent state, and a phase diffused coherent
state.
Phase fluctuations were obtained by applying a 400~Hz
sine waveform to the piezoelectric translator. For all the plots,
the phase space was scanned on a grid defined by 20 amplitudes and
40 phases. The scaling of the radial coordinate is obtained from the
average number of photons $n_{\prtext{\scriptsize
vac}}$ detected for the blocked
signal path. Thus the graphs are parameterized with the complex variable
$\beta=e^{i\varphi}n_{\prtext{\scriptsize
vac}}^{1/2}$, where $\varphi$ is the
phase shift generated by the phase modulator EOM2.  At each selected
point of the phase space, the photocount statistics $p_n(\beta)$
was determined from a sequence of $N=8000$ counting intervals, each
$\tau=30\mu$s long. The duration of the counting interval $\tau$ defines
the temporal envelope of the measured mode. The count statistics was
used to evaluate the alternating sum\footnote{Here we explicitly write
dependence of $\Pi$ on the phase space point $\beta$. The number
$s$ used in the previous chapter as a parameter of $\Pi$ is now fixed
and equal to zero. The operational phase space point $\beta$ can be
expressed by the amplitude $\alpha$ of the probe field as
$\beta = \sqrt{\eta(1-T)}\alpha$.}
\begin{equation}
\Pi(\beta) = \sum_{n=0}^{\infty} (-1)^{n} p_n(\beta),
\end{equation}
which, up to the normalization factor $2/\pi$ is equal to the Wigner
function of the measured state.
Statistical variance of this result can be estimated
by $\prtext{Var}[\Pi(\beta)] = \{1-[\Pi(\beta)]^{2}\}/N$,
according to the discussion in Sec.~\ref{Sec:Error}.
Thus, the statistical error of our measurement
reaches its maximum value, equal to $1/N^{1/2} \approx 1.1\%$, when
the value of the Wigner function is close to zero.

The Wigner functions of the vacuum and of the coherent state are
Gaussians centered at the average complex amplitude of the field, and
their widths characterize quantum fluctuations. It can be noticed that
the measured Wigner function of the coherent state is slightly lower than
that of the vacuum state. In the following, when discussing experimental
imperfections, we shall explain this as a result of non-unit interference
visibility.  In the plot of the Wigner function of the phase diffused
coherent state, one can clearly distinguish two outer peaks corresponding
to the turning points of the harmonically modulated phase.
Another set of results is presented in a colour plate at the end
of this chapter.

\section{Discussion}
\label{Sec:Discussions}

There are several experimental factors whose impact on the result of
the measurement needs to be analyzed. First, there are losses of the
signal field resulting from two main sources: the reflection from the
beam splitter BS2 and, what is more important, imperfect photodetection
characterized by the quantum efficiency $\eta$.  Analysis of these losses
performed in Sec.~\ref{Sec:Generalization} shows, that in such a case the
alternating series evaluated from photocount statistics is proportional
to a generalized, $s$-ordered quasidistribution function $W(\alpha; s)$,
with the ordering parameter equal $s=-(1-\eta T)/\eta T$.

In addition, the two modes interfered at the beam splitter BS2 are never
matched perfectly. The effects of the mode mismatch have been analysed
in Sec.~\ref{Sec:ModeMismatch}. We will now apply this analysis to
the experimental results, but first let us recall briefly the physical
picture of mode mismatch.  For this purpose, we need to consider the
normalized mode functions describing the transmitted signal field
and the reflected probe field. The squared overlap $\xi$ of these two
mode functions can be related to the interference visibility $v$ as
$\xi=v/(2-v)$. In order to describe the effects of the mode mismatch,
we need to decompose the probe mode function into a part that precisely
overlaps with the signal, and the orthogonal remainder.  The amplitude of
the probe field effectively interfering with the signal is thus multiplied
by $\xi^{1/2}$, and the remaining part of the probe field contributes to
independent Poissonian counts with the average number of detected photons
equal $(1-\xi)|\beta|^2$. Consequently, the full count statistics is given
by a convolution of the statistics generated by the interfering fields,
and the Poissonian statistics of mismatched photons. A simple calculation
shows, that the alternating sum evaluated from such a convolution can
be represented as a product of the contributions corresponding to the
two components of the probe field:
\begin{eqnarray}
\Pi (\beta)
& = & \exp[-2(1-\xi)|\beta|^2] \nonumber \\
& &
\times \frac{\pi}{ 2\eta T} 
W \left( \sqrt{\frac{\xi}{\eta T}} \beta ; - \frac{1-\eta T}{\eta T}
\right).
\end{eqnarray}
Here on the right-hand side we have made use of the theoretical results
for imperfect detection obtained in Secs.~\ref{Sec:Generalization} and
\ref{Sec:ModeMismatch}.  Specializing the above result to a coherent
signal state $|\alpha_0\rangle$ with the amplitude $\alpha_0$, yields:
\begin{equation}
\Pi(\beta) = \exp[-2|\beta - \sqrt{\xi \eta T } \alpha_0|^2
- 2 (1-\xi)\eta T |\alpha_0|^2].
\end{equation}
Thus, in a realistic case 
$\Pi(\beta)$ represents a Gaussian centered at the
attenuated amplitude $\sqrt{\xi \eta T }\alpha_0$, and the width
remains unchanged. This Gaussian function in multiplied by the constant
factor $\exp[- 2 (1-\xi)\eta T |\alpha_0|^2]$.
For our measurement, $\xi \approx 97\%$ and 
$\eta T |\alpha_0|^2 \approx 1.34$,
which gives the value of this factor equal 0.92. This result agrees
with the height of the
experimentally measured Wigner function of a coherent state.

\begin{figure}

\vspace*{5mm}

\centerline{\epsffile{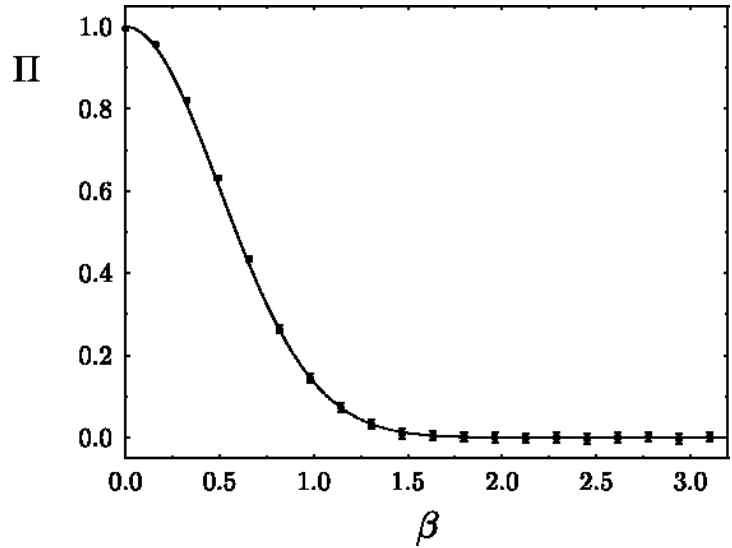}}
\caption{Comparison of the phase-averaged experimental Wigner function
for the vacuum state (points with error
bars) with the theoretical prediction (line),
given by the Gaussian $\exp(-2|\beta|^2)$.}
\label{Fig:ExpVacuum}
\end{figure}

In Fig.~\ref{Fig:ExpVacuum} we compare the phase-averaged measured Wigner
function for the vacuum state with theoretical predictions. Agreement
between the experimental points and the Gaussian curve is very
good. This plot can be used to estimate the amount of excess thermal
noise in the laser radiation. Let us assume that from the average
$n_{\prtext{\scriptsize vac}} = |\beta|^2$ registered photons a constant
fraction $\varkappa_{\prtext{\scriptsize th}}$ originates from thermal
noise. The $P$-representation of the field that is effectively detected
is given by
\begin{equation}
P(\gamma) = \frac{1}{\pi \varkappa_{\prtext{\scriptsize th}} |\beta|^2}
\exp \left( - \frac{|\gamma - \sqrt{1 -\varkappa_{\prtext{\scriptsize
th}}} \beta |^2 }{\varkappa_{\prtext{\scriptsize th}} |\beta|^2} \right).
\end{equation}
The average photon number parity measured for such a field reads:
\begin{equation}
\Pi(\beta) =
\frac{1}{2 \varkappa_{\prtext{\scriptsize th}} |\beta|^2 + 1}
\exp\left(
-\frac{2(1-\varkappa_{\prtext{\scriptsize th}} ) |\beta|^2}{2
\varkappa_{\prtext{\scriptsize th}}|\beta|^2 +1 }
\right)
\end{equation}
and in the presence of thermal noise exhibits departure from a pure
Gaussian shape. The perfect agreement observed in Fig.~\ref{Fig:ExpVacuum}
confirms that the contribution of thermal noise is negligibly small.

Concluding, let us compare the
demonstrated direct method for measuring the Wigner function with
the optical homodyne tomography approach. An important parameter
in experimental quantum state reconstruction is the detection
efficiency. Here better figures are exhibited by the homodyne technique,
which detects quantum fluctuations as a difference between two
rather intense fields. Such fields can be efficiently converted into
photocurrent signals with the help of {\it p-i-n} diodes. It should
be also noted that an avalanche photodiode is not capable of resolving
the number of simultaneously absorbed photons, and that it delivers a
signal proportional to  the light intensity only in the regime used in
our experiment. However, continuous progress in single photon detection
technology gives hope to overcome current limitations of photon counting
\cite{KwiaSteiPRA93}.  Alternatively, the displacement transformation
implemented in the photon counting technique can be combined with
efficient random phase homodyne detection. This yields the recently
proposed scheme for cascaded homodyning \cite{KisKissPRA99}.

The simplicity of the relation (\ref{Eq:Wdef}) linking the count
statistics with quasidistribution functions allows one to determine
the Wigner function at a given point from a relatively small sample of
experimental data.  This feature becomes particularly advantageous, when
we consider detection of multimode light.  Optical homodyne tomography
requires substantial numerical effort to reconstruct the multimode Wigner
function. In contrast, the photon counting method has a very elegant
generalization to the multimode case: after applying the displacement to
each of the involved modes, the Wigner function at the selected point
is simply given by the average parity of the total number of detected
photons. Moreover, the dichotomic outcome of such a measurement provides
a novel way of testing quantum nonlocality exhibited by correlated states
of optical radiation \cite{BanaWodkPRL99}.

\chapter{Statistical uncertainty in photodetection measurements}
\label{Chap:Statistical}

\markright{CHAPTER \thechapter . \uppercase{Statistical uncertainty\ldots}}

In this chapter, we shall study the problem of statistical uncertainty
from a more general point of view.
Over recent years, the set of tools for measuring quantum statistical
properties of optical radiation has substantially enlarged. 
In addition to the techniques discussed in this thesis: double homodyne
detection, optical homodyne tomography, and the direct method for measuring
quasidistribution functions, other novel schemes have been proposed
\cite{BardMayrPRA95,PaulTormPRL96,ZuccVogePRA96,%
RaymMcAlPRA96,JacoKnigJMO97}.
The quantum optical ``toolbox'' for measuring light contains now
experimentally established schemes for reconstructing various
representations of its quantum state: the $Q$ function,
the Wigner function, and the density matrix in the
quadrature and the Fock bases. A
device that is used in most of quantum optical schemes to convert the
quantum signal to a macroscopic level is the photodetector. Thus,
photodetection is a basic ingredient of quantum optical measurements.

The quantum state can be characterized using various representations:
quasidistribution functions or a density matrix in a specific basis. 
From a theoretical point of view, all these forms are equivalent. Each
representation contains complete information on the quantum state, and
they can be transformed from one to another. Any observable related to
the measured systems can be evaluated from an arbitrary representation
using an appropriate expression \cite{VogelWelsch}.

This simple picture becomes much more complicated when we deal with
real experimental data rather than analytical formulae. Each quantity
determined from a finite number of experimental runs is affected
by a statistical error. Consequently, the density matrix or the
quasidistribution function reconstructed from the experimental data
is known only with some statistical uncertainty. This uncertainty
is important when we further use the reconstructed information to
calculate other observables or to pass to another representation.
The crucial question is, whether determination of a certain representation
with sufficient accuracy guarantees that arbitrary observable can be
calculated from these data with a reasonably low statistical error. If this
is not the case, the reconstructed information on the quantum state of
the measured system turns out to be somewhat incomplete. Furthermore,
transformation between various representations of the quantum state
becomes a delicate matter.

These and related problems call for a rigorous statistical analysis of
quantum optical measurements. We shall provide here
complete statistical description of a measurement of quantum observables
in optical schemes based on photodetection. This approach
fully characterizes statistical properties of quantities determined
in a realistic measurement from a finite number of experimental
runs. It can be applied either to determination of a single quantum
observable, or to the reconstruction of the quantum state in a specific
representation. Within the presented framework we study, using  a simple
example, the completeness of the experimentally reconstructed information
on the quantum state. We demonstrate the pathological behavior suggested
above, when the reconstructed data cannot be used to calculate certain
observables due to rapidly exploding statistical errors.
Our example is motivated by the direct scheme for measuring
quasidistribution functions of light.

Compared to previous works on statistical noise,
our attention will be focused here
on two important statistical aspects of quantum state measurement.  First,
we shall go beyond the minimum second-order treatment of the statistical
error \cite{DAriMaccQSO97,LeonMunrOpC96,DAriPariPLA97} and provide,
in a closed mathematical form, a complete statistical description of
quantum observables determined from realistic measurements. This
result, derived from the first principles without using any
approximations, is an exact analytical solution to the problem which
so far has been approached only by means of Monte Carlo simulations
\cite{DAriMaccPRA94,DAriMaccPLA94}.  The second problem discussed
here will be the feasibility of reconstructing the quantum state from
realistic, finite data collected in a specific experimental scheme. It
is generally believed that a sufficient condition for successful
reconstruction is the existence of the covariance matrix with finite
elements \cite{KissHerzPRA95,DAriMaccPRA98,KissHerzPRA98}. We point out
that this belief misses an important issue resulting from statistical
uncertainty.  Suppose we have reconstructed a family of observables with
finite statistical variances. In order to be sure that the reconstructed
information on the quantum state is accurate and complete, we should ask
the following question: can we always use these observables to evaluate
any quantum property of the measured system that can be expressed in
terms of the reconstructed family?  We shall give a clear negative
answer to this question, based on a detailed discussion of carefully
selected counterexamples. Although all observables involved in these
examples are represented by bounded, well-defined operators, statistical
fluctuations are shown to be arbitrarily huge, when the reconstructed
family is further used to evaluate certain quantum expectation values.
This singular behaviour is quantitatively explained as a result of strong
statistical correlations between observables reconstructed from the same
sample of experimental data.  Consequently, the discussed examples clearly
demonstrate that statistical properties of reconstructed observables
depend in an essential way on a specific experimental scheme, which
effectively limits available information on the measured system.

This chapter is organized
as follows. The starting point of our analysis is the probability
distribution of obtaining a specific histogram from $N$ runs of the
experimental setup. This basic quantity determines all statistical
properties of quantum observables reconstructed from a finite sample of
experimental data. We characterize these properties using the
generating function, for which we derive an exact expression
directly from the probability distribution of the experimental
outcomes. These general results are presented in Sec.~\ref{Sec:Analysis}.
Then, in Sec.~\ref{Sec:Random}, we use the developed formalism to discuss
the reconstruction of the photon number distribution 
of a single light mode, and its subsequent utilization to
evaluate the parity operator $\hat{\Pi}$. We consider two experimental
schemes: direct photon counting using an imperfect detector, and
homodyne detection with random phase. In both the cases we
find that the evaluation of the parity operator from the reconstructed
photon statistics is a very delicate matter. For photon counting of a
thermal state, we show that neither the statistical mean value 
of $\hat{\Pi}$ nor its
variance have to exist when we take into account arbitrarily high count
numbers.  For random phase homodyne detection, the statistical error of
the parity operator is an interplay of the number of runs $N$ and the
specific regularization method used for its evaluation. The example
of the parity operator illustrates
difficulties related to the transformations
between various experimentally determined representations of the
quantum state, as the parity operator yields, up to a multiplicative
constant, the Wigner function at the phase space origin. Finally,
in Sec.~\ref{Sec:Consequences} we discuss consequences of statistical
uncertainty in the measurements of the quantum state.

\section{Statistical analysis of experiment}
\label{Sec:Analysis}

In photodetection measurements,
the raw quantity delivered by a single experimental run is the number of
photoelectrons ejected from the active material of the detectors. The
data recorded for further processing depends on a specific scheme. It
may be just the number of counts on a single detector, or a
difference of photocounts on a pair of photodetectors, which
is the case of balanced homodyne detection. It may also be a finite
sequence of integer numbers, e.g.\ for double homodyne detection. We
will denote in general this data by $n$, keeping in mind all the
possibilities.

The experimental scheme may have some external parameters $\theta$,
for example the phase of the local oscillator in homodyne detection.
The series of measurements are repeated for various settings
$\theta_i$ of these parameters. Thus, what is eventually obtained from the
experiment, is a set of histograms $\{k_n\}_{\theta_i}$, telling in how
many runs with the settings $\theta_i$ the outcome $n$ has been recorded.
We assume that for each setting the same total number of $N$ runs has
been performed.

The theoretical probability $p_n(\theta_i)$ of
obtaining the outcome $n$ in a run with settings $\theta_i$ is given by
the expectation value of a positive operator-valued measure (POVM)
$\hat{p}_{n}(\theta_i)$ acting in the Hilbert space of the measured
system. The reconstruction of an observable $\hat{A}$ is possible, if it
can be represented as a linear combination of the POVMs for the
settings used in the experiment:
\begin{equation}
\label{Eq:hatAdef} 
\hat{A} = \sum_{i} \sum_{n}
a_{n}(\theta_{i})\hat{p}_{n}(\theta_i),
\end{equation} 
where $a_{n}(\theta_i)$ are the kernel functions. 
The above formula allows
one to compute the quantum expectation value $\langle\hat{A}\rangle$
from the probability distributions $p_n(\theta_i)
= \langle \hat{p}_{n}(\theta_i)\rangle$.

This theoretical relation has to be applied now to the experimental
data. The simplest and the most commonly used strategy is to estimate
the probability distributions $p_n(\theta_i)$
by experimental relative frequencies
$(k_n/N)_{\theta_i}$. The relative frequencies integrated with the
appropriate kernel functions yield an estimate for the expectation
value of the operator $\hat{A}$. For simplicity, we will denote this
estimate just by $A$. Thus, the recipe for reconstructing the observable
$A$ from experimental data is given by 
the counterpart of Eq.~(\ref{Eq:hatAdef}):
\begin{equation}
\label{Eq:Adeterm}
A = \sum_{i} \sum_{n} a_{n}(\theta_{i}) \left(\frac{k_n}{N}
\right)_{\theta_i}.
\end{equation}

We will now analyse statistical properties of the
observable $A$ evaluated according to Eq.~(\ref{Eq:Adeterm})
from data collected in a finite number of experimental runs. 
Our goal is to characterize the statistical distribution $w(A)$
defining the probability that the experiment yields a specific
result $A$.
The fundamental object in this
analysis is the probability ${\cal P}(\{k_n\};\theta)$ of
obtaining a specific histogram $\{k_n\}$ for the settings $\theta$. 
In order to avoid convergence problems, we will
restrict the possible values of $n$ to a finite set by introducing
a cut-off. The probability
${\cal P}(\{k_n\};\theta)$ is then given by the multinomial distribution
\cite{EadieMultinomial}:
\begin{equation}
{\cal P}(\{k_n\};\theta)
=
\frac{N!}{(N-\sum_{n}'k_n)!}
\left( 1 - {\sum_{n}}'p_n(\theta) \right)^{N-\sum_{n}' k_n}
{\prod_{n}}'\frac{1}{k_n!} [p_n(\theta)]^{k_n},
\end{equation}
where prim in sums and products denotes the cut-off. This distribution
describing experimental histograms is derived from an assumption that 
all the detection events are statistically independent.

Let us first consider a contribution $A_i$ to the observable
$A$ calculated from the histogram $\theta_i$:
\begin{equation}
A_i = {\sum_{n}}' a_n (\theta_i) \left( \frac{k_n}{N} 
\right)_{\theta_i}.
\end{equation}
Its statistical distribution $w(A_i; \theta_i)$ is given by the 
following sum over all possible histograms that can be
obtained from $N$ experimental runs:
\begin{equation}
w(A_i;\theta_i) = \sum_{\{k_n\}} {\cal P}(\{k_n\}; \theta_i)
\delta\left( A_i - \frac{1}{N} {\sum_{n}}'a_n(\theta_i) k_n \right).
\end{equation}
Equivalently, the statistics of $A_i$ can be characterized by
the generating function $\tilde{w}(\lambda;\theta_i)$ for the moments,
which is the Fourier transform of the distribution $w(A_i;\theta_i)$:
\begin{eqnarray}
\tilde{w}(\lambda;\theta_i) & =  & 
\int \prtext{d}A_i \; e^{i\lambda A_i}
w(A_i;\theta_i)
\nonumber \\
& = & \sum_{\{k_n\}} {\cal P}(\{k_n\}; \theta_i)
\exp \left( \frac{i\lambda}{N}
{\sum_{n}}' a_n(\theta_i) k_n \right). 
\end{eqnarray}
An easy calculation yields the explicit form of the generating function:
\begin{equation}
\tilde{w}(\lambda; \theta_i) = \left(
1 + {\sum_{n}}' p_n(\theta_i)(e^{i\lambda a_n(\theta_i)/N}-1)
\right)^{N}.
\end{equation}
The observable $A$ is obtained via summation of the components $A_i$
corresponding to all settings of the external parameters $\theta_i$. As these
components are determined from disjoint subsets of the experimental
data, they are statistically independent. Consequently, the generating
function $\tilde{w}(\lambda)$ for the moments of the observable $A$
is given by the product:
\begin{eqnarray}
\tilde{w}(\lambda) & = & 
\int\prtext{d}A \; e^{i\lambda A} w(A) =
\prod_{i} \tilde{w}(\lambda; \theta_i)
\nonumber \\
\label{Eq:wtildefinal}
& = &
\prod_{i}
\left(
1 + {\sum_{n}}' p_n(\theta_i)(e^{i\lambda a_n(\theta_i)/N}-1)
\right)^{N}.
\end{eqnarray}
This expression contains the complete statistical information
on determination of the observable $A$ from a finite number
of runs of a specific experimental setup. The measuring apparatus is
included in this expression in the form of a family of POVMs
$\hat{p}_{n}(\theta_i)$. The quantum expectation value of these POVMs
over the state of the measured system yields the probability
distributions $p_n(\theta_i)$. Finally, the coefficients
$a_n(\theta_i)$ are given by the computational recipe for
reconstructing the observable $A$ from the measured distributions. The
analytical expression for the generating function given in
Eq.~(\ref{Eq:wtildefinal}), derived from the exact description of raw
experimental outcomes, provides a complete characterization of
statistical fluctuations in realistic measurements of quantum
observables. Let us note that in general the generating function
$\tilde{w}(\lambda)$ cannot be expressed by the operator
$\hat{A}$ alone. Both the POVMs $\hat{p}_{n}(\theta_i)$ and the kernel
functions $a_n(\theta_i)$ enter Eq.~(\ref{Eq:wtildefinal}) in a
nontrivial way, which clearly shows that statistical properties of the
reconstructed observable depend essentially on the specific measurement
scheme.

The basic characteristics of statistical properties of the observable $A$ is
provided by the mean value $\prtext{E}(A)$ 
and the variance $\prtext{Var}(A)$. These
two quantities can be easily found by differentiating the logarithm of the
generating operator:
\begin{eqnarray}
\prtext{E}(A) & := & \left. \frac{1}{i} \frac{\prtext{d}}{\prtext{d}\lambda}
\log \tilde{w}(\lambda) \right|_{\lambda=0} =
\sum_{i} {\sum_{n}}' a_n(\theta_i) p_{n}(\theta_i), \\
\prtext{Var}(A) & := & \left.
\frac{1}{i^2} \frac{\prtext{d}^2}{\prtext{d}\lambda^2}
\log\tilde{w}(\lambda) \right|_{\lambda=0} \nonumber \\
& = & 
\label{Eq:VarA}
\frac{1}{N}
\left[ \sum_{i}
{\sum_{n}}' a_n^{2} (\theta_i) p_{n}(\theta_i)
- 
\sum_{i}
\left( {\sum_{n}}' a_n(\theta_i) p_n(\theta_i) \right)^{2}
\right].
\end{eqnarray}
The statistical error is scaled with the inverse of the square root
of the number of runs $N$. Let us note that the second component in the
derived formula for $\prtext{Var}(A)$ differs from that used in 
the discussions of homodyne tomography in
Refs.~\cite{DAriMaccQSO97,DAriPariPLA97},
where it was equal just to $[\prtext{E}(A)]^2$.
This difference results from different assumptions about the local
oscillator phase: in Refs.~\cite{DAriMaccQSO97,DAriPariPLA97} 
it was considered to be a uniformly distributed
stochastic variable in order to avoid systematic
errors, whereas we have assumed that
the number of runs is fixed for each selected setting of the
external parameters.

The goal of quantum state measurements is to retrieve the maximum amount of
information on the quantum state available from the experimental data.
Therefore the experimental histograms are usually processed many times in
order to reconstruct a family of observables characterizing the quantum
state. Of course, quantities determined from the same set of 
experimental data are not
statistically independent, but in general exhibit correlations. 
The analysis presented above can be easily extended to the
evaluation of any number of observables from the same sample of
experimental data. If we restrict our
attention to the basic, second-order characterization of these correlations,
it is sufficient to discuss simultaneous determination of two observables. 
Let us suppose that
in addition to $A$, another observable $B$ has been calculated
from the histograms $\{k_n\}_{\theta_i}$ according to the formula:
\begin{equation}
B = \sum_{i} {\sum_{n}}' b_{n} (\theta_i) \left( \frac{k_n}{N}
\right)_{\theta_i}.
\end{equation}
The generating function $\tilde{w}(\lambda,\mu)$ corresponding
to the joint probability distribution $w(A,B)$ can be found analogously
to the calculations presented above. The final result is:
\begin{eqnarray}
\tilde{w}(\lambda,\mu) & = & \int \prtext{d}A \prtext{d}B \;
e^{i\lambda A + i \mu B} w(A,B) \nonumber \\
& = &  
\prod_{i} \left( 1 + {\sum_{n}}' p_n(\theta_i)
(e^{i\lambda a_n(\theta_i)/N + i\mu b_n(\theta_i)/N} - 1 )
\right)^{N}.
\end{eqnarray}
The covariance between the experimentally determined values
of $A$ and $B$ is given by:
\begin{eqnarray}
\lefteqn{\begin{array}{rcl} \prtext{Cov}(A,B) & := &
\displaystyle
\left.
\frac{1}{i^2} \frac{\prtext{d}^2}{\prtext{d}\lambda
\prtext{d}\mu} \log \tilde{w}(\lambda,\mu)
\right|_{\lambda,\mu=0}
\end{array} } & &  \nonumber \\
& = & 
\frac{1}{N}
\sum_{i} \left[ {\sum_{n}}' a_n (\theta_i) b_n (\theta_i)
p_n (\theta_i) - \left(
{\sum_{n}}' a_n(\theta_i) p_n(\theta_i) \right)
\left({\sum_{m}}' b_{m}(\theta_i) p_{m}(\theta_i) \right)
\right] . \nonumber \\
& &
\end{eqnarray}
The covariance can be normalized to the interval $[-1,1]$ using
$\prtext{Var}(A)$ and $\prtext{Var}(B)$, which yields the correlation
coefficient for the pair of observables $A$ and $B$:
\begin{equation}
\label{Eq:Corr}
\prtext{Corr}(A,B) := \frac{\prtext{Cov}(A,B)}{\sqrt{\prtext{Var}(A)
\prtext{Var}(B)}} .
\end{equation}
This quantity defines whether the statistical deviations of $A$ and
$B$ tend to have the same or opposite sign, which corresponds
respectively to the positive or negative value of $\prtext{Corr}(A,B)$. 

We have assumed that the histograms $k_n$ have been measured for
a finite number of external parameters settings $\theta_i$, which
is always the case in an experiment. However, in some schemes the
measurement of histograms is in principle necessary for all values of a
continuous parameter. For example, in optical homodyne tomography the full
information on the quantum state is contained in a family of quadrature
distributions for all local oscillator phases.  Restriction to a finite
set of phases introduces a systematic error to the measurement
\cite{DAriMaccQSO97,LeonMunrPRA96}.

\section{Phase-insensitive detection of a light mode}
\label{Sec:Random}

We will now apply the general formalism developed in the preceding
section to the reconstruction of phase-independent properties
of a single light
mode. The basic advantage of this exemplary system is that it
will allow us to discuss, in a very transparent way, pathologies
resulting from the statistical uncertainty.
All phase-independent properties of a single light mode are fully
characterized by its photon number distribution $\rho_\nu$.
Therefore, it is sufficient to apply a
phase-insensitive technique to measure the photon statistics of the field. 
We will consider two measurement schemes that can be used for this
purpose: direct photon counting and
random phase homodyne detection.

The photon number distribution is given by the expectation value of a
family of projection operators 
$\hat{\rho}_{\nu} = |\nu\rangle\langle\nu|$,
where $|\nu\rangle$ is the $\nu$th Fock state. In principle, 
knowledge of this
distribution enables us to evaluate any phase-independent
observable related to the measured field. A simple yet nontrivial
observable, which we will use to point out difficulties with the
completeness of the reconstructed information on the quantum state,
is the parity operator:
\begin{equation}
\label{Eq:ParityOp}
\hat{\Pi} = \sum_{\nu=0}^{\infty} (-1)^{\nu} |\nu\rangle\langle\nu|.
\end{equation}
This operator is bounded, and well defined on the complete Hilbert
space of a single light mode. Its expectation value is given by the
alternating series of the photon number distribution:
\begin{equation}
\label{Eq:Parity}
\langle\hat{\Pi}\rangle = \sum_{\nu=0}^{\infty} (-1)^{\nu} 
\langle\hat{\rho}_{\nu}\rangle,
\end{equation}
which is absolutely convergent for any quantum state. Therefore, any
pathologies connected to its determination
from experimental data, if there appear any, cannot be
ascribed to its singular analytical properties.

\subsection{Direct photon counting}

First, we will consider the reconstruction of phase insensitive
properties of a single light mode from data measured using a
realistic, imperfect photodetector. 
The positive operator-valued measure
$\hat{p}_n$ describing the probability of ejecting $n$ photoelectrons
from the detector is given by \cite{KellKleiPR64}:
\begin{equation}
\hat{p}_{n} = \; : \frac{(\eta\hat{a}^{\dagger}\hat{a})^{n}}{n!}
\exp(-\eta\hat{a}^{\dagger}\hat{a}):,
\end{equation}
where $\hat{a}$ is the annihilation operator of the light mode, and
$\eta$ is the quantum efficiency of the photodetector. 
In the limit $\eta\rightarrow 1$ we get directly $\hat{p}_{n}
= |n\rangle\langle n|$. In a general case, the probability
distribution for the photoelectron number 
is related to the photon statistics
via the Bernoulli transformation.
This relation can be analytically inverted
\cite{KissHerzPRA95},
which yields the expression:
\begin{equation}
\label{Eq:InvBernoulli}
\hat{\rho}_{\nu} = \sum_{n=0}^{\infty} r_{\nu n}^{(\eta)} \hat{p}_{n},
\end{equation}
where the kernel functions $r_{\nu n}^{(\eta)}$
are given by:
\begin{equation}
r_{\nu n}^{(\eta)} = \left\{
\begin{array}{ll}
0, & n < \nu, \\
\displaystyle
\frac{1}{\eta^{\nu}}{n \choose \nu} \left( 1 - \frac{1}{\eta}
\right)^{n-\nu},
& n \ge \nu. 
\end{array}
\right.
\end{equation}
The inversion formula has a remarkable property that $\rho_{\nu}$
depends only on the ``tail'' of the photocount statistics for
$n \ge \nu$.
It has been shown that the inverse transformation can be applied to
experimentally determined photocount statistics for an arbitrary state
of the field, provided that the detection efficiency is higher that
50\% \cite{KissHerzPRA95}.

We will now discuss statistical properties of the photon number
distribution determined by photon counting within the general framework
developed in Sec.~\ref{Sec:Analysis}. The case of perfect detection is
trivial for statistical analysis. Therefore we will consider nonunit
detection efficiency, which is numerically compensated 
in the reconstruction process using the inverse Bernoulli
transformation according to Eq.~(\ref{Eq:InvBernoulli}). For all
examples presented here, the efficiency is $\eta=80\%$, which is well
above the $50\%$ stability limit.

\begin{figure}[t]

\vspace*{5mm}

\begin{center}
\epsfig{file=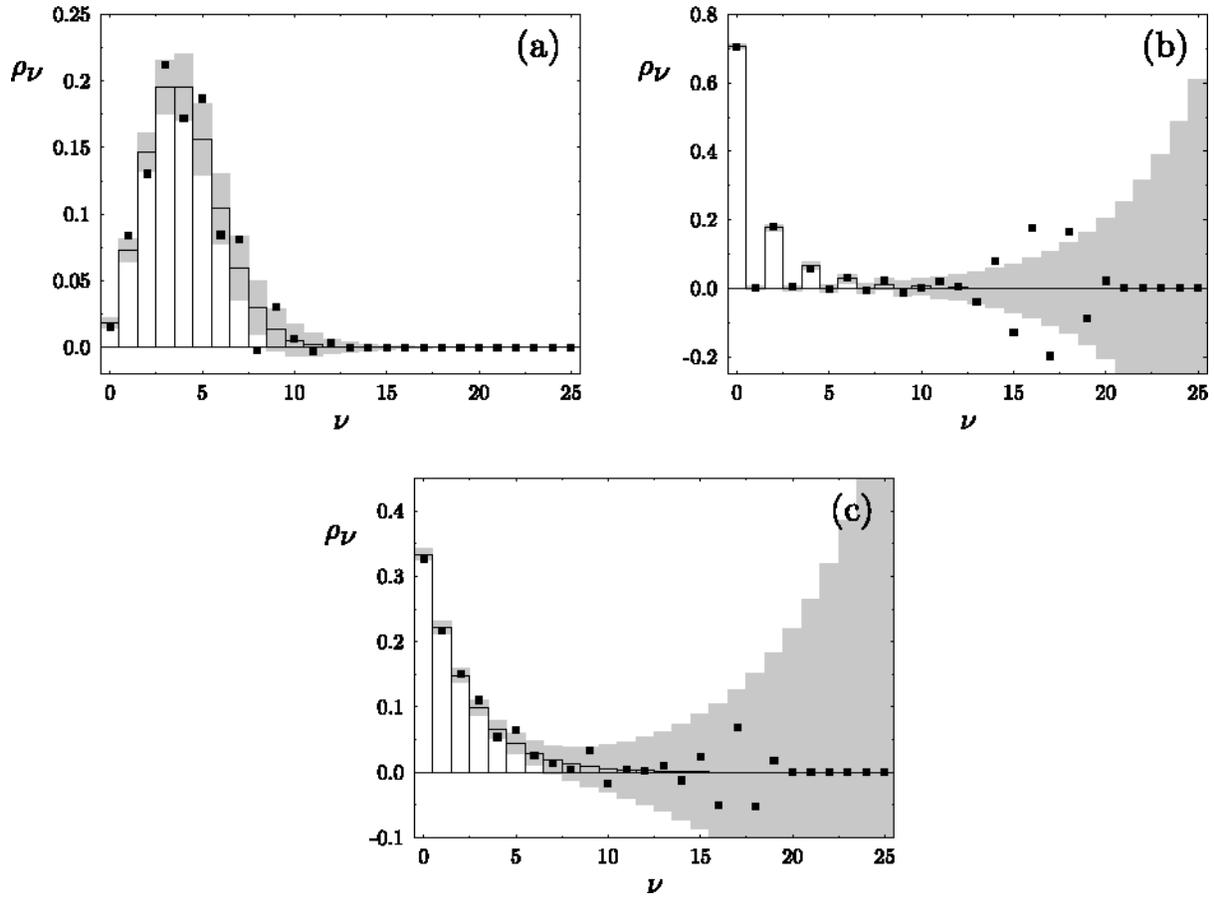}
\end{center}

\caption{Reconstruction of the photon number distribution from
photon counting for (a) a coherent state with
$\langle\hat{n}\rangle=4$, (b) a squeezed vacuum state with
$\langle\hat{n}\rangle=1$, and (c) a thermal state with
$\langle\hat{n}\rangle=2$, from $N=4000$ runs in each case.
Monte Carlo simulations of a photon counting experiment, depicted with
points, are compared with exact values (solid lines), with the
statistical errors $[\prtext{Var}(\rho_{\nu})]^{1/2}$ marked as grey
areas. The detection efficiency is $\eta=80\%$.} 
\label{Fig:PhotonDistributions}
\end{figure}

\begin{figure}
\vspace*{5mm}
\centerline{\epsffile{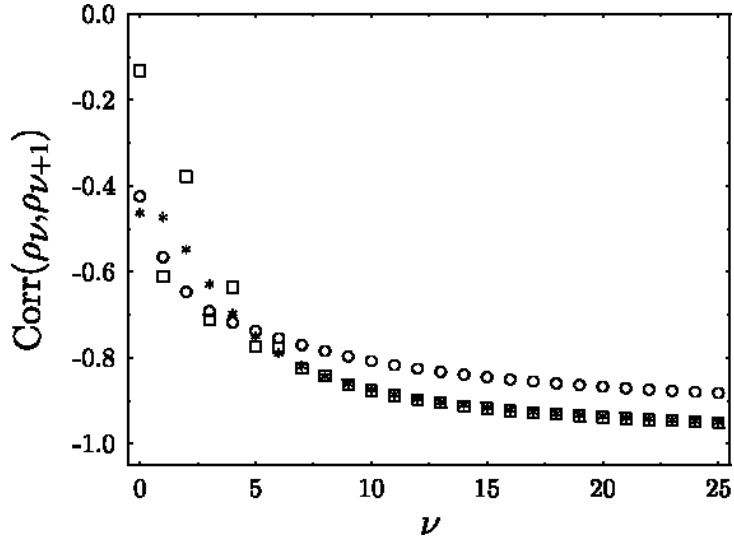}}

\caption{The correlation coefficient between the consecutive density
matrix elements $\prtext{Corr}(\rho_{\nu},\rho_{\nu+1})$, depicted for the
coherent state ($\circ$), the squeezed state ($\square$), and the
thermal state ($\ast$) from Fig.~\protect\ref{Fig:PhotonDistributions}.}
\label{Fig:Correlations}
\end{figure}

In Fig.~\ref{Fig:PhotonDistributions} we depict the reconstructed
photon number distributions for a coherent state, a thermal state, and
a squeezed vacuum state. The mean
values $\prtext{E}(\rho_\nu)$ along with their
statistical errors $[\prtext{Var}(\rho_\nu)]^{1/2}$ are compared with
Monte Carlo realizations of a photon counting experiment, with
the number of runs $N=4000$. It is seen that for a thermal state and a
squeezed vacuum state, the statistical error 
of the probabilities $\rho_\nu$
grows unlimitedly with the
photon number $\nu$. However, any experimental histogram 
obtained from a finite number of runs ends up
for a certain count number, and therefore the reconstructed photon
statistics is zero above this number. An important feature that is
evidently seen in the Monte Carlo simulations, are correlations between
the consecutive matrix elements. The reconstructed photon number
distribution clearly exhibits oscillations around the true values.
This property can be quantified using the correlation
coefficient defined in Eq.~(\ref{Eq:Corr}), which we plot for all three
states in Fig.~\ref{Fig:Correlations}. For large $\nu$'s,
$\prtext{Corr}(\rho_\nu,\rho_{\nu+1})$ is close to its minimum allowed
value $-1$, which acknowledges that statistical correlations are indeed
significant.

These correlations affect any quantity computed from the reconstructed
photon number distribution. The parity operator is here a good example:
since in Eq.~(\ref{Eq:Parity}) we sum up consecutive $\rho_{\nu}$'s with
opposite signs, their statistical deviations do not add randomly,
but rather contribute with the same sign. Consequently, the statistical
error of the evaluated parity operator may be huge. It is therefore
interesting to study this case in detail. In principle, we could 
obtain statistical properties of the reconstructed
parity using the covariance matrix for the photon distribution. 
However, it will be more instructive to express
the parity directly in terms of the photocount statistics, and then to
apply the statistical analysis to this reconstruction recipe. 
This route is completely
equivalent to studying evaluation of the parity
 via the photon number distribution, as all
transformations of the experimental data, which we consider here, are
linear. 

A simple calculation
combining Eqs.~(\ref{Eq:ParityOp}) and (\ref{Eq:InvBernoulli}) shows that:
\begin{equation}
\label{Eq:Paritypn}
\hat{\Pi} = \sum_{n=0}^{K} \left( 1 - \frac{2}{\eta} \right)^n 
\hat{p}_n,
\end{equation}
where we have introduced in the upper summation limit 
a cut-off parameter $K$ for the photocount
number.  This formula clearly demonstrates pathologies related to the
determination of the parity operator.
For any $\eta<1$, the factor $(1-2/\eta)^n$ is not bounded,
which makes the convergence of the whole series questionable 
in the limit $K\rightarrow\infty$.
Of course, for an experimental histogram the summation is
always finite, but the exploding factor amplifies contribution from
the ``tail'' of the histogram, where usually only few events are
recorded, and consequently statistical errors are significant. 

\begin{figure}

\vspace*{0.5cm}

\centerline{\epsfig{file=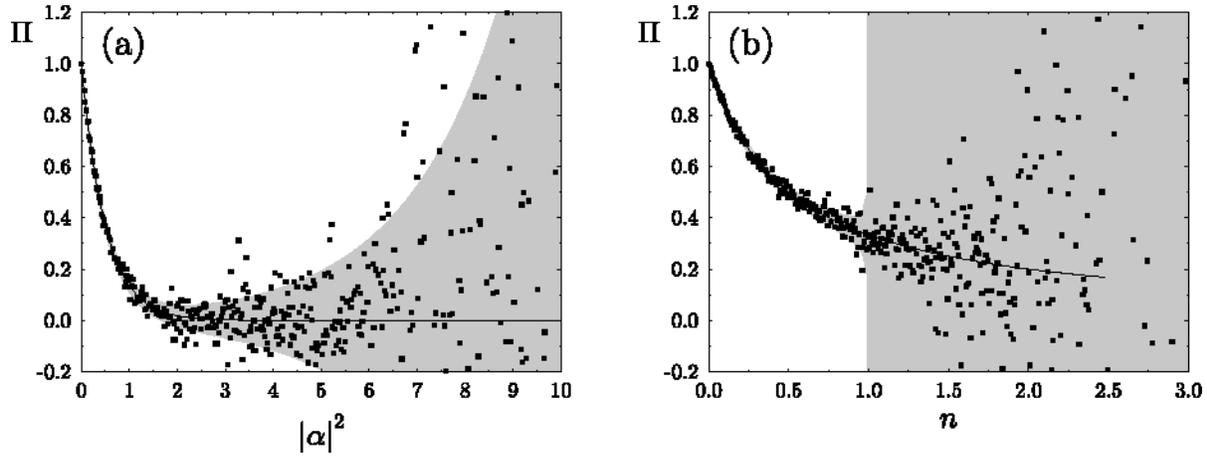}}

\caption{Determination of the parity operator for (a) coherent states
and (b) thermal states with the increasing average photon number,
assuming the photodetector efficiency $\eta=80\%$. Each
square represents the parity evaluated from Monte Carlo simulated
photon statistics with $N=4000$ runs
for a given average photon number. The solid lines and the grey areas
depict the mean value $\prtext{E}(\Pi)$ and
the error $[\prtext{Var}(\Pi)]^{1/2}$. For thermal states,
the variance diverges to infinity when $\bar{n} \ge 1$, and the mean
value $\prtext{E}(\Pi)$ does not exist above $\bar{n} \ge 2.5$.}
\label{Fig:ParityCounting}
\end{figure}

Let us study these pathologies more closely using examples of a
coherent state and a thermal state. For a coherent state
$|\alpha\rangle$, both the expressions for $\prtext{E}(\Pi)$ and
$\prtext{Var}(\Pi)$ are convergent with $K\rightarrow\infty$. However,
the variance, given by the formula:
\begin{equation}
\prtext{Var}(\Pi^{\prtext{\scriptsize 
coh}}) = \frac{1}{N} \left[ \exp \left(
\frac{4(1-\eta)}{\eta} |\alpha|^2 \right) - \exp(-4|\alpha|^2)
\right],
\end{equation}
grows very rapidly with the coherent state amplitude $\alpha$, when
the number of runs $N$ is fixed.
For a thermal state with the average photon number $\bar{n}$, the
matter becomes more delicate. When $K\rightarrow\infty$,
the series (\ref{Eq:Paritypn}) is
convergent only for $\bar{n} < 1/[2(1-\eta)]$, which for $\eta=80\%$
gives just $2.5$ photons. Even when the mean value exists, the
variance is finite only for $\bar{n}<\eta/[4(1-\eta)]$ and equals:
\begin{equation}
\prtext{Var}(\Pi^{\prtext{\scriptsize th}}) = 
\frac{1}{N}\left( \frac{\eta}{\eta - 4 \bar{n}(1-\eta)}
- \frac{1}{(1 + 2 \bar{n})^2} \right).
\end{equation}
We illustrate these results with Fig.~\ref{Fig:ParityCounting},
depicting Monte Carlo simulations for various average photon
numbers. For coherent states, statistical fluctuations can in
principle be suppressed by increasing the number of runs. For thermal
states, the situation is worse: when $\bar{n}\ge 1$, the variance
cannot even be used as a measure of statistical uncertainty. 

Let us recall that the bound $\eta > 50\%$ for the stability of the
inverse Bernoulli transformation is independent of the state to be
measured. It has been obtained from the requirement that in the limit
$K\rightarrow\infty$ both $\prtext{E}(\rho_\nu)$ and
$\prtext{Var}(\rho_\nu)$ should converge \cite{KissHerzPRA95}.
The example with the parity operator clearly shows, that the condition
$\eta>50\%$ does not guarantee that the reconstructed photon number
distribution can be safely used to determine an arbitrary well-behaved
phase independent observable. Thus, imperfect detection is inevitably
connected with some loss of the information on the measured quantum
state.

We have noted that as long as finite, experimental data are concerned,
evaluation of observables via intermediate
quantities is equivalent to expressing them directly in terms of the
measured probability distributions. 
One might try to circumvent the $\eta>50\%$ bound for
reconstructing
the photon statistics by applying the inverse Bernoulli transformation
in two or more steps,
and compensating in each step only a fraction of its inefficiency. Of
course, such a strategy must fail, as for any finite sample of
experimental data this treatment is equivalent to a single
transformation which is unstable. In many-step processing this
instability would be reflected in increasing correlations and
statistical errors exploding to infinity. 

\subsection{Random phase homodyne detection}

As we discussed in Sec.~\ref{Sec:RandomHomodyne},
random phase homodyne detection is a recently developed technique for
measuring phase-independent properties of optical radiation, which
goes beyond certain limitations of plain photon counting
\cite{MunrBoggPRA95}. Data recorded in this scheme is the difference of
counts on two photodetectors measuring superposition of the signal
field with a strong coherent local oscillator. The count difference is
rescaled by the local oscillator amplitude, and the resulting
stochastic variable $x$ can be treated with a good approximation as a
continuous one. The photon number distribution is reconstructed from
the random phase homodyne statistics $p_{\cal R}(x)$
by integrating it with pattern
functions $f_{\nu}(x)$ \cite{DAriMaccPRA94,DAriLeonPRA95,LeonPaulPRA95}:
\begin{equation}
\langle\hat{\rho}_{\nu}\rangle
 = \int_{-\infty}^{\infty} \prtext{d}x \, f_{\nu}(x)
p_{\cal R}(x).
\end{equation}
A convenient method for numerical evaluation of the pattern functions
has been described in Ref.~\cite{LeonMunrOpC96}. 

Let us now discuss statistical properties of the homodyne scheme in
its discretized version used in experiments, when the rescaled count
difference is divided into finite width bins. 
As the local oscillator phase is
random, the setup has no controllable parameters, and the statistics
of the observables is fully determined by $p_{\cal R}(x)$.
Statistical errors of the density matrix in the Fock basis
reconstructed via homodyne detection have been studied in
Ref.~\cite{DAriMaccQSO97}.
Here we will focus our attention on statistical
correlations exhibited by the diagonal 
density matrix elements, and their further
utilization for evaluating phase-independent observables. 

\begin{figure}[t]
\vspace*{5mm}
\begin{center}

\epsfig{file=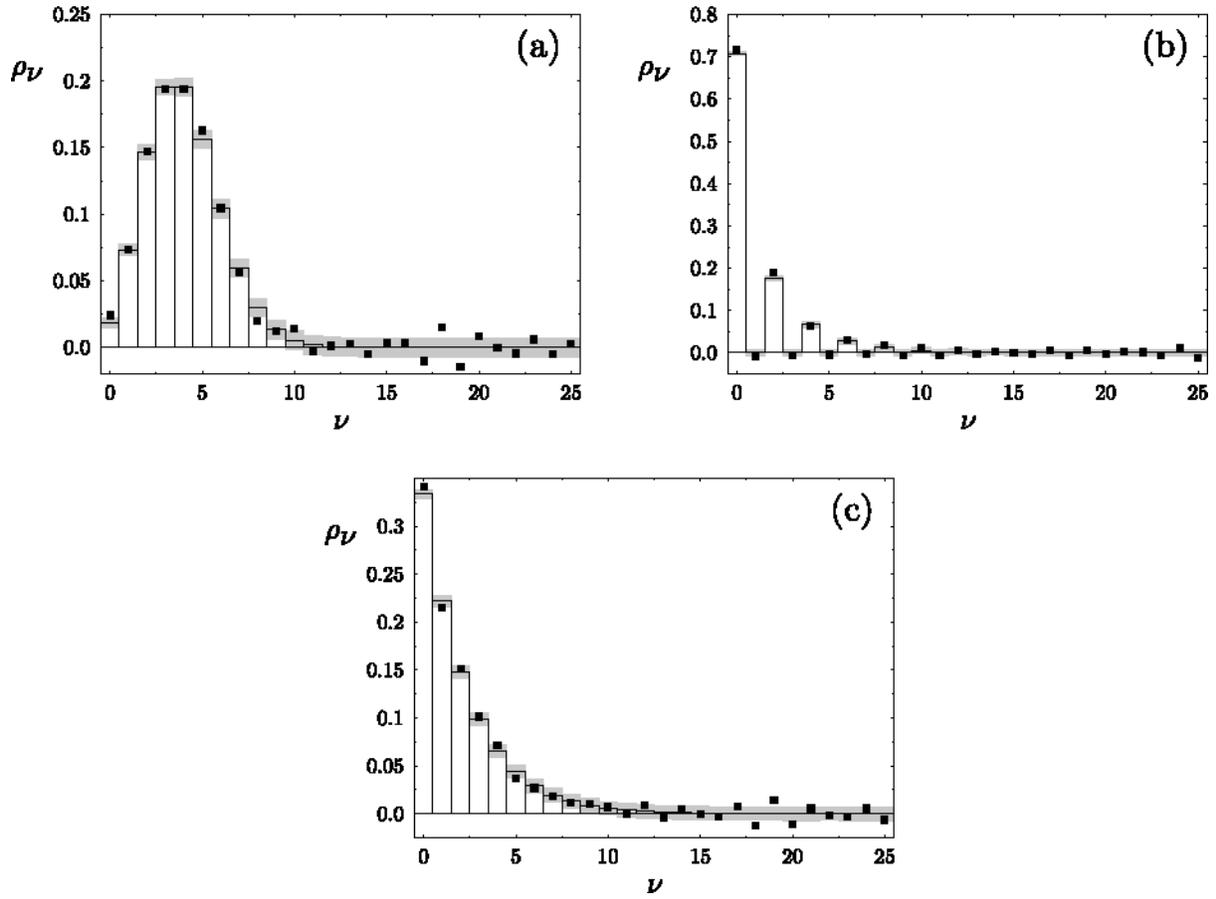}

\end{center}

\caption{Random phase homodyne reconstruction of the
photon number distribution for (a) the coherent state (b)
the squeezed state and (c) the thermal state, all states with the
same photon numbers as in Fig.~\protect\ref{Fig:PhotonDistributions}.
The range of the homodyne variable is restricted to the interval $-6\le x
\le 6$ divided into 1200 bins. The simulated homodyne statistics is obtained
from $N=4 \cdot 10^4$ Monte Carlo events.}
\label{Fig:HomoPhotonDist}
\end{figure}

\begin{figure}

\vspace*{5mm}

\centerline{\epsffile{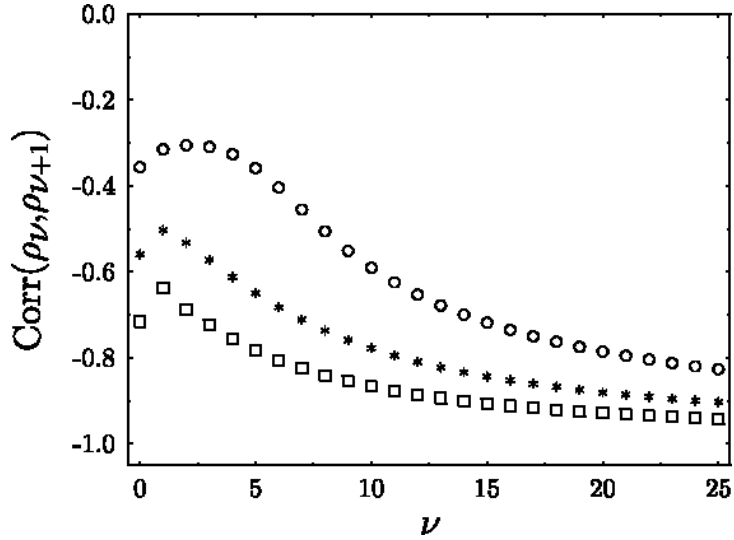}}
\caption{The correlation coefficient for the consecutive photon
number probabilities for the coherent state ($\circ$), the squeezed
state ($\square$) and the thermal state ($\ast$).}
\label{Fig:HomoCorr}
\end{figure}

We will consider the unit detection efficiency $\eta=1$, with no
compensation in the processing of the experimental data. 
This is the most regular case from the numerical point of view. When
$\eta < 1$ and the compensation is employed, the statistical errors are
known to increase dramatically \cite{DAriMaccQSO97}.
In Fig.~\ref{Fig:HomoPhotonDist} we depict 
the homodyne reconstruction of the photon
number distribution for the three states discussed in the previous
subsection. For large $\nu$, the statistical errors tend to a fixed
value $\sqrt{2/N}$, which has been explained by D'Ariano {\em et al.}
using the asymptotic form of the pattern functions \cite{DAriMaccQSO97}.
Again, Monte Carlo simulations suggest that the reconstructed density
matrix elements are correlated, which is confirmed by the correlation
coefficient for the consecutive photon number probabilities, plotted in
Fig.~\ref{Fig:HomoCorr}. A simple 
analytical calculation involving the asymptotic
form of the pattern functions shows, that for large $\nu$ this
coefficient tends to its minimum value $-1$. 

\begin{figure}

\vspace*{5mm}

\centerline{\epsffile{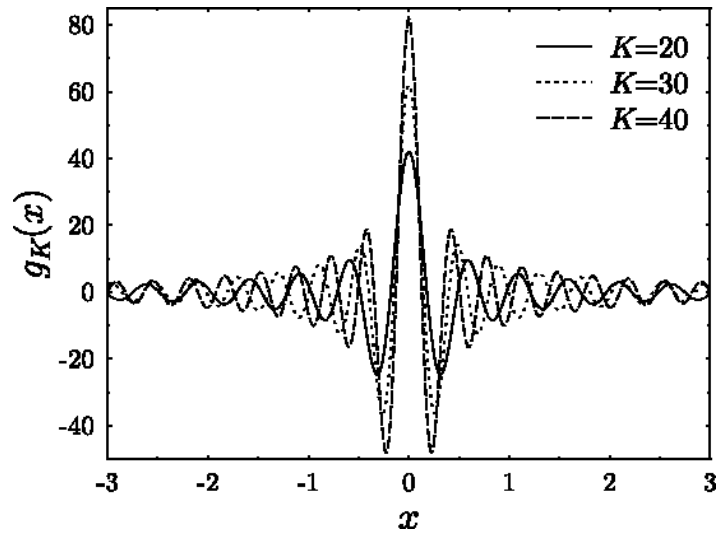}}
\caption{Regularized kernel functions for the parity operator
$g_K(x)$
evaluated as a finite sum of the Fock states pattern functions,
for increasing values of the cut-off parameter $K$.}
\label{Fig:ParityPatterns}
\end{figure}

\begin{figure}

\vspace*{5mm}

\centerline{\epsffile{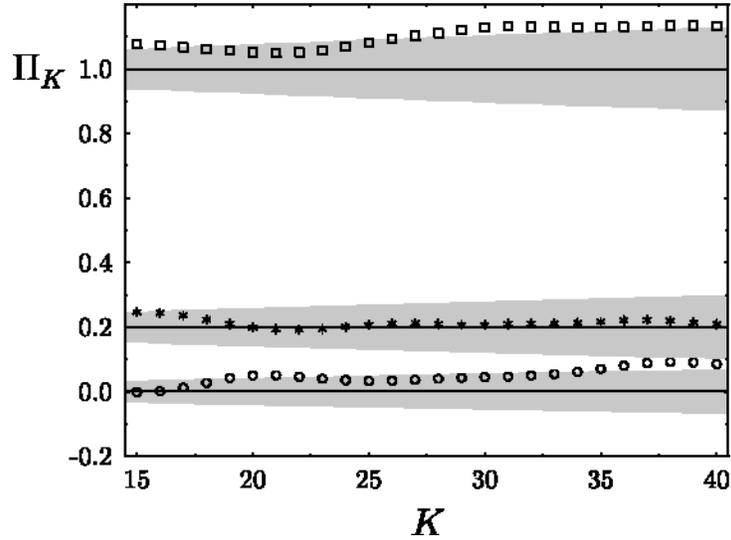}}
\caption{Reconstruction of the truncated parity operator 
$\hat{\Pi}_K$ for the coherent
state ($\circ$), the squeezed state ($\square$), and the thermal state
($\ast$) with various values of the cut-off parameter $K$,
using the same Monte Carlo homodyne statistics 
as in Fig.~\protect\ref{Fig:HomoPhotonDist}. The simulations are compared
with the corresponding mean values $\prtext{E}(\Pi_K)$ and errors
$[\prtext{Var}(\Pi_K)]^{1/2}$ plotted as solid lines surrounded by grey
areas.}
\label{Fig:HomoParity}
\end{figure}

One may now expect that no subtleties can be hidden in using the
reconstructed photon number distribution to evaluate the parity operator
according to Eq.~(\ref{Eq:Parity}).  However, let us recall that the
parity operator is equal, up to a 
multiplicative constant, to the Wigner function at
the phase space origin. 
The Wigner function is
related to the homodyne statistics via the inverse Radon transformation,
which is singular. In particular, applying this transformation
for the phase space point $(0,0)$ we obtain the following expression
for the parity operator in terms of the homodyne statistics
\cite{LeonJexPRA94}:
\begin{equation}
\langle\hat{\Pi}\rangle
 = \frac{1}{2} \int_{-\infty}^{\infty} \prtext{d}x \, p_{\cal R}(x)
\frac{\prtext{d}}{\prtext{d}x} P\frac{1}{x},
\end{equation}
where $P$ denotes the principal value. Due to the singularity of the
inverse Radon transform, its application to experimental data has to
be preceded by a special filtering procedure. This feature must
somehow show up, when we evaluate the parity
operator from the reconstructed photon statistics. In order to analyse
this problem in detail let us discuss evaluation of the 
truncated parity operator $\hat{\Pi}_K$
from a finite part of the photon number distribution:
\begin{equation}
\langle\hat{\Pi}_{K}\rangle
= \sum_{\nu=0}^{K} (-1)^\nu \langle\hat{\rho}_{\nu}\rangle
= \int_{-\infty}^{\infty} g_K(x) p(x),
\end{equation}
where 
\begin{equation}
g_{K}(x) = \sum_{\nu=0}^{K} (-1)^{\nu} f_{\nu}(x)
\end{equation}
can be considered to be a regularized kernel function for the parity
operator. In Fig.~\ref{Fig:ParityPatterns}
we plot this function for increasing
values of the cut-off parameter $K$. It is seen that the singularity
of the kernel function in the limit $K\rightarrow\infty$ is reflected
by an oscillatory behaviour around $x=0$ with growing both the
amplitude and the frequency. 
This amplifies the statistical uncertainty of the
experimental homodyne data. 
In Fig.~\ref{Fig:HomoParity} we show determination
of the parity operator for the three states discussed before, using
increasing values of the cut-off parameter $K$. Though we are in the
region where the true photon number distribution is negligibly small,
addition of subsequent matrix elements increases the statistical
error in an approximately linear manner. This is easily understood,
if we look again at the reconstructed photon number distributions:
increasing $K$ by one means a contribution of the order of
$\sqrt{2/N}$ added to the statistical uncertainty, and, moreover,
these contributions tend to have the same sign due to correlations
between the consecutive matrix elements. Thus, determination of the
parity operator from homodyne statistics requires an application of a
certain regularization procedure. It may be either the filtering used in
tomographic back-projection algorithms,
or the cut-off of the photon number
distribution. The statistical uncertainty of the final outcome is
eventually a result of an interplay between the number of experimental
runs and the applied regularization scheme. 

Finally, let us briefly comment on the compensation for the nonunit
efficiency of the homodyne detector. First, one might think of applying
a two-mode inverse Bernoulli transformation directly to the joint count
statistics on the detectors. However, it is impossible in the homodyne
scheme to resolve contributions from single absorbed photons due to
high intensity of the detected fields.  The inverse Bernoulli
transformation has no continuous limit, as consecutive count
probabilities are added with opposite signs.
Nevertheless, the
nonunit detection efficiency can be taken into account in the pattern
functions \cite{DAriLeonPRA95,LeonPaulPRA95}. 
In this case the statistical errors increase dramatically,
and explode with $\nu\rightarrow\infty$, which makes determination of
the parity operator even more problematic. This is easily understood
within the phase space picture: the distributions measured by an
imperfect homodyne detector are smeared-out by a convolution with a
Gaussian function 
\cite{VogeGrabPRA93,LeonPaulPRA93}. Evaluation of the parity
operator, or equivalently, the Wigner function at the phase space
origin requires application of a deconvolution procedure, which
enormously amplifies the statistical error \cite{LeonPaulJMO94}. 

\section{Consequences for quantum state measurement}
\label{Sec:Consequences}

We have presented a complete statistical analysis of determining
quantum observables in optical measurement schemes based on
photodetection. We have derived an exact expression for the generating
function characterizing statistical moments of the reconstructed
observables, which, in particular, provides formulae for statistical errors
and correlations between the determined quantities. 
These general results have been applied to the detection of
phase-independent properties of a single light mode using two schemes:
direct photon counting, and random phase homodyne detection. This study
has revealed difficulties related to the completeness of the
reconstructed information on the quantum state: in some cases the
parity observable, which is a well-behaved bounded operator,
effectively cannot be evaluated from the reconstructed data due to
the exploding statistical error.

We have recalled that the parity operator is directly related to the
value of the Wigner function at the phase space origin.  Thus, our
example can also be interpreted as a particular case of the
transformation between two representations of the quantum state: in
fact, we have considered evaluation of the Wigner function at a
specific point $(0,0)$ from the relevant elements of the density matrix
in the Fock basis.  Therefore, our discussion exemplifies subtleties
related to the transition between various quantum state
representations, when we deal with data reconstructed in experiments.
Though a certain representation can be determined with the statistical
uncertainty which seems to be reasonably small, it effectively cannot
be converted to another one due to accumulating statistical errors.

The presented study suggests that the notion of completeness in quantum
state measurements should inherently take into account the statistical
uncertainty. From a theoretical point of view, the quantum state can be
characterized in many different ways which are equivalent as long as
expectation values of quantum operators are known with perfect
accuracy.  In a real experiment, however, we always have to keep in
mind the specific experimental scheme used to perform the measurement.
This scheme defines statistical properties of the reconstructed
quantities, and may effectively limit the available information on the
quantum state. Determination of a family of observables does not
automatically guarantee the feasibility of reconstructing the
expectation value of an arbitrary well behaved operator. Reconstruction
of any observable should be preceded by an analysis how significantly
the final result is affected by 
statistical noise corrupting the raw experimental data.

\chapter{Conclusions}
\label{Chap:Conclusions}

\markright{CHAPTER \thechapter . \uppercase{Conclusions}}

In this thesis, we have developed a direct method for measuring
quasidistribution functions of light. Quasidistribution functions provide
a complete characterization of the quantum state in the form which is
analogous to a classical phase space distribution. We have shown that
quasidistributions, in particular the Wigner function, can be determined
using an optical scheme based on photon counting. We have reported an
experimental realisation of this scheme, and presented measurements of
the Wigner function for several classical-like states.

There is a wide range of problems arising from a closer look at various
aspects of the proposed scheme. We have discussed the deleterious
effects of imperfect detection, and shown that they cannot be in general
compensated in numerical processing of the experimental data. Further,
we have developed a general multimode theory of the scheme, and
extended the principle of the measurement to multimode quasidistributions.
The developed multimode theory has been a useful tool in studying the
role of experimental imperfections. We have also seen that in order to analyse
the feasibility of the measurement, it is necessary to take into
account the statistical uncertainty. We have derived estimates for the
statistical error, and shown that in some cases the statistical noise
effectively limits available information on the quantum state.

The field of quantum state measurement is continuously
developing. Currently, an interesting direction of research is
related to fundamental aspects of retrieving information on the
quantum state. Generally, our source of information on the quantum
state is the result of a measurement performed on an ensemble of a
finite number of copies. From these data we want to infer the quantum
state of the ensemble. The inference can be performed in different
ways, based on various statistical methodologies. Throughout this
thesis, we have used the linear approach, where quantum probability
distributions are estimated by relative frequency histograms. This is
the most straightforward and so far the most commonly used strategy,
but it is not necessarily the most efficient one. Recently, other
approaches to data processing have been proposed, based on the
maximum entropy principle \cite{BuzeAdamPRA96}, the least-squares
inversion \cite{OpatWelsPRA97}, and the maximum-likelihood estimation
\cite{HradPRA97,BanaPRA98,BanaPRA99,%
BanaDAriPreprint}.  These methods enhance the amount
of information which can be obtained from a realistic measurement. From
a more fundamental point of view, one may ask what is the ultimate bound
for estimating the quantum state from finite ensembles. This question
can be reformulated as a problem of designing the optimal quantum state
measurement on finite ensembles. Such a problem is in general extremely
difficult, and so far it has been studied only in selected yet highly
nontrivial cases \cite{MassPopePRL95,DerkBuzePRL98,VidaLatoPRA99}. The
optimal strategy has been shown to be of the form of a collective
measurement performed jointly on all the available copies.

The research on quantum state measurement significantly contributes
to the understanding of the foundations of quantum theory. Quantum
mechanics still unveils surprising consequences of its principles.
An important recent example is the observation that quantum mechanics
opens up completely new ways of processing, storing, and transmitting
information. Exploration of these these possibilities can result in novel
information technologies, such us quantum computing \cite{DeutEkerPW98}
and quantum cryptography \cite{HughAldeCoP95,PhoeTownCoP95}. Some issues
in quantum state measurement are shared with the quantum information
theory, for example methods of quantum estimation and the effect of
decoherence. On the other hand, quantum information technologies need
carefully prepared quantum systems, and measuring the quantum state
is here an indispensable diagnostic tool. With no doubt, the field of
quantum state measurement plays a prominent role at the forefront of
contemporary physics.

\end{document}